\documentclass[twocolumn,prb,showpacs,preprintnumbers,amsmath,amssymb,superscriptaddress,bibliography]{revtex4-1}
\usepackage[dvipdfmx]{graphicx}
\usepackage{setspace}
\usepackage{color}
\usepackage{ulem}
\graphicspath{{./figures/}}

\usepackage{amsmath,amsthm,amssymb}
\usepackage{mathrsfs}
\usepackage{braket,comment}
\usepackage{bm}
\usepackage[colorlinks,linkcolor=blue,citecolor=blue,urlcolor=blue]{hyperref}

\newcommand{\br}{{\bf r}}

\newcommand{\bA}{{\bm A}}

\newcommand{\bT}{{\bf T}}

\newcommand{\bL}{{\bm L}}

\newcommand{\hc}{\hat{c}}

\newcommand{\bd}{{\bf d}}

\newcommand{\hH}{\hat{H}}

\newcommand{\bh}{{\bf h}}
\newcommand{\hn}{\hat{n}}

\newcommand{\bU}{{\bf U}}

\newcommand{\beps}{\boldsymbol{\epsilon}}
\newcommand{\brho}{\boldsymbol{\rho}}

\newcommand{\hbpsi}{\hat{\boldsymbol{\psi}}}

\newcommand{\bk}{{\bm k}}

\newcommand{\eqq}[1]{\begin{align} #1 \end{align}}

\begin{document}
\title{Efficient Control of High Harmonic Terahertz Generation in Carbon Nanotubes using the Aharonov-Bohm Effect}

\author{Yuta Murakami}
\affiliation{Center for Emergent Matter Science, RIKEN, Wako, Saitama 351-0198, Japan}
\author{Kohei Nagai}
\affiliation{Department of Physics, Tokyo Institute of Technology, Meguro, Tokyo 152-8551, Japan}
\author{Akihisa Koga}
\affiliation{Department of Physics, Tokyo Institute of Technology, Meguro, Tokyo 152-8551, Japan}
\date{\today}

\begin{abstract} 
We show that high-harmonic generation (HHG) in carbon nanotubes (CNTs) can be efficiently controlled  using the Aharonov-Bohm (AB) effect.
When a static magnetic field ($B$) is applied along the tube, electronic wave functions acquire complex phases along the circumferential direction (the AB effect), which modifies the band structure.
When the magnetic field is applied to metallic CNTs, which can be regarded as one-dimensional massless Dirac systems, realistic values of $B$ lead to a nonzero gap in the THz regime.
We demonstrate that such change from gapless to gapped Dirac systems drastically increases the HHG intensity in the THz regime.
In the gapless Dirac system, the velocity of each electron never changes under the electric field, and thus there is no HHG.
On the other hand, the gap opening activates both the interband and itraband currents, which strongly contribute to HHG.
Our work demonstrates a unique way to manipulate HHG in nanotubes by tuning electronic wave functions using the magnetic field and the tube structure.
\end{abstract}

\maketitle

The recent development of strong lasers in the terahertz (THz) and mid-infrared regimes enables us to study various nonlinear and nonequilibrium phenomena in condensed matter~\cite{Kruchinin2018,Giannetti2016review,Basov2017review,Cavalleri2018review,Oka2019review,Sentef2021nonthermal,Koshihara2022review}.
Among them, high-harmonic generation (HHG), where higher harmonics of the laser injected to materials are radiated, is a fundamental and technologically important example~\cite{Corkum2007,Krausz2009RMP,Ghimire2019}.
HHG was originally studied in gas systems~\cite{Ferray_1988,Corkum1993PRL,Lewenstein1994}. Recently, HHG in condensed matter has been observed, and the scope of the HHG research is widely extended to conventional semiconductors, band insulators~\cite{Ghimire2011NatPhys,Schubert2014,Vampa2014PRL,Luu2015,Vampa2015Nature,Langer2016Nature,Hohenleutner2015Nature,Ndabashimiye2016,Otobe2016,Tancogne-Dejean2017,Ikemachi2017,Liu2017,You2017,Kaneshima2018,Sekiguchi2022PRB}, topological materials~\cite{Giorgianni2016,Yoshikawa2017Science,Hafez2018,Silva2019,Chacon2020PRB,Matsunaga2020PRL,Schmid2021,Baykusheva2021} and strongly correlated systems~\cite{Silva2018NatPhoton,Murakami2018PRL,Tancogne-Dejean2018,Ishihara2020,Murakami2021PRB,Orthodoxou2021,Udono2022PRB,Bionta2021PRR,Shao2022PRL,Hansen2022PRA,Uchida2022PRL,Murakami2022PRL,Hansen2022,Granas2022PRR}. 

Condensed matter is considered as an interesting playground of the HHG research, at least, for two good reasons.
Firstly, the periodic potential from atoms, which is absent in gaseous systems, can lead to peculiar motion of electrons under strong electric field.
An interesting example is the Dirac system, which is characterized by the non-parabolic band dispersion.
Due to the nonlinear dynamics of electrons in the vicinity of the Dirac node, such systems are theoretically expected to exhibit large nonlinearity~\cite{Slepyan1999PRA,Mikhailov_2007,Wright2009,Dignam2014PRB,Dignam2015PRB,Sipe2015PRB,Chizhova2017PRB,Dixit2021PRB,Sato2021PRB,Baykusheva2021PRA}.
HHG was experimentally studied for two-dimensional (2D) Dirac systems realized in graphene~\cite{Yoshikawa2017Science,Hafez2018} and in the surface of topological insulators~\cite{Giorgianni2016,Schmid2021,Baykusheva2021}, and for the 3D Dirac system in Cd$_3$As$_2$~\cite{Matsunaga2020PRL,Kovalev2020}.
It was reported that the efficiency of HHG is indeed large compared to other systems~\cite{Hafez2018,Matsunaga2020PRL,Schmid2021}, although the origin of HHG is not necessary processes theoretically expected~\cite{Paul_2013,Hafez2018}. 
The second reason is the controllability of HHG.
Generally, properties of condensed matter can be tuned by system parameters such as temperature and doping level,
which can be also applicable to HHG.
For example, peculiar HHG dependence on crystal orientations~\cite{Ghimire2011NatPhys,You2017}, electronic phases~\cite{Silva2018NatPhoton,Uchida2022PRL,Murakami2022PRL,Bionta2021PRR,Shao2022PRL}, physical pressure~\cite{Tamaya2021PRB} and doping~\cite{Nishidome2020,Javier2020PRR,Hansen2022} has been proposed theoretically or been observed experimentally.

\begin{figure}[t]
  \centering
    \hspace{-0.cm}
    \vspace{0.0cm}
\includegraphics[width=80mm]{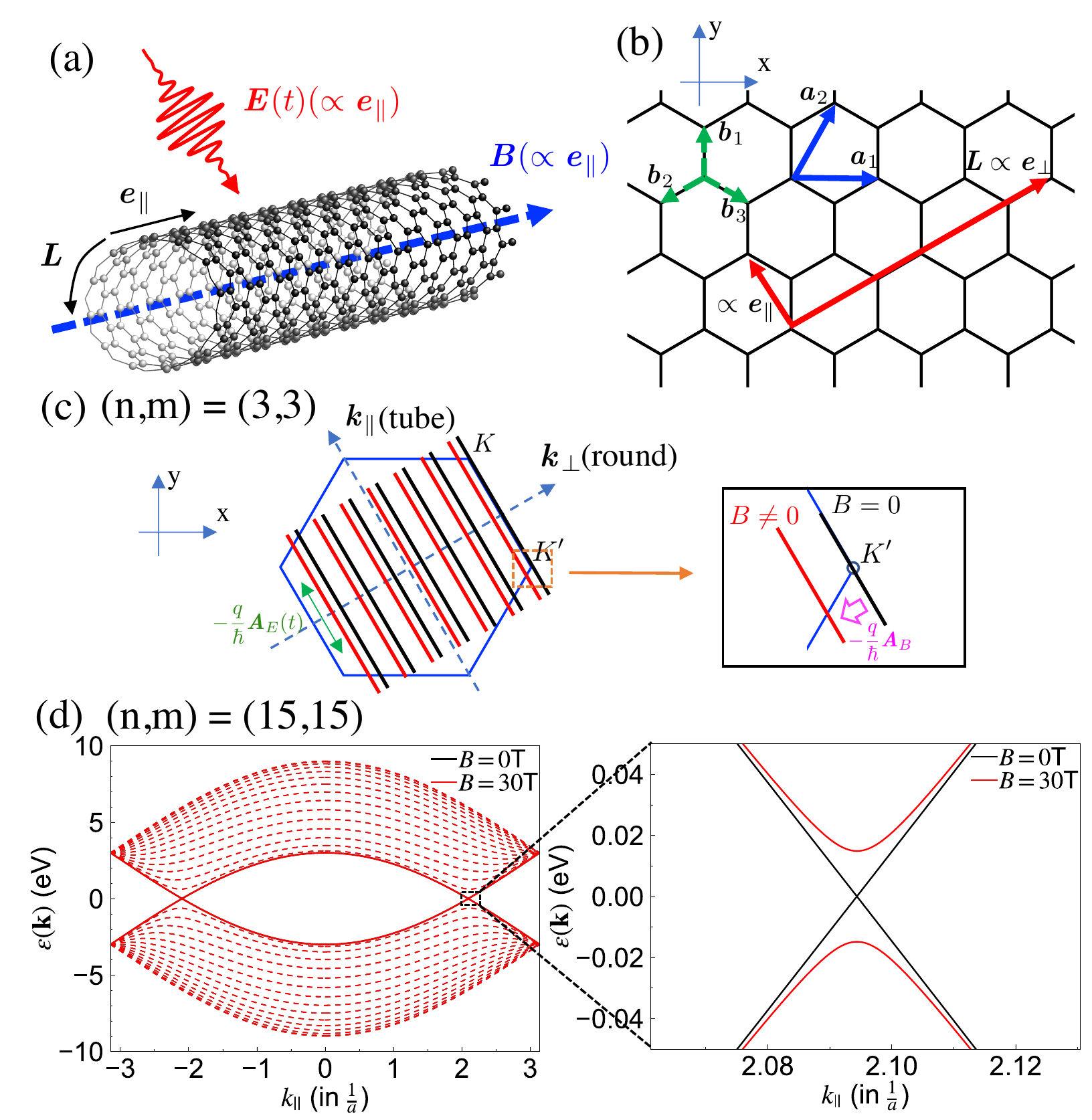} 
  \caption{(a) A CNT with the static magnetic field (${\bm B}=B{\bm e}_\parallel$) and the AC electric field (${\bm E}(t)=E(t){\bm e}_\parallel$) along the tube.
  (b) A graphite sheet. ${\bm b}_i$ represents the vector from a site of the $A$ sublattice to neighboring sites of the $B$ sublattice. ${\bf a}_1$ and ${\bf a}_2$ are primitive translation vectors. 
  ${\bm L}\equiv n{\bf a}_1 + m{\bf a}_2$($\propto{\bm e}_\perp$) is the chiral vector charactering CNTs. $ {\bm e}_\parallel$ and $ {\bm e}_\perp$  are unit vectors of the tube and circumferential directions.
(c) Correspondence of the momentum of an armchair CNT and that of graphene. The blue hexagon is the BZ of graphene, where $K$ and $K'$ are the Dirac points.
  (d) The band structure of the armchair CNT with $(n,m)=(15,15)$ with and without the static magnetic field. The sold line indicates the bands closest to $K$ and $K'$, which form 1D Dirac systems.}
  \label{fig:schematic}
\end{figure}

In this work, we propose a unique way to efficiently control HHG in carbon nanotubes (CNTs) combining physics of the Dirac system and the Aharonov-Bohm (AB) effect, see Fig.~\ref{fig:schematic}(a).
The proposed setup provides an ideal platform to study HHG in Dirac systems, systematically. 
When the static magnetic field ($B$) is applied along the tube, electrons in the CNT gain complex phases along the circumferential direction,
which modifies the band structure (the AB effect)~\cite{Ando1993JPSJ,Ando1994,Wei2004Science,Matsunaga2008PRL}. 
In particular, metallic CNTs, which can be regarded as 1D massless Dirac systems, become massive Dirac systems when $B\neq0$.
This change qualitatively modifies the electron dynamics, and drastically enhances HHG.

In the dipolar gauge~\cite{Li2020PRB,Michael2021PRB,Murakami2022PRB}, the tight-binding Hamiltonian for the CNT with the magnetic field and the AC electric field ($E(t)$) along the tube is expressed as
\eqq{
\hH(t) = -t_{\rm hop} \sum_{\langle ij\rangle} e^{i\frac{q}{\hbar}{\bm A}(t)\cdot {\bf r}_{ij}} \hc^\dagger_{i} \hc_{j} -\mu \sum_i \hn_i ~\label{eq:Hamiltonian}.
}
$\hc^\dagger_{i}$ is the creation operator of an electron at the $i$th site and we neglect the spin degrees of freedom.
$\langle ij\rangle$ indicates the nearest neighbor sites, $t_{\rm hop}$ is the hopping parameter, $q$ is the charge of an electron, $\hbar$ is the Dirac constant, $\mu$ is the chemical potential and $\hn_i = \hc^\dagger_i \hc_i$.
The model is defined on the graphite sheet imposing  the periodic boundary condition with respect to the chiral vector ${\bm L}$, which is characterized by the chiral index $(n,m)$ as ${\bm L}=n{\bm a}_1 + m{\bm a}_2$, see Fig.~\ref{fig:schematic}(b)~\cite{Hamada1992PRL}.
The effects of the magnetic and electric fields are included via the Peierls phase through the vector potential ${\bm A}(t)=\bA_B+\bA_E(t)$.
The contribution from the static magnetic field is $\bA_B =  A_B {\bm e}_\perp$ and that from the electric field is $\bA_E(t) = A_E(t) {\bm e}_\parallel$.
$ A_B=\frac{B\cdot |{\bm L}|}{4\pi}$ and $E(t) = -\partial_t A_E(t)$. $ {\bm e}_\parallel$ and $ {\bm e}_\perp$  are unit vectors of the tube and circumferential directions, respectively, see Fig.~\ref{fig:schematic}(b).
In the following, we focus on the armchair CNT with the chiral index $(n,m)$ with $n=m$.

The Fourier transform of the Hamiltonian \eqref{eq:Hamiltonian} is  $\hH(t)=\sum_\bk \hbpsi^\dagger_\bk {\bf h}(\bk(t)) \hbpsi_\bk $,
where
\eqq{
 {\bf h}(\bk) = 
\begin{bmatrix}
-\mu  & F( \bk) \\
 F^*( \bk) & -\mu
\end{bmatrix},  \label{eq:Graphene_TB_k}
}
and  $\bk(t) \equiv \bk-\frac{q}{\hbar}{\bm A}(t)$, see the Supplementary Material (SM) 
\footnote{See Supplemental Material at  [URL will be inserted by publisher] for the details of the model and numerical method; the derivation of Eqs.~\eqref{eq:J_t} and ~\eqref{eq:sig_3_2}; the doping effects; the dependence of the excitation frequency; the dependence on the chiral index; the effects of the relaxation times. The Supplemental Material also contains Refs.~\cite{Khosravi2009,Dimitrovsk2017,Dvuzhilova2021,Wilhelm2021PRB}.}
 for details. Here, $\hbpsi^\dagger_\bk = [\hc^\dagger_{\bk A},  \hc^\dagger_{\bk B}]$, and $\hc^\dagger_{\bk A}$ and $\hc^\dagger_{\bk B}$ are the Fourier transform of $\hc^\dagger_{i}$ on the A and B sublattices, respectively.
 $F( \bk)  =-t_{\rm hop}[e^{i\bk\cdot {\bm b}_1} + e^{i\bk\cdot {\bm b}_2} + e^{i\bk\cdot {\bm b}_3}]$, see Fig.~\ref{fig:schematic}(b) for  ${\bm b}_i$.
The Hamiltonian~\eqref{eq:Graphene_TB_k} is the same as that of graphene, and the set of $\bk_B(\equiv \bk-\frac{q}{\hbar}{\bm A}_B)$ for the CNT corresponds to some lines along $\bk_\perp$ on the BZ of graphene, see Fig.~\ref{fig:schematic}(c). Note that $\bk(t)=\bk_B-\frac{q}{\hbar}{\bm A}_E(t)$, which moves along $\bk_{\parallel}$ in time.
For $B=0$,  $\bk_B$'s for the CNT pass the Dirac points $K$ and $K'$ of graphene, hence the CNT bands host gapless Dirac bands, see Figs.~\ref{fig:schematic}(c) and (d).
Meanwhile, for $B\neq 0$, $\bk_B$'s are shifted by $-\frac{q}{\hbar}\bA_B$ and do not passes the Dirac points. Hence the CNT bands host gapful Dirac bands.
Around the Dirac points, the Hamiltonian~\eqref{eq:Graphene_TB_k} is effectively described by the Dirac model, see Fig.~\ref{fig:schematic} (d) and SM~\cite{Note1} for details.
In our setup, the dynamics of electrons around the Dirac points leads to HHG, which is captured by this model.

Diameters of single-wall CNTs are typically $0.7\sim 2$nm~\cite{Yanagi_2008}. The static magnetic field applied to CNTs in optical experiments can be several tens of Tesla~\cite{Wei2004Science}.
In the following, we set $t_{\rm hop}=3$eV and $a=0.246$nm and focus on the CNT with $(n,m)=(15,15)$, whose diameter is $2.04$nm. The band gap for $B=30$T becomes $29.7$meV ($\simeq 7.5$THz), see Fig.~\ref{fig:schematic} (d).
We mainly focus on half filling ($\mu=0$) and apply the gaussian electric field described by $A_E(t)=\frac{E_0}{\hbar\Omega} \exp[-\frac{(t-t_0)^2}{2\sigma^2}] \sin(\Omega(t-t_0))$. 
We set the electric-field frequency  $\hbar \Omega =10$meV ($\simeq 2.5$THz), the pulse center $t_0=3.95$ps and the pulse width  $\sigma={658}$fs, in the main text.

We simulate the time evolution of the tight-binding and Dirac models using the semiconductor Bloch equation (SBE) ~\cite{Huttner2017,Yue2022tutorial}
\eqq{
\partial_t \brho_\bk(t) &= \frac{i}{\hbar}[\brho_\bk(t),\bh(\bk(t))]  - \frac{\brho_\bk(t) - \brho_{{\rm eq},\bk(t)}}{T_1} \nonumber\\
& + \Bigl(\frac{1}{T_1}-\frac{1}{T_2}\Bigl) {\bf U}(\bk(t))\brho^{\rm H}_{{\rm off},\bk}(t){\bf U}^\dagger(\bk(t)), \label{eq:vN_eq_T1T2_DW}
}
where $\brho_\bk(t)$ is the single-particle density matrix consisting of $\rho_{\alpha\beta,\bk}(t) = \langle \hc^\dagger_{\bk \beta}(t) \hc_{\bk \alpha}(t)\rangle$ with $\alpha,\beta = A,B$.
The last two terms express the relaxation processes within  the relaxation-time approximation.
$\brho_{{\rm eq,}\bk}$ is the equilibrium density matrix for the initial temperature ($T$). ${\bf U}(\bk)$ is a unitary matrix diagonalizing ${\bf h}(\bk)$ as ${\bf U}^\dagger(\bk){\bf h}(\bk) {\bf U}(\bk) = {\boldsymbol \epsilon}(\bk)$ 
with $ {\boldsymbol \epsilon}(\bk)= {\rm diag}[\epsilon_0(\bk),\epsilon_1(\bk)]$.
It offers the descrition in the band (Houston) basis. In this basis, the density matrix becomes $\brho^{\rm H}_{\bk}(t)\equiv {\bf U}^\dagger(\bk(t))\brho_\bk(t) {\bf U}(\bk(t))$, whose off-diagonal components are $\brho^{\rm H}_{{\rm off},\bk}(t)$~\cite{Murakami2022PRB}. 
The diagonal components of $\brho^{\rm H}_\bk$ indicate the occupation of the bands, and the off-diagonal components indicate the interband hybridization.
Thus, $T_1$ represents the relaxation time of the band occupation, while $T_2$ represents the dephasing time of the interband coherence.
We set $T_1=98.8$fs and $T_2=19.8$fs to be consistent with previous studies for graphene and CNTs ~\cite{Habenicht2007,George2008,Tani2012PRL,Heide2021,Sato2021PRB}.

The light radiation originates from the oscillation of electric polarization (or corresponding current)~\cite{Jackson1998Book}.
Since the current along the circumferential direction does not yield polarization, 
the HHG spectrum (normalized by the number of sites) is evaluated from $J_\parallel(t)$ (the current along ${\bm e}_{\parallel}$) as $I_{\rm HHG}(\omega) = |\omega J_\parallel(\omega)|^2$.
Here, $J_\parallel(\omega) = \int dt e^{i\omega t} J_\parallel(t)$, and $J_\parallel(t)$ is evaluated from $\rho_{\alpha\beta,\bk}(t)$ using the current operator $\hat{J}_\parallel(t)\equiv -\frac{1}{N}\frac{\delta \hH(t)}{\delta A_E(t)}$.
 $2N$ is the number of atoms (sites) for the CNT.
In the conditions studied here, HHG mainly originates from the bands closest to the Dirac points, i.e. solid lines in Fig.~\ref{fig:schematic}(d).

 \begin{figure}[t]
  \centering
    \hspace{-0.cm}
    \vspace{0.0cm}
\includegraphics[width=75mm]{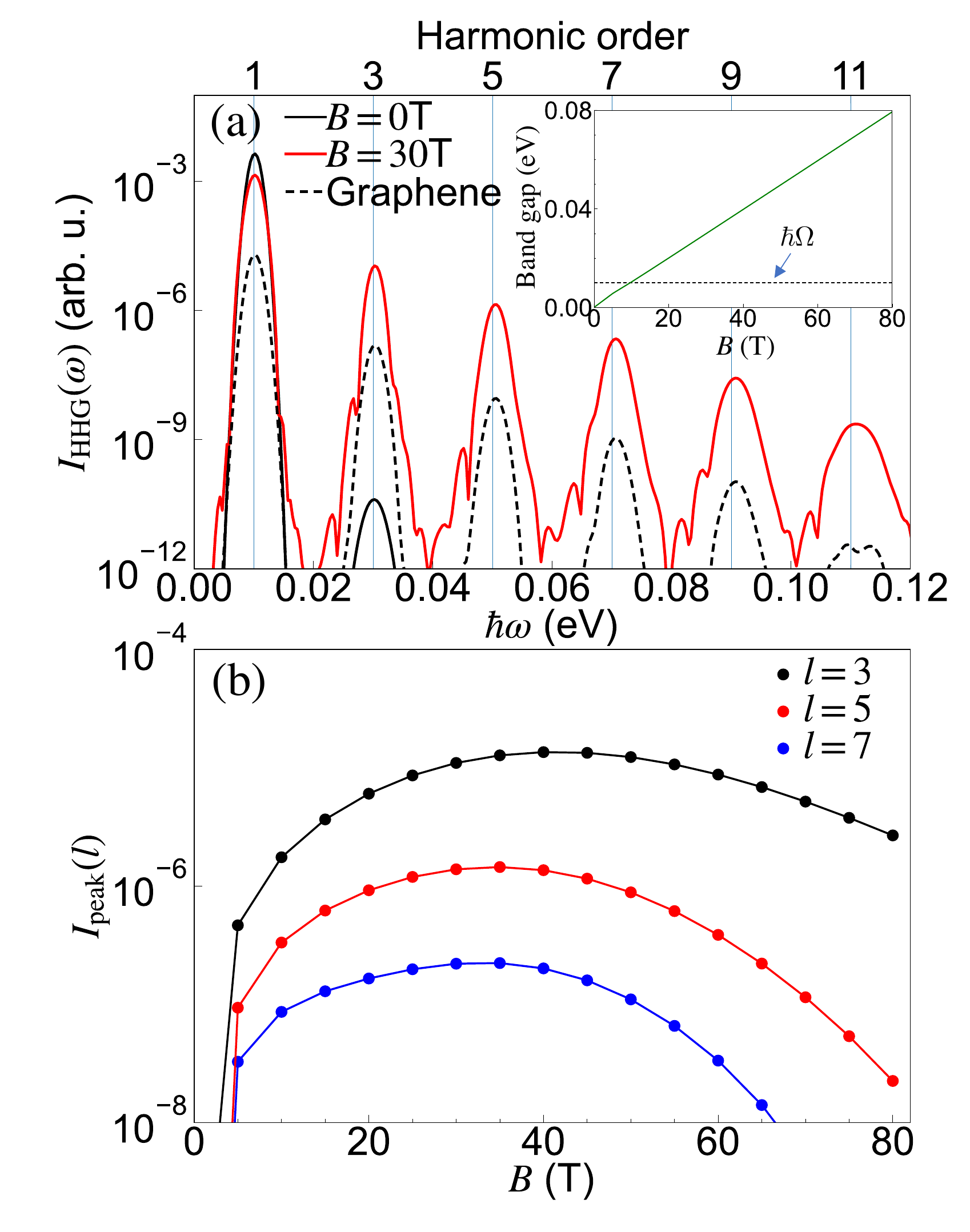} 
  \caption{(a) HHG spectra $I_{\rm HHG}$ of the armchair CNT with and without the static magnetic field simulated with the tight-binding model. $I_{\rm HHG}$  for graphene is also shown. 
  The inset show the bandgap induced by the magnetic field $B$. (b)The intensity at the HHG peaks $I_{\rm peak}(l)$ as a function of $B$.
  We set the chiral index $(n,m)=(15,15)$, $\mu=0$, $T=11.6$K, $T_1=98.8$fs and $T_2=19.8$fs. The parameters of the electric field are $\hbar\Omega = 10$meV and $E_0=30$kV/cm.} 
  \label{fig:HHG_total}
\end{figure}

 \begin{figure}[t]
  \centering
    \hspace{-0.cm}
    \vspace{0.0cm}
\includegraphics[width=85mm]{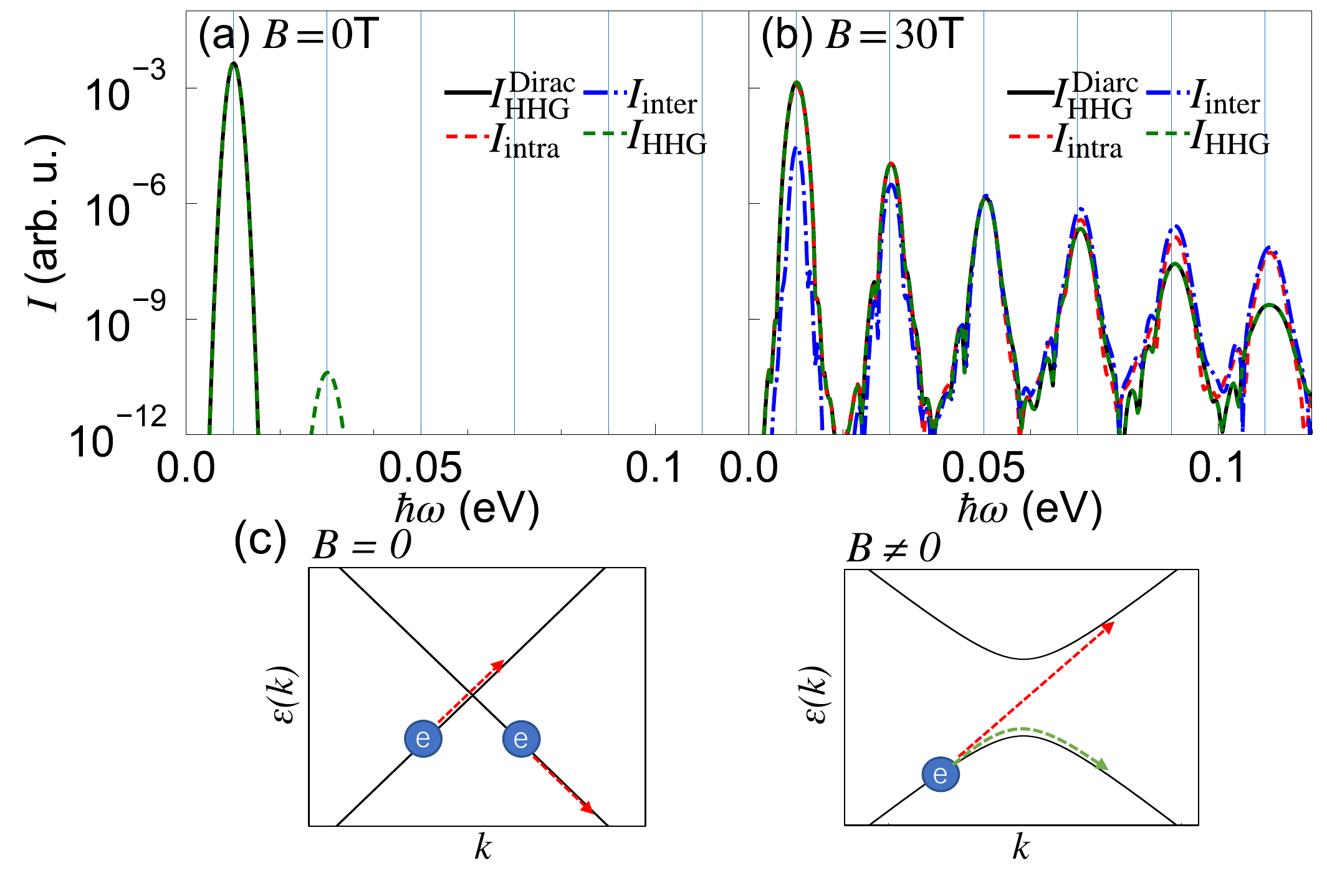} 
  \caption{(a)(b) Comparison of the HHG spectra of the armchair CNT obtained in different ways for (a) $B=0$T and (b) $B=30$T. $I_{\rm HHG}(\omega)$ is evaluated from the tight-binding model, while  $I_{\rm HHG}^{\rm Dirac}(\omega)$, $I_{\rm intra}(\omega)$ and $I_{\rm inter}(\omega)$  are evaluated from the Dirac model considering only bands closest to $K$ or $K'$. The parameters of the system and the electric field are the same as Fig.\ref{fig:HHG_total}. (c) The dynamics of electrons with and without the magnetic field. The momentum shift originates from the intraband acceleration by the electric field.
  }
  \label{fig:HHG_dec_total}
\end{figure}

Figure~\ref{fig:HHG_total}(a) shows $I_{\rm HHG}$ of the armchair CNT with and without the magnetic field obtained from the tight-binding model.
Only odd harmonics components appear in $I_{\rm HHG}(\omega)$, because the system~\eqref{eq:Hamiltonian} is symmetric under a time-space operation, $t\rightarrow t+\frac{\pi}{\Omega}$ and ${\bf e}_\parallel \rightarrow -{\bf e}_\parallel$ during the time-periodic excitation, see Ref.~\cite{Cohen2019NatCom} for general argument.
When $B=0$ the intensity for $l(\geq3)$th harmonics is very small.
When $B\neq 0$ the HHG intensity for $l\geq 3$ is strongly enhanced by more than several orders of magnitude.
Figure~\ref{fig:HHG_total}(b) shows the intensity at the HHG peaks $I_{\rm peak}(l)$ as a function of $B$ for $l=3,5,7$.
$I_{\rm peak}(l)$ is defined as the maximum value of $I_{\rm HHG}(\omega)$ for $\omega \in [l\Omega-\Omega,l\Omega+\Omega]$. 
$I_{\rm peak}(l)$ is non-monotonic against $B$, since a gap too large compared to $\Omega$  suppresses the excitation and thus HHG, see the inset of Fig.~\ref{fig:HHG_total}(a).
Namely, there exists the optimal value of $B$ maximizing $I_{\rm HHG}(\omega)$ for a given $E_0$.
Furthermore, the HHG intensity in the CNT per atom for this optimal $B$ is much larger than that in graphene per atom, see Fig.~\ref{fig:HHG_total}(a).
This is understood as follows.
In the CNT, we can tune $B$ such that the bands closest to the Dirac points yield HHG most efficiently for a given $E_0$, and the contributions from the rest bands are negligible.
Meanwhile, in graphene, which is identical to a CNT with $n,m\rightarrow \infty$, the number of bands along $\bk_\parallel$ contributing to HHG increases. 
However, the contribution from each band are not optimal for a given electric-field condition, and they may interfere and cancel each other on average. 
We also note that the drastic enhancement of HHG with $B$ can be observed even for doped CNTs or for higher-frequency excitations, see SM~\cite{Note1}.

The efficiency of the enhancement depends on the chiral index $n(=m)$.
Firstly, the momentum shift due to the AB effect is $\frac{q}{\hbar}A_B$, which is proportional to $B$ and $L(\equiv |{\bm L}|\propto n)$.
Therefore, in order to realize a given value of $A_B$ (or the gap), smaller $B$ is required for larger $n$, see SM~\cite{Note1} for examples.
Thus, as far as only the bands closest to the Dirac points are involved in HHG, larger $n$ is favorable for the efficient control of HHG.
Meanwhile, with increasing $n$, the number of bands along $\bk_\parallel$ increases, and many bands start to contribute to HHG.
Then, the impact of the AB effect on HHG becomes less prominent.
In the extreme limit of $n\rightarrow \infty$ (graphene), the AB effect has no impact on physical properties.
%

 \begin{figure}[t]
  \centering
    \hspace{-0.cm}
    \vspace{0.0cm}
\includegraphics[width=85mm]{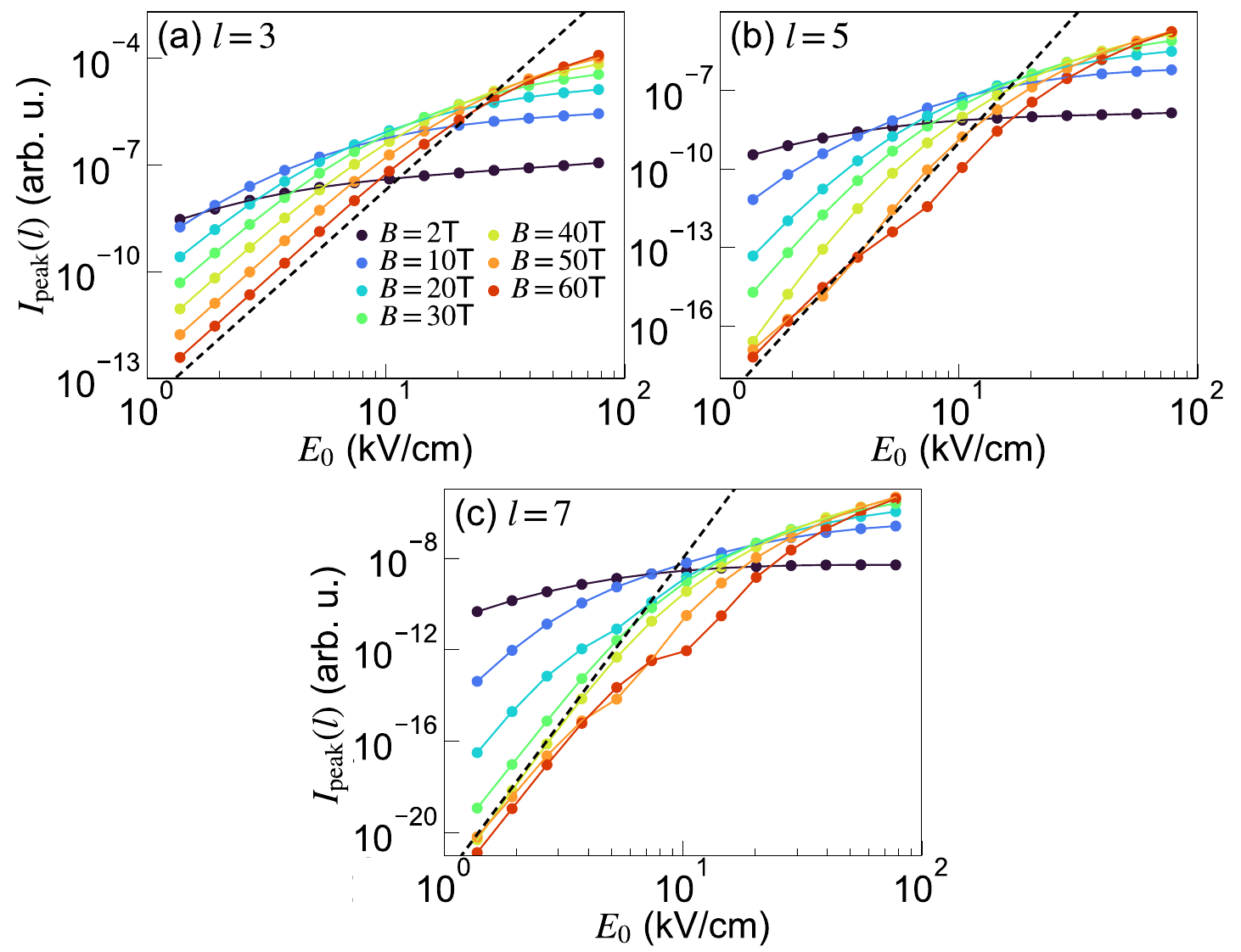} 
  \caption{The intensity at the HHG peaks $I_{\rm peak}(l)$ for the armchair CNT as a function of the strength of the AC electric field $E_0$ for (a) $l=3$, (b) $l=5$ and (c) $l=7$.
  The tight-binding model is used. The dashed lines are proportional to $(E_0)^{2l}$ for each panels. The parameters of the system and the electric field are the same as Fig.\ref{fig:HHG_total}.} 
  \label{fig:HHG_Edep}
\end{figure}

To understand the origin of the enhancement of HHG with the magnetic field, we compare the HHG intensity evaluated from the tight-binding model ($I_{\rm HHG}$) and that from the Dirac model ($I_{\rm HHG}^{\rm Dirac}$), see Fig.~\ref{fig:HHG_dec_total}(a)(b).
We only consider the contributions from the bands closest to $K$ and $K'$ for the Dirac model.
The results from the Dirac model match well those from the tight-binding model both for $B=0$ and $B=30$T.
The only exception is  the 3rd harmonic generation at $B=0$, which originates from the nonlinear part of the band dispersion.
Namely, the 1D gapless Dirac model shows no HHG at $B=0$.
Simply speaking, this is because the transition of electrons between bands in the 1D gapless Dirac system under electric field is always perfect and electrons never change the velocity, see Fig.~\ref{fig:HHG_dec_total}(c).
Without the relaxation, this leads to $J_{||}(t) \propto A_{E}(t)$~\cite{Ishikawa2010PRB,Zurron-Cifuentes:20}.
If we take into account the relaxation, we obtain 
\eqq{
\partial_t J_{||}(t) = \frac{1}{n} \frac{t_{\rm hop}}{\hbar}\sigma_0E(t) -\frac{1}{T_1}  J_{||}(t)~\label{eq:J_t},
}
from the SBE~\eqref{eq:vN_eq_T1T2_DW}, where $\sigma_0 =  \frac{\sqrt{3}a^2q^2}{2\pi\hbar}$, see SM~\cite{Note1}.
This is a linear differential equation, leading to no HHG.
The last term of Eq.\eqref{eq:J_t}, which we call $[\partial_t J_{\parallel}(t)]_{\rm corr}$, originates from the relaxation term of the SBE.
The present form originates from the relaxation-time approximation in Eq.~\eqref{eq:vN_eq_T1T2_DW}.
In principle, if $[\partial_t J_{\parallel}(t)]_{\rm corr}$ depends on $J_{\parallel}$ nonlinearly, the 1D gapless Dirac system can show HHG. 
Still, in a previous experiment of HHG in CNTs~\cite{Nishidome2020}, metallic CNTs show very weak HHG, which is consistent with the consequence from Eq.~\eqref{eq:J_t}.

To obtain further insight, we evaluate the contribution from the intraband current ($I_{\rm intra}=|\omega J_{\parallel,\rm intra}(\omega)|^2$) and  that from the interband current ($I_{\rm inter}=|\omega J_{\parallel,\rm inter}(\omega)|^2$) using the Dirac model. We define the intraband current as $J_{\parallel,\rm intra}(t)\equiv \frac{q}{\hbar N} \sum_\bk {\rm Tr}[\brho^H_\bk(t)\partial_{k_\parallel}{\boldsymbol \epsilon}(\bk(t))]$
and the interband current as $J_{\parallel,\rm inter}(t)\equiv J_{\parallel}(t)-J_{\parallel,\rm intra}(t)$. 
The former is associated with the change in the band occupation, while the latter is associated with that of the interband polarization (the off-diagonal component of $\brho^H_\bk(t)$), see SM~\cite{Note1} for details.
When $B=0$, the perfect transition between the conduction and valence bands happens without inducing the interband polarization, leading to $I_{\rm HHG} = I_{\rm intra}$, see Fig.~\ref{fig:HHG_dec_total}(c).
Meanwhile, when $B\neq0$, the gap opens and the dipole moments between the bands become finite, which allows electrons to stay in the same band and tunnel to the other band, see Fig.~\ref{fig:HHG_dec_total}(c). Thus, both the intraband and interband currents are activated to contribute to HHG, see Fig.~\ref{fig:HHG_dec_total}(b).
In the present case, the gap and the excitation frequency are comparable, which makes the contributions from these currents comparable.
The HHG contributions from $J_{\parallel,\rm intra}(t)$ and $J_{\parallel,\rm inter}(t)$  cancel each other due to their interference.

Finally, we discuss the dependence on the strength of the applied AC electric field $E_0$, see Fig.~\ref{fig:schematic}(a).
In Fig.~\ref{fig:HHG_Edep}, we show the intensity at the HHG peaks $I_{\rm peak}(l)$ as a function of $E_0$ for $l=3,5,7$.
In the perturbative regime with respect to $E_0$, we expect $I_{\rm peak}(l)\propto (E_0)^{2l}$.
(For small $B$, this perturbative behavior is recovered with $E_0$ below the range shown in Fig.~\ref{fig:HHG_Edep}.)
For all harmonics, the deviation from the perturbative regime happens at larger $E_0$ for larger $B$, i.e. larger gap.
$I_{\rm peak}(l)$ tends to be saturated due to the higher order corrections.
Because of this, the value of $B$ maximizing $I_{\rm peak}(l)$ for a given $E_0$ depends on the value of $E_0$.
Namely, when $E_0$ is small, the optimal $B$ is small, while for larger $E_0$, the optimal $B$ becomes large.
We note that these properties are insensitive to the choice of $T_1$ and $T_2$, see SM~\cite{Note1}.

One interesting observation is that the optimal value of $B$ continuously decreases with decreasing $E_0$ and it becomes a few Tesla for $E_0\sim 1$ kV/cm.
This indicates that  $\sigma^{(l)}$ monotonically increases with decreasing $B$.
Here, $\sigma^{(l)}$ represents the $l$th order optical conductivity, i.e. $J_{\parallel} (l\Omega)\simeq \sigma^{(l)} (E_0)^l$.
However, this peculiar behavior turns out to be sensitive to the choice of  $T_1$ and $T_2$, see SM~\cite{Note1}.
When $T_1=T_2$, the optimal $B$ for small $E_0$ remains around $20$T, where the gap is almost twice of $\Omega$.
The crucial dependence on $T_1$ and $T_2$ can be attributed to the change in cancelation between the interaband and intraband currents.
This point can be directly seen in the analytic expression of  $\sigma^{(3)}$.
If we express contributions from $J_{\rm inter}$ and $J_{\rm intra}$ as $\sigma^{(3)}_{\rm inter}$ and $\sigma^{(3)}_{\rm intra}$,
 the leading terms in the limit of $B\rightarrow 0$ read
\eqq{
& \sigma^{(3)}_{\rm intra} \simeq -F \frac{1}{A_B^2} \frac{1}{\zeta_3}, \;\;
\sigma^{(3)}_{\rm inter} \simeq  F \frac{1}{A_B^2} \frac{1}{\xi_3}, \nonumber \\
& F\equiv i \sigma_0 t_{\rm hop} \hbar^2 \frac{1}{n} \frac{1}{ \xi_1} \Bigl(\frac{\lambda_0}{2\zeta_2} + \frac{\lambda_2}{\xi_2}\Bigl) \label{eq:sig_3_2}
}
at $T=0$.
Here $\zeta_m = m\hbar\Omega +  i \hbar/T_1 $ and $\xi_m = m\hbar\Omega +  i \hbar/T_2 $ and $\lambda_m = \int dx \frac{x^m}{\sqrt{1+x^2}^7}$.
The expression tells that $\sigma^{(3)}$ diverges for $B\rightarrow 0$ when $T_1\neq T_2$, while the divergence is absent when $T_1= T_2$ due to the cancelation between $\sigma^{(3)}_{\rm inter}$  and $\sigma^{(3)}_{\rm intra}$.
Our numerical simulation suggests that the same happens also for higher harmonics with $l\geq5$.
Namely, for $T_1 = T_2$, the contributions from the interband and intraband currents cancel each other efficiently, while for $T_1 \neq T_2$, the cancelation becomes weaker, see SM~\cite{Note1}.
Although we used $T_1=98.8$fs and $T_2=19.8$fs to be consistent with previous studies, 
to fully determine the realistic choice of $T_1$ and $T_2$ for the present energy scale, microscopic evaluation of relaxation and dephasing processes is required~\cite{Malic2011,Kemper2013NJP,stefanucci_nonequilibrium_2013,Nessi2020,ridley2022manybody,Rostami2021PRB}.

To summarize, we showed that HHG in metallic CNTs can be drastically manipulated in the THz regime using the AB effect.
Tuning the strength of the magnetic field, the HHG intensity can be maximized, where HHG per atom is more efficient than in graphene. 
Such controllability of HHG may be useful to develop an efficient HHG device in the THz regime.
In addition, with realistic parameters of CNTs and external fields, a metallic CNT under magnetic fields can be regarded as a tunable 1D Dirac system for THz excitations.
Therefore, it serves as a useful platform to systematically study HHG in Dirac systems.
Furthermore, the unique idea of manipulating HHG via the AB effect can be applied for all kinds of nanotubes including semiconducting CNTs, boron nitride nanotubes~\cite{Rubio1994PRB,Nasreen1995} and transition-metal dichalcogenide nanotubes~\cite{Tenne1992}.

\begin{acknowledgments}
We would like to acknowledge fruitful discussions with  Hiroyuki Nishidome and Kazuhiro Yanagi.
This work is supported by Grant-in-Aid for Scientific Research from JSPS, KAKENHI Grant Nos. JP20K14412, JP21H05017 (Y. M.), JP22K03525, JP21H01025, JP19H05821 (A.K.), JST CREST Grant No. JPMJCR1901 (Y. M.).
\end{acknowledgments}

\bibliography{HHG_Ref}

\begin{thebibliography}{98}%
\makeatletter
\providecommand \@ifxundefined [1]{%
 \@ifx{#1\undefined}
}%
\providecommand \@ifnum [1]{%
 \ifnum #1\expandafter \@firstoftwo
 \else \expandafter \@secondoftwo
 \fi
}%
\providecommand \@ifx [1]{%
 \ifx #1\expandafter \@firstoftwo
 \else \expandafter \@secondoftwo
 \fi
}%
\providecommand \natexlab [1]{#1}%
\providecommand \enquote  [1]{``#1''}%
\providecommand \bibnamefont  [1]{#1}%
\providecommand \bibfnamefont [1]{#1}%
\providecommand \citenamefont [1]{#1}%
\providecommand \href@noop [0]{\@secondoftwo}%
\providecommand \href [0]{\begingroup \@sanitize@url \@href}%
\providecommand \@href[1]{\@@startlink{#1}\@@href}%
\providecommand \@@href[1]{\endgroup#1\@@endlink}%
\providecommand \@sanitize@url [0]{\catcode `\\12\catcode `\$12\catcode
  `\&12\catcode `\#12\catcode `\^12\catcode `\_12\catcode `\%12\relax}%
\providecommand \@@startlink[1]{}%
\providecommand \@@endlink[0]{}%
\providecommand \url  [0]{\begingroup\@sanitize@url \@url }%
\providecommand \@url [1]{\endgroup\@href {#1}{\urlprefix }}%
\providecommand \urlprefix  [0]{URL }%
\providecommand \Eprint [0]{\href }%
\providecommand \doibase [0]{http://dx.doi.org/}%
\providecommand \selectlanguage [0]{\@gobble}%
\providecommand \bibinfo  [0]{\@secondoftwo}%
\providecommand \bibfield  [0]{\@secondoftwo}%
\providecommand \translation [1]{[#1]}%
\providecommand \BibitemOpen [0]{}%
\providecommand \bibitemStop [0]{}%
\providecommand \bibitemNoStop [0]{.\EOS\space}%
\providecommand \EOS [0]{\spacefactor3000\relax}%
\providecommand \BibitemShut  [1]{\csname bibitem#1\endcsname}%
\let\auto@bib@innerbib\@empty
\bibitem [{\citenamefont {Kruchinin}\ \emph {et~al.}(2018)\citenamefont
  {Kruchinin}, \citenamefont {Krausz},\ and\ \citenamefont
  {Yakovlev}}]{Kruchinin2018}%
  \BibitemOpen
  \bibfield  {author} {\bibinfo {author} {\bibfnamefont {S.~Y.}\ \bibnamefont
  {Kruchinin}}, \bibinfo {author} {\bibfnamefont {F.}~\bibnamefont {Krausz}}, \
  and\ \bibinfo {author} {\bibfnamefont {V.~S.}\ \bibnamefont {Yakovlev}},\
  }\href {\doibase 10.1103/RevModPhys.90.021002} {\bibfield  {journal}
  {\bibinfo  {journal} {Rev. Mod. Phys.}\ }\textbf {\bibinfo {volume} {90}},\
  \bibinfo {pages} {021002} (\bibinfo {year} {2018})}\BibitemShut {NoStop}%
\bibitem [{\citenamefont {Giannetti}\ \emph {et~al.}(2016)\citenamefont
  {Giannetti}, \citenamefont {Capone}, \citenamefont {Fausti}, \citenamefont
  {Fabrizio}, \citenamefont {Parmigiani},\ and\ \citenamefont
  {Mihailovic}}]{Giannetti2016review}%
  \BibitemOpen
  \bibfield  {author} {\bibinfo {author} {\bibfnamefont {C.}~\bibnamefont
  {Giannetti}}, \bibinfo {author} {\bibfnamefont {M.}~\bibnamefont {Capone}},
  \bibinfo {author} {\bibfnamefont {D.}~\bibnamefont {Fausti}}, \bibinfo
  {author} {\bibfnamefont {M.}~\bibnamefont {Fabrizio}}, \bibinfo {author}
  {\bibfnamefont {F.}~\bibnamefont {Parmigiani}}, \ and\ \bibinfo {author}
  {\bibfnamefont {D.}~\bibnamefont {Mihailovic}},\ }\href {\doibase
  10.1080/00018732.2016.1194044} {\bibfield  {journal} {\bibinfo  {journal}
  {Advances in Physics}\ }\textbf {\bibinfo {volume} {65}},\ \bibinfo {pages}
  {58} (\bibinfo {year} {2016})}\BibitemShut {NoStop}%
\bibitem [{\citenamefont {Basov}\ \emph {et~al.}(2017)\citenamefont {Basov},
  \citenamefont {Averitt},\ and\ \citenamefont {Hsieh}}]{Basov2017review}%
  \BibitemOpen
  \bibfield  {author} {\bibinfo {author} {\bibfnamefont {D.~N.}\ \bibnamefont
  {Basov}}, \bibinfo {author} {\bibfnamefont {R.~D.}\ \bibnamefont {Averitt}},
  \ and\ \bibinfo {author} {\bibfnamefont {D.}~\bibnamefont {Hsieh}},\ }\href
  {\doibase 10.1038/nmat5017} {\bibfield  {journal} {\bibinfo  {journal}
  {Nature Materials}\ }\textbf {\bibinfo {volume} {16}},\ \bibinfo {pages}
  {1077} (\bibinfo {year} {2017})}\BibitemShut {NoStop}%
\bibitem [{\citenamefont {Cavalleri}(2018)}]{Cavalleri2018review}%
  \BibitemOpen
  \bibfield  {author} {\bibinfo {author} {\bibfnamefont {A.}~\bibnamefont
  {Cavalleri}},\ }\href {\doibase 10.1080/00107514.2017.1406623} {\bibfield
  {journal} {\bibinfo  {journal} {Contemporary Physics}\ }\textbf {\bibinfo
  {volume} {59}},\ \bibinfo {pages} {31} (\bibinfo {year} {2018})}\BibitemShut
  {NoStop}%
\bibitem [{\citenamefont {Oka}\ and\ \citenamefont
  {Kitamura}(2019)}]{Oka2019review}%
  \BibitemOpen
  \bibfield  {author} {\bibinfo {author} {\bibfnamefont {T.}~\bibnamefont
  {Oka}}\ and\ \bibinfo {author} {\bibfnamefont {S.}~\bibnamefont {Kitamura}},\
  }\href {\doibase 10.1146/annurev-conmatphys-031218-013423} {\bibfield
  {journal} {\bibinfo  {journal} {Annual Review of Condensed Matter Physics}\
  }\textbf {\bibinfo {volume} {10}},\ \bibinfo {pages} {387} (\bibinfo {year}
  {2019})}\BibitemShut {NoStop}%
\bibitem [{\citenamefont {de~la Torre}\ \emph {et~al.}(2021)\citenamefont
  {de~la Torre}, \citenamefont {Kennes}, \citenamefont {Claassen},
  \citenamefont {Gerber}, \citenamefont {McIver},\ and\ \citenamefont
  {Sentef}}]{Sentef2021nonthermal}%
  \BibitemOpen
  \bibfield  {author} {\bibinfo {author} {\bibfnamefont {A.}~\bibnamefont
  {de~la Torre}}, \bibinfo {author} {\bibfnamefont {D.~M.}\ \bibnamefont
  {Kennes}}, \bibinfo {author} {\bibfnamefont {M.}~\bibnamefont {Claassen}},
  \bibinfo {author} {\bibfnamefont {S.}~\bibnamefont {Gerber}}, \bibinfo
  {author} {\bibfnamefont {J.~W.}\ \bibnamefont {McIver}}, \ and\ \bibinfo
  {author} {\bibfnamefont {M.~A.}\ \bibnamefont {Sentef}},\ }\href {\doibase
  10.1103/RevModPhys.93.041002} {\bibfield  {journal} {\bibinfo  {journal}
  {Rev. Mod. Phys.}\ }\textbf {\bibinfo {volume} {93}},\ \bibinfo {pages}
  {041002} (\bibinfo {year} {2021})}\BibitemShut {NoStop}%
\bibitem [{\citenamefont {Koshihara}\ \emph {et~al.}(2022)\citenamefont
  {Koshihara}, \citenamefont {Ishikawa}, \citenamefont {Okimoto}, \citenamefont
  {Onda}, \citenamefont {Fukaya}, \citenamefont {Hada}, \citenamefont
  {Hayashi}, \citenamefont {Ishihara},\ and\ \citenamefont
  {Luty}}]{Koshihara2022review}%
  \BibitemOpen
  \bibfield  {author} {\bibinfo {author} {\bibfnamefont {S.}~\bibnamefont
  {Koshihara}}, \bibinfo {author} {\bibfnamefont {T.}~\bibnamefont {Ishikawa}},
  \bibinfo {author} {\bibfnamefont {Y.}~\bibnamefont {Okimoto}}, \bibinfo
  {author} {\bibfnamefont {K.}~\bibnamefont {Onda}}, \bibinfo {author}
  {\bibfnamefont {R.}~\bibnamefont {Fukaya}}, \bibinfo {author} {\bibfnamefont
  {M.}~\bibnamefont {Hada}}, \bibinfo {author} {\bibfnamefont {Y.}~\bibnamefont
  {Hayashi}}, \bibinfo {author} {\bibfnamefont {S.}~\bibnamefont {Ishihara}}, \
  and\ \bibinfo {author} {\bibfnamefont {T.}~\bibnamefont {Luty}},\ }\href
  {\doibase https://doi.org/10.1016/j.physrep.2021.10.003} {\bibfield
  {journal} {\bibinfo  {journal} {Physics Reports}\ }\textbf {\bibinfo {volume}
  {942}},\ \bibinfo {pages} {1} (\bibinfo {year} {2022})}\BibitemShut {NoStop}%
\bibitem [{\citenamefont {Corkum}\ and\ \citenamefont
  {Krausz}(2007)}]{Corkum2007}%
  \BibitemOpen
  \bibfield  {author} {\bibinfo {author} {\bibfnamefont {P.~B.}\ \bibnamefont
  {Corkum}}\ and\ \bibinfo {author} {\bibfnamefont {F.}~\bibnamefont
  {Krausz}},\ }\href {\doibase 10.1038/nphys620} {\bibfield  {journal}
  {\bibinfo  {journal} {Nature Physics}\ }\textbf {\bibinfo {volume} {3}},\
  \bibinfo {pages} {381} (\bibinfo {year} {2007})}\BibitemShut {NoStop}%
\bibitem [{\citenamefont {Krausz}\ and\ \citenamefont
  {Ivanov}(2009)}]{Krausz2009RMP}%
  \BibitemOpen
  \bibfield  {author} {\bibinfo {author} {\bibfnamefont {F.}~\bibnamefont
  {Krausz}}\ and\ \bibinfo {author} {\bibfnamefont {M.}~\bibnamefont
  {Ivanov}},\ }\href {\doibase 10.1103/RevModPhys.81.163} {\bibfield  {journal}
  {\bibinfo  {journal} {Rev. Mod. Phys.}\ }\textbf {\bibinfo {volume} {81}},\
  \bibinfo {pages} {163} (\bibinfo {year} {2009})}\BibitemShut {NoStop}%
\bibitem [{\citenamefont {Ghimire}\ and\ \citenamefont
  {Reis}(2019)}]{Ghimire2019}%
  \BibitemOpen
  \bibfield  {author} {\bibinfo {author} {\bibfnamefont {S.}~\bibnamefont
  {Ghimire}}\ and\ \bibinfo {author} {\bibfnamefont {D.~A.}\ \bibnamefont
  {Reis}},\ }\href {\doibase 10.1038/s41567-018-0315-5} {\bibfield  {journal}
  {\bibinfo  {journal} {Nat. Phys.}\ }\textbf {\bibinfo {volume} {15}},\
  \bibinfo {pages} {10} (\bibinfo {year} {2019})}\BibitemShut {NoStop}%
\bibitem [{\citenamefont {Ferray}\ \emph {et~al.}(1988)\citenamefont {Ferray},
  \citenamefont {L'Huillier}, \citenamefont {Li}, \citenamefont {Lompre},
  \citenamefont {Mainfray},\ and\ \citenamefont {Manus}}]{Ferray_1988}%
  \BibitemOpen
  \bibfield  {author} {\bibinfo {author} {\bibfnamefont {M.}~\bibnamefont
  {Ferray}}, \bibinfo {author} {\bibfnamefont {A.}~\bibnamefont {L'Huillier}},
  \bibinfo {author} {\bibfnamefont {X.~F.}\ \bibnamefont {Li}}, \bibinfo
  {author} {\bibfnamefont {L.~A.}\ \bibnamefont {Lompre}}, \bibinfo {author}
  {\bibfnamefont {G.}~\bibnamefont {Mainfray}}, \ and\ \bibinfo {author}
  {\bibfnamefont {C.}~\bibnamefont {Manus}},\ }\href {\doibase
  10.1088/0953-4075/21/3/001} {\bibfield  {journal} {\bibinfo  {journal}
  {Journal of Physics B: Atomic, Molecular and Optical Physics}\ }\textbf
  {\bibinfo {volume} {21}},\ \bibinfo {pages} {L31} (\bibinfo {year}
  {1988})}\BibitemShut {NoStop}%
\bibitem [{\citenamefont {Corkum}(1993)}]{Corkum1993PRL}%
  \BibitemOpen
  \bibfield  {author} {\bibinfo {author} {\bibfnamefont {P.~B.}\ \bibnamefont
  {Corkum}},\ }\href {\doibase 10.1103/PhysRevLett.71.1994} {\bibfield
  {journal} {\bibinfo  {journal} {Phys. Rev. Lett.}\ }\textbf {\bibinfo
  {volume} {71}},\ \bibinfo {pages} {1994} (\bibinfo {year}
  {1993})}\BibitemShut {NoStop}%
\bibitem [{\citenamefont {Lewenstein}\ \emph {et~al.}(1994)\citenamefont
  {Lewenstein}, \citenamefont {Balcou}, \citenamefont {Ivanov}, \citenamefont
  {L'Huillier},\ and\ \citenamefont {Corkum}}]{Lewenstein1994}%
  \BibitemOpen
  \bibfield  {author} {\bibinfo {author} {\bibfnamefont {M.}~\bibnamefont
  {Lewenstein}}, \bibinfo {author} {\bibfnamefont {P.}~\bibnamefont {Balcou}},
  \bibinfo {author} {\bibfnamefont {M.~Y.}\ \bibnamefont {Ivanov}}, \bibinfo
  {author} {\bibfnamefont {A.}~\bibnamefont {L'Huillier}}, \ and\ \bibinfo
  {author} {\bibfnamefont {P.~B.}\ \bibnamefont {Corkum}},\ }\href {\doibase
  10.1103/PhysRevA.49.2117} {\bibfield  {journal} {\bibinfo  {journal} {Phys.
  Rev. A}\ }\textbf {\bibinfo {volume} {49}},\ \bibinfo {pages} {2117}
  (\bibinfo {year} {1994})}\BibitemShut {NoStop}%
\bibitem [{\citenamefont {Ghimire}\ \emph {et~al.}(2011)\citenamefont
  {Ghimire}, \citenamefont {DiChiara}, \citenamefont {Sistrunk}, \citenamefont
  {Agostini}, \citenamefont {DiMauro},\ and\ \citenamefont
  {Reis}}]{Ghimire2011NatPhys}%
  \BibitemOpen
  \bibfield  {author} {\bibinfo {author} {\bibfnamefont {S.}~\bibnamefont
  {Ghimire}}, \bibinfo {author} {\bibfnamefont {A.~D.}\ \bibnamefont
  {DiChiara}}, \bibinfo {author} {\bibfnamefont {E.}~\bibnamefont {Sistrunk}},
  \bibinfo {author} {\bibfnamefont {P.}~\bibnamefont {Agostini}}, \bibinfo
  {author} {\bibfnamefont {L.~F.}\ \bibnamefont {DiMauro}}, \ and\ \bibinfo
  {author} {\bibfnamefont {D.~A.}\ \bibnamefont {Reis}},\ }\href
  {https://doi.org/10.1038/nphys1847} {\bibfield  {journal} {\bibinfo
  {journal} {Nat. Phys.}\ }\textbf {\bibinfo {volume} {7}},\ \bibinfo {pages}
  {138} (\bibinfo {year} {2011})}\BibitemShut {NoStop}%
\bibitem [{\citenamefont {Schubert}\ \emph {et~al.}(2014)\citenamefont
  {Schubert}, \citenamefont {Hohenleutner}, \citenamefont {Langer},
  \citenamefont {Urbanek}, \citenamefont {Lange}, \citenamefont {Huttner},
  \citenamefont {Golde}, \citenamefont {Meier}, \citenamefont {Kira},
  \citenamefont {Koch},\ and\ \citenamefont {Huber}}]{Schubert2014}%
  \BibitemOpen
  \bibfield  {author} {\bibinfo {author} {\bibfnamefont {O.}~\bibnamefont
  {Schubert}}, \bibinfo {author} {\bibfnamefont {M.}~\bibnamefont
  {Hohenleutner}}, \bibinfo {author} {\bibfnamefont {F.}~\bibnamefont
  {Langer}}, \bibinfo {author} {\bibfnamefont {B.}~\bibnamefont {Urbanek}},
  \bibinfo {author} {\bibfnamefont {C.}~\bibnamefont {Lange}}, \bibinfo
  {author} {\bibfnamefont {U.}~\bibnamefont {Huttner}}, \bibinfo {author}
  {\bibfnamefont {D.}~\bibnamefont {Golde}}, \bibinfo {author} {\bibfnamefont
  {T.}~\bibnamefont {Meier}}, \bibinfo {author} {\bibfnamefont
  {M.}~\bibnamefont {Kira}}, \bibinfo {author} {\bibfnamefont {S.~W.}\
  \bibnamefont {Koch}}, \ and\ \bibinfo {author} {\bibfnamefont
  {R.}~\bibnamefont {Huber}},\ }\href
  {http://dx.doi.org/10.1038/nphoton.2013.349} {\bibfield  {journal} {\bibinfo
  {journal} {Nat. Photon.}\ }\textbf {\bibinfo {volume} {8}},\ \bibinfo {pages}
  {119} (\bibinfo {year} {2014})}\BibitemShut {NoStop}%
\bibitem [{\citenamefont {Vampa}\ \emph {et~al.}(2014)\citenamefont {Vampa},
  \citenamefont {McDonald}, \citenamefont {Orlando}, \citenamefont {Klug},
  \citenamefont {Corkum},\ and\ \citenamefont {Brabec}}]{Vampa2014PRL}%
  \BibitemOpen
  \bibfield  {author} {\bibinfo {author} {\bibfnamefont {G.}~\bibnamefont
  {Vampa}}, \bibinfo {author} {\bibfnamefont {C.~R.}\ \bibnamefont {McDonald}},
  \bibinfo {author} {\bibfnamefont {G.}~\bibnamefont {Orlando}}, \bibinfo
  {author} {\bibfnamefont {D.~D.}\ \bibnamefont {Klug}}, \bibinfo {author}
  {\bibfnamefont {P.~B.}\ \bibnamefont {Corkum}}, \ and\ \bibinfo {author}
  {\bibfnamefont {T.}~\bibnamefont {Brabec}},\ }\href {\doibase
  10.1103/PhysRevLett.113.073901} {\bibfield  {journal} {\bibinfo  {journal}
  {Phys. Rev. Lett.}\ }\textbf {\bibinfo {volume} {113}},\ \bibinfo {pages}
  {073901} (\bibinfo {year} {2014})}\BibitemShut {NoStop}%
\bibitem [{\citenamefont {Luu}\ \emph {et~al.}(2015)\citenamefont {Luu},
  \citenamefont {Garg}, \citenamefont {Kruchinin}, \citenamefont {Moulet},
  \citenamefont {Hassan},\ and\ \citenamefont {Goulielmakis}}]{Luu2015}%
  \BibitemOpen
  \bibfield  {author} {\bibinfo {author} {\bibfnamefont {T.~T.}\ \bibnamefont
  {Luu}}, \bibinfo {author} {\bibfnamefont {M.}~\bibnamefont {Garg}}, \bibinfo
  {author} {\bibfnamefont {S.~Y.}\ \bibnamefont {Kruchinin}}, \bibinfo {author}
  {\bibfnamefont {A.}~\bibnamefont {Moulet}}, \bibinfo {author} {\bibfnamefont
  {M.~T.}\ \bibnamefont {Hassan}}, \ and\ \bibinfo {author} {\bibfnamefont
  {E.}~\bibnamefont {Goulielmakis}},\ }\href
  {http://www.nature.com/doifinder/10.1038/nature14456} {\bibfield  {journal}
  {\bibinfo  {journal} {Nature (London)}\ }\textbf {\bibinfo {volume} {521}},\
  \bibinfo {pages} {498} (\bibinfo {year} {2015})}\BibitemShut {NoStop}%
\bibitem [{\citenamefont {Vampa}\ \emph {et~al.}(2015)\citenamefont {Vampa},
  \citenamefont {Hammond}, \citenamefont {Thire}, \citenamefont {Schmidt},
  \citenamefont {Legare}, \citenamefont {McDonald}, \citenamefont {Brabec},\
  and\ \citenamefont {Corkum}}]{Vampa2015Nature}%
  \BibitemOpen
  \bibfield  {author} {\bibinfo {author} {\bibfnamefont {G.}~\bibnamefont
  {Vampa}}, \bibinfo {author} {\bibfnamefont {T.~J.}\ \bibnamefont {Hammond}},
  \bibinfo {author} {\bibfnamefont {N.}~\bibnamefont {Thire}}, \bibinfo
  {author} {\bibfnamefont {B.~E.}\ \bibnamefont {Schmidt}}, \bibinfo {author}
  {\bibfnamefont {F.}~\bibnamefont {Legare}}, \bibinfo {author} {\bibfnamefont
  {C.~R.}\ \bibnamefont {McDonald}}, \bibinfo {author} {\bibfnamefont
  {T.}~\bibnamefont {Brabec}}, \ and\ \bibinfo {author} {\bibfnamefont {P.~B.}\
  \bibnamefont {Corkum}},\ }\href {http://dx.doi.org/10.1038/nature14517}
  {\bibfield  {journal} {\bibinfo  {journal} {Nature (London)}\ }\textbf
  {\bibinfo {volume} {522}},\ \bibinfo {pages} {462} (\bibinfo {year}
  {2015})}\BibitemShut {NoStop}%
\bibitem [{\citenamefont {Langer}\ \emph {et~al.}(2016)\citenamefont {Langer},
  \citenamefont {Hohenleutner}, \citenamefont {Schmid}, \citenamefont
  {P{\"o}llmann}, \citenamefont {Nagler}, \citenamefont {Korn}, \citenamefont
  {Sch{\"u}ller}, \citenamefont {Sherwin}, \citenamefont {Huttner},
  \citenamefont {Steiner}, \citenamefont {Koch}, \citenamefont {Kira},\ and\
  \citenamefont {Huber}}]{Langer2016Nature}%
  \BibitemOpen
  \bibfield  {author} {\bibinfo {author} {\bibfnamefont {F.}~\bibnamefont
  {Langer}}, \bibinfo {author} {\bibfnamefont {M.}~\bibnamefont
  {Hohenleutner}}, \bibinfo {author} {\bibfnamefont {C.~P.}\ \bibnamefont
  {Schmid}}, \bibinfo {author} {\bibfnamefont {C.}~\bibnamefont
  {P{\"o}llmann}}, \bibinfo {author} {\bibfnamefont {P.}~\bibnamefont
  {Nagler}}, \bibinfo {author} {\bibfnamefont {T.}~\bibnamefont {Korn}},
  \bibinfo {author} {\bibfnamefont {C.}~\bibnamefont {Sch{\"u}ller}}, \bibinfo
  {author} {\bibfnamefont {M.}~\bibnamefont {Sherwin}}, \bibinfo {author}
  {\bibfnamefont {U.}~\bibnamefont {Huttner}}, \bibinfo {author} {\bibfnamefont
  {J.}~\bibnamefont {Steiner}}, \bibinfo {author} {\bibfnamefont
  {S.}~\bibnamefont {Koch}}, \bibinfo {author} {\bibfnamefont {M.}~\bibnamefont
  {Kira}}, \ and\ \bibinfo {author} {\bibfnamefont {R.}~\bibnamefont {Huber}},\
  }\href {http://www.nature.com/doifinder/10.1038/nature17958} {\bibfield
  {journal} {\bibinfo  {journal} {Nature (London)}\ }\textbf {\bibinfo {volume}
  {533}},\ \bibinfo {pages} {225} (\bibinfo {year} {2016})}\BibitemShut
  {NoStop}%
\bibitem [{\citenamefont {Hohenleutner}\ \emph {et~al.}(2015)\citenamefont
  {Hohenleutner}, \citenamefont {Langer}, \citenamefont {Schubert},
  \citenamefont {Knorr}, \citenamefont {Huttner}, \citenamefont {Koch},
  \citenamefont {Kira},\ and\ \citenamefont {Huber}}]{Hohenleutner2015Nature}%
  \BibitemOpen
  \bibfield  {author} {\bibinfo {author} {\bibfnamefont {M.}~\bibnamefont
  {Hohenleutner}}, \bibinfo {author} {\bibfnamefont {F.}~\bibnamefont
  {Langer}}, \bibinfo {author} {\bibfnamefont {O.}~\bibnamefont {Schubert}},
  \bibinfo {author} {\bibfnamefont {M.}~\bibnamefont {Knorr}}, \bibinfo
  {author} {\bibfnamefont {U.}~\bibnamefont {Huttner}}, \bibinfo {author}
  {\bibfnamefont {S.}~\bibnamefont {Koch}}, \bibinfo {author} {\bibfnamefont
  {M.}~\bibnamefont {Kira}}, \ and\ \bibinfo {author} {\bibfnamefont
  {R.}~\bibnamefont {Huber}},\ }\href {https://doi.org/10.1038/nature14652}
  {\bibfield  {journal} {\bibinfo  {journal} {Nature (London)}\ }\textbf
  {\bibinfo {volume} {523}},\ \bibinfo {pages} {572} (\bibinfo {year}
  {2015})}\BibitemShut {NoStop}%
\bibitem [{\citenamefont {Ndabashimiye}\ \emph {et~al.}(2016)\citenamefont
  {Ndabashimiye}, \citenamefont {Ghimire}, \citenamefont {Wu}, \citenamefont
  {Browne}, \citenamefont {Schafer}, \citenamefont {Gaarde},\ and\
  \citenamefont {Reis}}]{Ndabashimiye2016}%
  \BibitemOpen
  \bibfield  {author} {\bibinfo {author} {\bibfnamefont {G.}~\bibnamefont
  {Ndabashimiye}}, \bibinfo {author} {\bibfnamefont {S.}~\bibnamefont
  {Ghimire}}, \bibinfo {author} {\bibfnamefont {M.}~\bibnamefont {Wu}},
  \bibinfo {author} {\bibfnamefont {D.~A.}\ \bibnamefont {Browne}}, \bibinfo
  {author} {\bibfnamefont {K.~J.}\ \bibnamefont {Schafer}}, \bibinfo {author}
  {\bibfnamefont {M.~B.}\ \bibnamefont {Gaarde}}, \ and\ \bibinfo {author}
  {\bibfnamefont {D.~A.}\ \bibnamefont {Reis}},\ }\href
  {http://www.nature.com/doifinder/10.1038/nature17660} {\bibfield  {journal}
  {\bibinfo  {journal} {Nature (London)}\ }\textbf {\bibinfo {volume} {534}},\
  \bibinfo {pages} {520} (\bibinfo {year} {2016})}\BibitemShut {NoStop}%
\bibitem [{\citenamefont {Otobe}(2016)}]{Otobe2016}%
  \BibitemOpen
  \bibfield  {author} {\bibinfo {author} {\bibfnamefont {T.}~\bibnamefont
  {Otobe}},\ }\href {\doibase 10.1103/PhysRevB.94.235152} {\bibfield  {journal}
  {\bibinfo  {journal} {Phys. Rev. B}\ }\textbf {\bibinfo {volume} {94}},\
  \bibinfo {pages} {235152} (\bibinfo {year} {2016})}\BibitemShut {NoStop}%
\bibitem [{\citenamefont {Tancogne-Dejean}\ \emph {et~al.}(2017)\citenamefont
  {Tancogne-Dejean}, \citenamefont {M{\"u}cke}, \citenamefont {K{\"a}rtner},\
  and\ \citenamefont {Rubio}}]{Tancogne-Dejean2017}%
  \BibitemOpen
  \bibfield  {author} {\bibinfo {author} {\bibfnamefont {N.}~\bibnamefont
  {Tancogne-Dejean}}, \bibinfo {author} {\bibfnamefont {O.~D.}\ \bibnamefont
  {M{\"u}cke}}, \bibinfo {author} {\bibfnamefont {F.~X.}\ \bibnamefont
  {K{\"a}rtner}}, \ and\ \bibinfo {author} {\bibfnamefont {A.}~\bibnamefont
  {Rubio}},\ }\href {\doibase 10.1038/s41467-017-00764-5} {\bibfield  {journal}
  {\bibinfo  {journal} {Nat. Comm.}\ }\textbf {\bibinfo {volume} {8}},\
  \bibinfo {pages} {745} (\bibinfo {year} {2017})}\BibitemShut {NoStop}%
\bibitem [{\citenamefont {Ikemachi}\ \emph {et~al.}(2017)\citenamefont
  {Ikemachi}, \citenamefont {Shinohara}, \citenamefont {Sato}, \citenamefont
  {Yumoto}, \citenamefont {Kuwata-Gonokami},\ and\ \citenamefont
  {Ishikawa}}]{Ikemachi2017}%
  \BibitemOpen
  \bibfield  {author} {\bibinfo {author} {\bibfnamefont {T.}~\bibnamefont
  {Ikemachi}}, \bibinfo {author} {\bibfnamefont {Y.}~\bibnamefont {Shinohara}},
  \bibinfo {author} {\bibfnamefont {T.}~\bibnamefont {Sato}}, \bibinfo {author}
  {\bibfnamefont {J.}~\bibnamefont {Yumoto}}, \bibinfo {author} {\bibfnamefont
  {M.}~\bibnamefont {Kuwata-Gonokami}}, \ and\ \bibinfo {author} {\bibfnamefont
  {K.~L.}\ \bibnamefont {Ishikawa}},\ }\href {\doibase
  10.1103/PhysRevA.95.043416} {\bibfield  {journal} {\bibinfo  {journal} {Phys.
  Rev. A}\ }\textbf {\bibinfo {volume} {95}},\ \bibinfo {pages} {043416}
  (\bibinfo {year} {2017})}\BibitemShut {NoStop}%
\bibitem [{\citenamefont {Liu}\ \emph {et~al.}(2017)\citenamefont {Liu},
  \citenamefont {Li}, \citenamefont {You}, \citenamefont {Ghimire},
  \citenamefont {Heinz},\ and\ \citenamefont {Reis}}]{Liu2017}%
  \BibitemOpen
  \bibfield  {author} {\bibinfo {author} {\bibfnamefont {H.}~\bibnamefont
  {Liu}}, \bibinfo {author} {\bibfnamefont {Y.}~\bibnamefont {Li}}, \bibinfo
  {author} {\bibfnamefont {Y.~S.}\ \bibnamefont {You}}, \bibinfo {author}
  {\bibfnamefont {S.}~\bibnamefont {Ghimire}}, \bibinfo {author} {\bibfnamefont
  {T.~F.}\ \bibnamefont {Heinz}}, \ and\ \bibinfo {author} {\bibfnamefont
  {D.~A.}\ \bibnamefont {Reis}},\ }\href
  {https://www.nature.com/articles/nphys3946} {\bibfield  {journal} {\bibinfo
  {journal} {Nat. Phys.}\ }\textbf {\bibinfo {volume} {13}},\ \bibinfo {pages}
  {262} (\bibinfo {year} {2017})}\BibitemShut {NoStop}%
\bibitem [{\citenamefont {You}\ \emph {et~al.}(2017)\citenamefont {You},
  \citenamefont {Reis},\ and\ \citenamefont {Ghimire}}]{You2017}%
  \BibitemOpen
  \bibfield  {author} {\bibinfo {author} {\bibfnamefont {Y.~S.}\ \bibnamefont
  {You}}, \bibinfo {author} {\bibfnamefont {D.}~\bibnamefont {Reis}}, \ and\
  \bibinfo {author} {\bibfnamefont {S.}~\bibnamefont {Ghimire}},\ }\href
  {\doibase 10.1038/nphys3955} {\bibfield  {journal} {\bibinfo  {journal}
  {Nature Physics}\ }\textbf {\bibinfo {volume} {13}},\ \bibinfo {pages} {345}
  (\bibinfo {year} {2017})}\BibitemShut {NoStop}%
\bibitem [{\citenamefont {Kaneshima}\ \emph {et~al.}(2018)\citenamefont
  {Kaneshima}, \citenamefont {Shinohara}, \citenamefont {Takeuchi},
  \citenamefont {Ishii}, \citenamefont {Imasaka}, \citenamefont {Kaji},
  \citenamefont {Ashihara}, \citenamefont {Ishikawa},\ and\ \citenamefont
  {Itatani}}]{Kaneshima2018}%
  \BibitemOpen
  \bibfield  {author} {\bibinfo {author} {\bibfnamefont {K.}~\bibnamefont
  {Kaneshima}}, \bibinfo {author} {\bibfnamefont {Y.}~\bibnamefont
  {Shinohara}}, \bibinfo {author} {\bibfnamefont {K.}~\bibnamefont {Takeuchi}},
  \bibinfo {author} {\bibfnamefont {N.}~\bibnamefont {Ishii}}, \bibinfo
  {author} {\bibfnamefont {K.}~\bibnamefont {Imasaka}}, \bibinfo {author}
  {\bibfnamefont {T.}~\bibnamefont {Kaji}}, \bibinfo {author} {\bibfnamefont
  {S.}~\bibnamefont {Ashihara}}, \bibinfo {author} {\bibfnamefont {K.~L.}\
  \bibnamefont {Ishikawa}}, \ and\ \bibinfo {author} {\bibfnamefont
  {J.}~\bibnamefont {Itatani}},\ }\href {\doibase
  10.1103/PhysRevLett.120.243903} {\bibfield  {journal} {\bibinfo  {journal}
  {Phys. Rev. Lett.}\ }\textbf {\bibinfo {volume} {120}},\ \bibinfo {pages}
  {243903} (\bibinfo {year} {2018})}\BibitemShut {NoStop}%
\bibitem [{\citenamefont {Sekiguchi}\ \emph {et~al.}(2022)\citenamefont
  {Sekiguchi}, \citenamefont {Yumoto}, \citenamefont {Hirori},\ and\
  \citenamefont {Kanemitsu}}]{Sekiguchi2022PRB}%
  \BibitemOpen
  \bibfield  {author} {\bibinfo {author} {\bibfnamefont {F.}~\bibnamefont
  {Sekiguchi}}, \bibinfo {author} {\bibfnamefont {G.}~\bibnamefont {Yumoto}},
  \bibinfo {author} {\bibfnamefont {H.}~\bibnamefont {Hirori}}, \ and\ \bibinfo
  {author} {\bibfnamefont {Y.}~\bibnamefont {Kanemitsu}},\ }\href {\doibase
  10.1103/PhysRevB.106.L241201} {\bibfield  {journal} {\bibinfo  {journal}
  {Phys. Rev. B}\ }\textbf {\bibinfo {volume} {106}},\ \bibinfo {pages}
  {L241201} (\bibinfo {year} {2022})}\BibitemShut {NoStop}%
\bibitem [{\citenamefont {Giorgianni}\ \emph {et~al.}(2016)\citenamefont
  {Giorgianni}, \citenamefont {Chiadroni}, \citenamefont {Rovere},
  \citenamefont {Cestelli-Guidi}, \citenamefont {Perucchi}, \citenamefont
  {Bellaveglia}, \citenamefont {Castellano}, \citenamefont {Di~Giovenale},
  \citenamefont {Di~Pirro}, \citenamefont {Ferrario}, \citenamefont {Pompili},
  \citenamefont {Vaccarezza}, \citenamefont {Villa}, \citenamefont {Cianchi},
  \citenamefont {Mostacci}, \citenamefont {Petrarca}, \citenamefont {Brahlek},
  \citenamefont {Koirala}, \citenamefont {Oh},\ and\ \citenamefont
  {Lupi}}]{Giorgianni2016}%
  \BibitemOpen
  \bibfield  {author} {\bibinfo {author} {\bibfnamefont {F.}~\bibnamefont
  {Giorgianni}}, \bibinfo {author} {\bibfnamefont {E.}~\bibnamefont
  {Chiadroni}}, \bibinfo {author} {\bibfnamefont {A.}~\bibnamefont {Rovere}},
  \bibinfo {author} {\bibfnamefont {M.}~\bibnamefont {Cestelli-Guidi}},
  \bibinfo {author} {\bibfnamefont {A.}~\bibnamefont {Perucchi}}, \bibinfo
  {author} {\bibfnamefont {M.}~\bibnamefont {Bellaveglia}}, \bibinfo {author}
  {\bibfnamefont {M.}~\bibnamefont {Castellano}}, \bibinfo {author}
  {\bibfnamefont {D.}~\bibnamefont {Di~Giovenale}}, \bibinfo {author}
  {\bibfnamefont {G.}~\bibnamefont {Di~Pirro}}, \bibinfo {author}
  {\bibfnamefont {M.}~\bibnamefont {Ferrario}}, \bibinfo {author}
  {\bibfnamefont {R.}~\bibnamefont {Pompili}}, \bibinfo {author} {\bibfnamefont
  {C.}~\bibnamefont {Vaccarezza}}, \bibinfo {author} {\bibfnamefont
  {F.}~\bibnamefont {Villa}}, \bibinfo {author} {\bibfnamefont
  {A.}~\bibnamefont {Cianchi}}, \bibinfo {author} {\bibfnamefont
  {A.}~\bibnamefont {Mostacci}}, \bibinfo {author} {\bibfnamefont
  {M.}~\bibnamefont {Petrarca}}, \bibinfo {author} {\bibfnamefont
  {M.}~\bibnamefont {Brahlek}}, \bibinfo {author} {\bibfnamefont
  {N.}~\bibnamefont {Koirala}}, \bibinfo {author} {\bibfnamefont
  {S.}~\bibnamefont {Oh}}, \ and\ \bibinfo {author} {\bibfnamefont
  {S.}~\bibnamefont {Lupi}},\ }\href {\doibase 10.1038/ncomms11421} {\bibfield
  {journal} {\bibinfo  {journal} {Nature Communications}\ }\textbf {\bibinfo
  {volume} {7}},\ \bibinfo {pages} {11421} (\bibinfo {year}
  {2016})}\BibitemShut {NoStop}%
\bibitem [{\citenamefont {Yoshikawa}\ \emph {et~al.}(2017)\citenamefont
  {Yoshikawa}, \citenamefont {Tamaya},\ and\ \citenamefont
  {Tanaka}}]{Yoshikawa2017Science}%
  \BibitemOpen
  \bibfield  {author} {\bibinfo {author} {\bibfnamefont {N.}~\bibnamefont
  {Yoshikawa}}, \bibinfo {author} {\bibfnamefont {T.}~\bibnamefont {Tamaya}}, \
  and\ \bibinfo {author} {\bibfnamefont {K.}~\bibnamefont {Tanaka}},\ }\href
  {\doibase 10.1126/science.aam8861} {\bibfield  {journal} {\bibinfo  {journal}
  {Science}\ }\textbf {\bibinfo {volume} {356}},\ \bibinfo {pages} {736}
  (\bibinfo {year} {2017})}\BibitemShut {NoStop}%
\bibitem [{\citenamefont {Hafez}\ \emph {et~al.}(2018)\citenamefont {Hafez},
  \citenamefont {Kovalev}, \citenamefont {Deinert}, \citenamefont {Mics},
  \citenamefont {Green}, \citenamefont {Awari}, \citenamefont {Chen},
  \citenamefont {Germanskiy}, \citenamefont {Lehnert}, \citenamefont
  {Teichert}, \citenamefont {Wang}, \citenamefont {Tielrooij}, \citenamefont
  {Liu}, \citenamefont {Chen}, \citenamefont {Narita}, \citenamefont {Mullen},
  \citenamefont {Bonn}, \citenamefont {Gensch},\ and\ \citenamefont
  {Turchinovich}}]{Hafez2018}%
  \BibitemOpen
  \bibfield  {author} {\bibinfo {author} {\bibfnamefont {H.~A.}\ \bibnamefont
  {Hafez}}, \bibinfo {author} {\bibfnamefont {S.}~\bibnamefont {Kovalev}},
  \bibinfo {author} {\bibfnamefont {J.-C.}\ \bibnamefont {Deinert}}, \bibinfo
  {author} {\bibfnamefont {Z.}~\bibnamefont {Mics}}, \bibinfo {author}
  {\bibfnamefont {B.}~\bibnamefont {Green}}, \bibinfo {author} {\bibfnamefont
  {N.}~\bibnamefont {Awari}}, \bibinfo {author} {\bibfnamefont
  {M.}~\bibnamefont {Chen}}, \bibinfo {author} {\bibfnamefont {S.}~\bibnamefont
  {Germanskiy}}, \bibinfo {author} {\bibfnamefont {U.}~\bibnamefont {Lehnert}},
  \bibinfo {author} {\bibfnamefont {J.}~\bibnamefont {Teichert}}, \bibinfo
  {author} {\bibfnamefont {Z.}~\bibnamefont {Wang}}, \bibinfo {author}
  {\bibfnamefont {K.-J.}\ \bibnamefont {Tielrooij}}, \bibinfo {author}
  {\bibfnamefont {Z.}~\bibnamefont {Liu}}, \bibinfo {author} {\bibfnamefont
  {Z.}~\bibnamefont {Chen}}, \bibinfo {author} {\bibfnamefont {A.}~\bibnamefont
  {Narita}}, \bibinfo {author} {\bibfnamefont {K.}~\bibnamefont {Mullen}},
  \bibinfo {author} {\bibfnamefont {M.}~\bibnamefont {Bonn}}, \bibinfo {author}
  {\bibfnamefont {M.}~\bibnamefont {Gensch}}, \ and\ \bibinfo {author}
  {\bibfnamefont {D.}~\bibnamefont {Turchinovich}},\ }\href
  {http://dx.doi.org/10.1038/s41586-018-0508-1} {\bibfield  {journal} {\bibinfo
   {journal} {Nature (London)}\ }\textbf {\bibinfo {volume} {561}},\ \bibinfo
  {pages} {507} (\bibinfo {year} {2018})}\BibitemShut {NoStop}%
\bibitem [{\citenamefont {Silva}\ \emph {et~al.}(2019)\citenamefont {Silva},
  \citenamefont {Jim{\'e}nez-Gal{\'a}n}, \citenamefont {Amorim}, \citenamefont
  {Smirnova},\ and\ \citenamefont {Ivanov}}]{Silva2019}%
  \BibitemOpen
  \bibfield  {author} {\bibinfo {author} {\bibfnamefont {R.~E.~F.}\
  \bibnamefont {Silva}}, \bibinfo {author} {\bibfnamefont {{\'A}.}~\bibnamefont
  {Jim{\'e}nez-Gal{\'a}n}}, \bibinfo {author} {\bibfnamefont {B.}~\bibnamefont
  {Amorim}}, \bibinfo {author} {\bibfnamefont {O.}~\bibnamefont {Smirnova}}, \
  and\ \bibinfo {author} {\bibfnamefont {M.}~\bibnamefont {Ivanov}},\ }\href
  {\doibase 10.1038/s41566-019-0516-1} {\bibfield  {journal} {\bibinfo
  {journal} {Nature Photonics}\ }\textbf {\bibinfo {volume} {13}},\ \bibinfo
  {pages} {849} (\bibinfo {year} {2019})}\BibitemShut {NoStop}%
\bibitem [{\citenamefont {Chac{\'o}n}\ \emph {et~al.}(2020)\citenamefont
  {Chac{\'o}n}, \citenamefont {Kim}, \citenamefont {Zhu}, \citenamefont
  {Kelly}, \citenamefont {Dauphin}, \citenamefont {Pisanty}, \citenamefont
  {Maxwell}, \citenamefont {Pic\'on}, \citenamefont {Ciappina}, \citenamefont
  {Kim}, \citenamefont {Ticknor}, \citenamefont {Saxena},\ and\ \citenamefont
  {Lewenstein}}]{Chacon2020PRB}%
  \BibitemOpen
  \bibfield  {author} {\bibinfo {author} {\bibfnamefont {A.}~\bibnamefont
  {Chac{\'o}n}}, \bibinfo {author} {\bibfnamefont {D.}~\bibnamefont {Kim}},
  \bibinfo {author} {\bibfnamefont {W.}~\bibnamefont {Zhu}}, \bibinfo {author}
  {\bibfnamefont {S.~P.}\ \bibnamefont {Kelly}}, \bibinfo {author}
  {\bibfnamefont {A.}~\bibnamefont {Dauphin}}, \bibinfo {author} {\bibfnamefont
  {E.}~\bibnamefont {Pisanty}}, \bibinfo {author} {\bibfnamefont {A.~S.}\
  \bibnamefont {Maxwell}}, \bibinfo {author} {\bibfnamefont {A.}~\bibnamefont
  {Pic\'on}}, \bibinfo {author} {\bibfnamefont {M.~F.}\ \bibnamefont
  {Ciappina}}, \bibinfo {author} {\bibfnamefont {D.~E.}\ \bibnamefont {Kim}},
  \bibinfo {author} {\bibfnamefont {C.}~\bibnamefont {Ticknor}}, \bibinfo
  {author} {\bibfnamefont {A.}~\bibnamefont {Saxena}}, \ and\ \bibinfo {author}
  {\bibfnamefont {M.}~\bibnamefont {Lewenstein}},\ }\href {\doibase
  10.1103/PhysRevB.102.134115} {\bibfield  {journal} {\bibinfo  {journal}
  {Phys. Rev. B}\ }\textbf {\bibinfo {volume} {102}},\ \bibinfo {pages}
  {134115} (\bibinfo {year} {2020})}\BibitemShut {NoStop}%
\bibitem [{\citenamefont {Cheng}\ \emph {et~al.}(2020)\citenamefont {Cheng},
  \citenamefont {Kanda}, \citenamefont {Ikeda}, \citenamefont {Matsuda},
  \citenamefont {Xia}, \citenamefont {Schumann}, \citenamefont {Stemmer},
  \citenamefont {Itatani}, \citenamefont {Armitage},\ and\ \citenamefont
  {Matsunaga}}]{Matsunaga2020PRL}%
  \BibitemOpen
  \bibfield  {author} {\bibinfo {author} {\bibfnamefont {B.}~\bibnamefont
  {Cheng}}, \bibinfo {author} {\bibfnamefont {N.}~\bibnamefont {Kanda}},
  \bibinfo {author} {\bibfnamefont {T.~N.}\ \bibnamefont {Ikeda}}, \bibinfo
  {author} {\bibfnamefont {T.}~\bibnamefont {Matsuda}}, \bibinfo {author}
  {\bibfnamefont {P.}~\bibnamefont {Xia}}, \bibinfo {author} {\bibfnamefont
  {T.}~\bibnamefont {Schumann}}, \bibinfo {author} {\bibfnamefont
  {S.}~\bibnamefont {Stemmer}}, \bibinfo {author} {\bibfnamefont
  {J.}~\bibnamefont {Itatani}}, \bibinfo {author} {\bibfnamefont {N.~P.}\
  \bibnamefont {Armitage}}, \ and\ \bibinfo {author} {\bibfnamefont
  {R.}~\bibnamefont {Matsunaga}},\ }\href {\doibase
  10.1103/PhysRevLett.124.117402} {\bibfield  {journal} {\bibinfo  {journal}
  {Phys. Rev. Lett.}\ }\textbf {\bibinfo {volume} {124}},\ \bibinfo {pages}
  {117402} (\bibinfo {year} {2020})}\BibitemShut {NoStop}%
\bibitem [{\citenamefont {Schmid}\ \emph {et~al.}(2021)\citenamefont {Schmid},
  \citenamefont {Weigl}, \citenamefont {Gr{\"o}ssing}, \citenamefont {Junk},
  \citenamefont {Gorini}, \citenamefont {Schlauderer}, \citenamefont {Ito},
  \citenamefont {Meierhofer}, \citenamefont {Hofmann}, \citenamefont
  {Afanasiev}, \citenamefont {Crewse}, \citenamefont {Kokh}, \citenamefont
  {Tereshchenko}, \citenamefont {G{\"u}dde}, \citenamefont {Evers},
  \citenamefont {Wilhelm}, \citenamefont {Richter}, \citenamefont {H{\"o}fer},\
  and\ \citenamefont {Huber}}]{Schmid2021}%
  \BibitemOpen
  \bibfield  {author} {\bibinfo {author} {\bibfnamefont {C.~P.}\ \bibnamefont
  {Schmid}}, \bibinfo {author} {\bibfnamefont {L.}~\bibnamefont {Weigl}},
  \bibinfo {author} {\bibfnamefont {P.}~\bibnamefont {Gr{\"o}ssing}}, \bibinfo
  {author} {\bibfnamefont {V.}~\bibnamefont {Junk}}, \bibinfo {author}
  {\bibfnamefont {C.}~\bibnamefont {Gorini}}, \bibinfo {author} {\bibfnamefont
  {S.}~\bibnamefont {Schlauderer}}, \bibinfo {author} {\bibfnamefont
  {S.}~\bibnamefont {Ito}}, \bibinfo {author} {\bibfnamefont {M.}~\bibnamefont
  {Meierhofer}}, \bibinfo {author} {\bibfnamefont {N.}~\bibnamefont {Hofmann}},
  \bibinfo {author} {\bibfnamefont {D.}~\bibnamefont {Afanasiev}}, \bibinfo
  {author} {\bibfnamefont {J.}~\bibnamefont {Crewse}}, \bibinfo {author}
  {\bibfnamefont {K.~A.}\ \bibnamefont {Kokh}}, \bibinfo {author}
  {\bibfnamefont {O.~E.}\ \bibnamefont {Tereshchenko}}, \bibinfo {author}
  {\bibfnamefont {J.}~\bibnamefont {G{\"u}dde}}, \bibinfo {author}
  {\bibfnamefont {F.}~\bibnamefont {Evers}}, \bibinfo {author} {\bibfnamefont
  {J.}~\bibnamefont {Wilhelm}}, \bibinfo {author} {\bibfnamefont
  {K.}~\bibnamefont {Richter}}, \bibinfo {author} {\bibfnamefont
  {U.}~\bibnamefont {H{\"o}fer}}, \ and\ \bibinfo {author} {\bibfnamefont
  {R.}~\bibnamefont {Huber}},\ }\href {\doibase 10.1038/s41586-021-03466-7}
  {\bibfield  {journal} {\bibinfo  {journal} {Nature}\ }\textbf {\bibinfo
  {volume} {593}},\ \bibinfo {pages} {385} (\bibinfo {year}
  {2021})}\BibitemShut {NoStop}%
\bibitem [{\citenamefont {Baykusheva}\ \emph
  {et~al.}(2021{\natexlab{a}})\citenamefont {Baykusheva}, \citenamefont
  {Chac{\'o}n}, \citenamefont {Lu}, \citenamefont {Bailey}, \citenamefont
  {Sobota}, \citenamefont {Soifer}, \citenamefont {Kirchmann}, \citenamefont
  {Rotundu}, \citenamefont {Uher}, \citenamefont {Heinz}, \citenamefont
  {Reis},\ and\ \citenamefont {Ghimire}}]{Baykusheva2021}%
  \BibitemOpen
  \bibfield  {author} {\bibinfo {author} {\bibfnamefont {D.}~\bibnamefont
  {Baykusheva}}, \bibinfo {author} {\bibfnamefont {A.}~\bibnamefont
  {Chac{\'o}n}}, \bibinfo {author} {\bibfnamefont {J.}~\bibnamefont {Lu}},
  \bibinfo {author} {\bibfnamefont {T.~P.}\ \bibnamefont {Bailey}}, \bibinfo
  {author} {\bibfnamefont {J.~A.}\ \bibnamefont {Sobota}}, \bibinfo {author}
  {\bibfnamefont {H.}~\bibnamefont {Soifer}}, \bibinfo {author} {\bibfnamefont
  {P.~S.}\ \bibnamefont {Kirchmann}}, \bibinfo {author} {\bibfnamefont
  {C.}~\bibnamefont {Rotundu}}, \bibinfo {author} {\bibfnamefont
  {C.}~\bibnamefont {Uher}}, \bibinfo {author} {\bibfnamefont {T.~F.}\
  \bibnamefont {Heinz}}, \bibinfo {author} {\bibfnamefont {D.~A.}\ \bibnamefont
  {Reis}}, \ and\ \bibinfo {author} {\bibfnamefont {S.}~\bibnamefont
  {Ghimire}},\ }\href {\doibase 10.1021/acs.nanolett.1c02145} {\bibfield
  {journal} {\bibinfo  {journal} {Nano Letters}\ }\textbf {\bibinfo {volume}
  {21}},\ \bibinfo {pages} {8970} (\bibinfo {year}
  {2021}{\natexlab{a}})}\BibitemShut {NoStop}%
\bibitem [{\citenamefont {Silva}\ \emph {et~al.}(2018)\citenamefont {Silva},
  \citenamefont {Blinov}, \citenamefont {Rubtsov}, \citenamefont {Smirnova},\
  and\ \citenamefont {Ivanov}}]{Silva2018NatPhoton}%
  \BibitemOpen
  \bibfield  {author} {\bibinfo {author} {\bibfnamefont {R.~E.~F.}\
  \bibnamefont {Silva}}, \bibinfo {author} {\bibfnamefont {I.~V.}\ \bibnamefont
  {Blinov}}, \bibinfo {author} {\bibfnamefont {A.~N.}\ \bibnamefont {Rubtsov}},
  \bibinfo {author} {\bibfnamefont {O.}~\bibnamefont {Smirnova}}, \ and\
  \bibinfo {author} {\bibfnamefont {M.}~\bibnamefont {Ivanov}},\ }\href
  {https://doi.org/10.1038/s41566-018-0129-0} {\bibfield  {journal} {\bibinfo
  {journal} {Nat. Photon.}\ }\textbf {\bibinfo {volume} {12}},\ \bibinfo
  {pages} {266} (\bibinfo {year} {2018})}\BibitemShut {NoStop}%
\bibitem [{\citenamefont {Murakami}\ \emph {et~al.}(2018)\citenamefont
  {Murakami}, \citenamefont {Eckstein},\ and\ \citenamefont
  {Werner}}]{Murakami2018PRL}%
  \BibitemOpen
  \bibfield  {author} {\bibinfo {author} {\bibfnamefont {Y.}~\bibnamefont
  {Murakami}}, \bibinfo {author} {\bibfnamefont {M.}~\bibnamefont {Eckstein}},
  \ and\ \bibinfo {author} {\bibfnamefont {P.}~\bibnamefont {Werner}},\ }\href
  {\doibase 10.1103/PhysRevLett.121.057405} {\bibfield  {journal} {\bibinfo
  {journal} {Phys. Rev. Lett.}\ }\textbf {\bibinfo {volume} {121}},\ \bibinfo
  {pages} {057405} (\bibinfo {year} {2018})}\BibitemShut {NoStop}%
\bibitem [{\citenamefont {Tancogne-Dejean}\ \emph {et~al.}(2018)\citenamefont
  {Tancogne-Dejean}, \citenamefont {Sentef},\ and\ \citenamefont
  {Rubio}}]{Tancogne-Dejean2018}%
  \BibitemOpen
  \bibfield  {author} {\bibinfo {author} {\bibfnamefont {N.}~\bibnamefont
  {Tancogne-Dejean}}, \bibinfo {author} {\bibfnamefont {M.~A.}\ \bibnamefont
  {Sentef}}, \ and\ \bibinfo {author} {\bibfnamefont {A.}~\bibnamefont
  {Rubio}},\ }\href {\doibase 10.1103/PhysRevLett.121.097402} {\bibfield
  {journal} {\bibinfo  {journal} {Phys. Rev. Lett.}\ }\textbf {\bibinfo
  {volume} {121}},\ \bibinfo {pages} {097402} (\bibinfo {year}
  {2018})}\BibitemShut {NoStop}%
\bibitem [{\citenamefont {Imai}\ \emph {et~al.}(2020)\citenamefont {Imai},
  \citenamefont {Ono},\ and\ \citenamefont {Ishihara}}]{Ishihara2020}%
  \BibitemOpen
  \bibfield  {author} {\bibinfo {author} {\bibfnamefont {S.}~\bibnamefont
  {Imai}}, \bibinfo {author} {\bibfnamefont {A.}~\bibnamefont {Ono}}, \ and\
  \bibinfo {author} {\bibfnamefont {S.}~\bibnamefont {Ishihara}},\ }\href
  {\doibase 10.1103/PhysRevLett.124.157404} {\bibfield  {journal} {\bibinfo
  {journal} {Phys. Rev. Lett.}\ }\textbf {\bibinfo {volume} {124}},\ \bibinfo
  {pages} {157404} (\bibinfo {year} {2020})}\BibitemShut {NoStop}%
\bibitem [{\citenamefont {Murakami}\ \emph {et~al.}(2021)\citenamefont
  {Murakami}, \citenamefont {Takayoshi}, \citenamefont {Koga},\ and\
  \citenamefont {Werner}}]{Murakami2021PRB}%
  \BibitemOpen
  \bibfield  {author} {\bibinfo {author} {\bibfnamefont {Y.}~\bibnamefont
  {Murakami}}, \bibinfo {author} {\bibfnamefont {S.}~\bibnamefont {Takayoshi}},
  \bibinfo {author} {\bibfnamefont {A.}~\bibnamefont {Koga}}, \ and\ \bibinfo
  {author} {\bibfnamefont {P.}~\bibnamefont {Werner}},\ }\href {\doibase
  10.1103/PhysRevB.103.035110} {\bibfield  {journal} {\bibinfo  {journal}
  {Phys. Rev. B}\ }\textbf {\bibinfo {volume} {103}},\ \bibinfo {pages}
  {035110} (\bibinfo {year} {2021})}\BibitemShut {NoStop}%
\bibitem [{\citenamefont {Orthodoxou}\ \emph {et~al.}(2021)\citenamefont
  {Orthodoxou}, \citenamefont {Za{\"i}r},\ and\ \citenamefont
  {Booth}}]{Orthodoxou2021}%
  \BibitemOpen
  \bibfield  {author} {\bibinfo {author} {\bibfnamefont {C.}~\bibnamefont
  {Orthodoxou}}, \bibinfo {author} {\bibfnamefont {A.}~\bibnamefont
  {Za{\"i}r}}, \ and\ \bibinfo {author} {\bibfnamefont {G.~H.}\ \bibnamefont
  {Booth}},\ }\href {\doibase 10.1038/s41535-021-00377-8} {\bibfield  {journal}
  {\bibinfo  {journal} {npj Quantum Materials}\ }\textbf {\bibinfo {volume}
  {6}},\ \bibinfo {pages} {76} (\bibinfo {year} {2021})}\BibitemShut {NoStop}%
\bibitem [{\citenamefont {Udono}\ \emph {et~al.}(2022)\citenamefont {Udono},
  \citenamefont {Sugimoto}, \citenamefont {Kaneko},\ and\ \citenamefont
  {Ohta}}]{Udono2022PRB}%
  \BibitemOpen
  \bibfield  {author} {\bibinfo {author} {\bibfnamefont {M.}~\bibnamefont
  {Udono}}, \bibinfo {author} {\bibfnamefont {K.}~\bibnamefont {Sugimoto}},
  \bibinfo {author} {\bibfnamefont {T.}~\bibnamefont {Kaneko}}, \ and\ \bibinfo
  {author} {\bibfnamefont {Y.}~\bibnamefont {Ohta}},\ }\href {\doibase
  10.1103/PhysRevB.105.L241108} {\bibfield  {journal} {\bibinfo  {journal}
  {Phys. Rev. B}\ }\textbf {\bibinfo {volume} {105}},\ \bibinfo {pages}
  {L241108} (\bibinfo {year} {2022})}\BibitemShut {NoStop}%
\bibitem [{\citenamefont {Bionta}\ \emph {et~al.}(2021)\citenamefont {Bionta},
  \citenamefont {Haddad}, \citenamefont {Leblanc}, \citenamefont {Gruson},
  \citenamefont {Lassonde}, \citenamefont {Ibrahim}, \citenamefont {Chaillou},
  \citenamefont {\'Emond}, \citenamefont {Otto}, \citenamefont
  {Jim\'enez-Gal\'an}, \citenamefont {Silva}, \citenamefont {Ivanov},
  \citenamefont {Siwick}, \citenamefont {Chaker},\ and\ \citenamefont
  {L\'egar\'e}}]{Bionta2021PRR}%
  \BibitemOpen
  \bibfield  {author} {\bibinfo {author} {\bibfnamefont {M.~R.}\ \bibnamefont
  {Bionta}}, \bibinfo {author} {\bibfnamefont {E.}~\bibnamefont {Haddad}},
  \bibinfo {author} {\bibfnamefont {A.}~\bibnamefont {Leblanc}}, \bibinfo
  {author} {\bibfnamefont {V.}~\bibnamefont {Gruson}}, \bibinfo {author}
  {\bibfnamefont {P.}~\bibnamefont {Lassonde}}, \bibinfo {author}
  {\bibfnamefont {H.}~\bibnamefont {Ibrahim}}, \bibinfo {author} {\bibfnamefont
  {J.}~\bibnamefont {Chaillou}}, \bibinfo {author} {\bibfnamefont
  {N.}~\bibnamefont {\'Emond}}, \bibinfo {author} {\bibfnamefont {M.~R.}\
  \bibnamefont {Otto}}, \bibinfo {author} {\bibfnamefont {A.}~\bibnamefont
  {Jim\'enez-Gal\'an}}, \bibinfo {author} {\bibfnamefont {R.~E.~F.}\
  \bibnamefont {Silva}}, \bibinfo {author} {\bibfnamefont {M.}~\bibnamefont
  {Ivanov}}, \bibinfo {author} {\bibfnamefont {B.~J.}\ \bibnamefont {Siwick}},
  \bibinfo {author} {\bibfnamefont {M.}~\bibnamefont {Chaker}}, \ and\ \bibinfo
  {author} {\bibfnamefont {F.~m.~c.}\ \bibnamefont {L\'egar\'e}},\ }\href
  {\doibase 10.1103/PhysRevResearch.3.023250} {\bibfield  {journal} {\bibinfo
  {journal} {Phys. Rev. Research}\ }\textbf {\bibinfo {volume} {3}},\ \bibinfo
  {pages} {023250} (\bibinfo {year} {2021})}\BibitemShut {NoStop}%
\bibitem [{\citenamefont {Shao}\ \emph {et~al.}(2022)\citenamefont {Shao},
  \citenamefont {Lu}, \citenamefont {Zhang}, \citenamefont {Yu}, \citenamefont
  {Tohyama},\ and\ \citenamefont {Lu}}]{Shao2022PRL}%
  \BibitemOpen
  \bibfield  {author} {\bibinfo {author} {\bibfnamefont {C.}~\bibnamefont
  {Shao}}, \bibinfo {author} {\bibfnamefont {H.}~\bibnamefont {Lu}}, \bibinfo
  {author} {\bibfnamefont {X.}~\bibnamefont {Zhang}}, \bibinfo {author}
  {\bibfnamefont {C.}~\bibnamefont {Yu}}, \bibinfo {author} {\bibfnamefont
  {T.}~\bibnamefont {Tohyama}}, \ and\ \bibinfo {author} {\bibfnamefont
  {R.}~\bibnamefont {Lu}},\ }\href {\doibase 10.1103/PhysRevLett.128.047401}
  {\bibfield  {journal} {\bibinfo  {journal} {Phys. Rev. Lett.}\ }\textbf
  {\bibinfo {volume} {128}},\ \bibinfo {pages} {047401} (\bibinfo {year}
  {2022})}\BibitemShut {NoStop}%
\bibitem [{\citenamefont {Hansen}\ \emph {et~al.}(2022)\citenamefont {Hansen},
  \citenamefont {Jensen},\ and\ \citenamefont {Madsen}}]{Hansen2022PRA}%
  \BibitemOpen
  \bibfield  {author} {\bibinfo {author} {\bibfnamefont {T.}~\bibnamefont
  {Hansen}}, \bibinfo {author} {\bibfnamefont {S.~V.~B.}\ \bibnamefont
  {Jensen}}, \ and\ \bibinfo {author} {\bibfnamefont {L.~B.}\ \bibnamefont
  {Madsen}},\ }\href {\doibase 10.1103/PhysRevA.105.053118} {\bibfield
  {journal} {\bibinfo  {journal} {Phys. Rev. A}\ }\textbf {\bibinfo {volume}
  {105}},\ \bibinfo {pages} {053118} (\bibinfo {year} {2022})}\BibitemShut
  {NoStop}%
\bibitem [{\citenamefont {Uchida}\ \emph {et~al.}(2022)\citenamefont {Uchida},
  \citenamefont {Mattoni}, \citenamefont {Yonezawa}, \citenamefont {Nakamura},
  \citenamefont {Maeno},\ and\ \citenamefont {Tanaka}}]{Uchida2022PRL}%
  \BibitemOpen
  \bibfield  {author} {\bibinfo {author} {\bibfnamefont {K.}~\bibnamefont
  {Uchida}}, \bibinfo {author} {\bibfnamefont {G.}~\bibnamefont {Mattoni}},
  \bibinfo {author} {\bibfnamefont {S.}~\bibnamefont {Yonezawa}}, \bibinfo
  {author} {\bibfnamefont {F.}~\bibnamefont {Nakamura}}, \bibinfo {author}
  {\bibfnamefont {Y.}~\bibnamefont {Maeno}}, \ and\ \bibinfo {author}
  {\bibfnamefont {K.}~\bibnamefont {Tanaka}},\ }\href {\doibase
  10.1103/PhysRevLett.128.127401} {\bibfield  {journal} {\bibinfo  {journal}
  {Phys. Rev. Lett.}\ }\textbf {\bibinfo {volume} {128}},\ \bibinfo {pages}
  {127401} (\bibinfo {year} {2022})}\BibitemShut {NoStop}%
\bibitem [{\citenamefont {Murakami}\ \emph {et~al.}(2022)\citenamefont
  {Murakami}, \citenamefont {Uchida}, \citenamefont {Koga}, \citenamefont
  {Tanaka},\ and\ \citenamefont {Werner}}]{Murakami2022PRL}%
  \BibitemOpen
  \bibfield  {author} {\bibinfo {author} {\bibfnamefont {Y.}~\bibnamefont
  {Murakami}}, \bibinfo {author} {\bibfnamefont {K.}~\bibnamefont {Uchida}},
  \bibinfo {author} {\bibfnamefont {A.}~\bibnamefont {Koga}}, \bibinfo {author}
  {\bibfnamefont {K.}~\bibnamefont {Tanaka}}, \ and\ \bibinfo {author}
  {\bibfnamefont {P.}~\bibnamefont {Werner}},\ }\href {\doibase
  10.1103/PhysRevLett.129.157401} {\bibfield  {journal} {\bibinfo  {journal}
  {Phys. Rev. Lett.}\ }\textbf {\bibinfo {volume} {129}},\ \bibinfo {pages}
  {157401} (\bibinfo {year} {2022})}\BibitemShut {NoStop}%
\bibitem [{\citenamefont {Hansen}\ and\ \citenamefont
  {Madsen}(2022)}]{Hansen2022}%
  \BibitemOpen
  \bibfield  {author} {\bibinfo {author} {\bibfnamefont {T.}~\bibnamefont
  {Hansen}}\ and\ \bibinfo {author} {\bibfnamefont {L.~B.}\ \bibnamefont
  {Madsen}},\ }\href {\doibase 10.1103/PhysRevB.106.235142} {\bibfield
  {journal} {\bibinfo  {journal} {Phys. Rev. B}\ }\textbf {\bibinfo {volume}
  {106}},\ \bibinfo {pages} {235142} (\bibinfo {year} {2022})}\BibitemShut
  {NoStop}%
\bibitem [{\citenamefont {Gr\aa{}n\"as}\ \emph {et~al.}(2022)\citenamefont
  {Gr\aa{}n\"as}, \citenamefont {Vaskivskyi}, \citenamefont {Wang},
  \citenamefont {Thunstr\"om}, \citenamefont {Ghimire}, \citenamefont {Knut},
  \citenamefont {S\"oderstr\"om}, \citenamefont {Kjellsson}, \citenamefont
  {Turenne}, \citenamefont {Engel}, \citenamefont {Beye}, \citenamefont {Lu},
  \citenamefont {Higley}, \citenamefont {Reid}, \citenamefont {Schlotter},
  \citenamefont {Coslovich}, \citenamefont {Hoffmann}, \citenamefont {Kolesov},
  \citenamefont {Sch\"u\ss{}ler-Langeheine}, \citenamefont {Styervoyedov},
  \citenamefont {Tancogne-Dejean}, \citenamefont {Sentef}, \citenamefont
  {Reis}, \citenamefont {Rubio}, \citenamefont {Parkin}, \citenamefont {Karis},
  \citenamefont {Rubensson}, \citenamefont {Eriksson},\ and\ \citenamefont
  {D\"urr}}]{Granas2022PRR}%
  \BibitemOpen
  \bibfield  {author} {\bibinfo {author} {\bibfnamefont {O.}~\bibnamefont
  {Gr\aa{}n\"as}}, \bibinfo {author} {\bibfnamefont {I.}~\bibnamefont
  {Vaskivskyi}}, \bibinfo {author} {\bibfnamefont {X.}~\bibnamefont {Wang}},
  \bibinfo {author} {\bibfnamefont {P.}~\bibnamefont {Thunstr\"om}}, \bibinfo
  {author} {\bibfnamefont {S.}~\bibnamefont {Ghimire}}, \bibinfo {author}
  {\bibfnamefont {R.}~\bibnamefont {Knut}}, \bibinfo {author} {\bibfnamefont
  {J.}~\bibnamefont {S\"oderstr\"om}}, \bibinfo {author} {\bibfnamefont
  {L.}~\bibnamefont {Kjellsson}}, \bibinfo {author} {\bibfnamefont
  {D.}~\bibnamefont {Turenne}}, \bibinfo {author} {\bibfnamefont {R.~Y.}\
  \bibnamefont {Engel}}, \bibinfo {author} {\bibfnamefont {M.}~\bibnamefont
  {Beye}}, \bibinfo {author} {\bibfnamefont {J.}~\bibnamefont {Lu}}, \bibinfo
  {author} {\bibfnamefont {D.~J.}\ \bibnamefont {Higley}}, \bibinfo {author}
  {\bibfnamefont {A.~H.}\ \bibnamefont {Reid}}, \bibinfo {author}
  {\bibfnamefont {W.}~\bibnamefont {Schlotter}}, \bibinfo {author}
  {\bibfnamefont {G.}~\bibnamefont {Coslovich}}, \bibinfo {author}
  {\bibfnamefont {M.}~\bibnamefont {Hoffmann}}, \bibinfo {author}
  {\bibfnamefont {G.}~\bibnamefont {Kolesov}}, \bibinfo {author} {\bibfnamefont
  {C.}~\bibnamefont {Sch\"u\ss{}ler-Langeheine}}, \bibinfo {author}
  {\bibfnamefont {A.}~\bibnamefont {Styervoyedov}}, \bibinfo {author}
  {\bibfnamefont {N.}~\bibnamefont {Tancogne-Dejean}}, \bibinfo {author}
  {\bibfnamefont {M.~A.}\ \bibnamefont {Sentef}}, \bibinfo {author}
  {\bibfnamefont {D.~A.}\ \bibnamefont {Reis}}, \bibinfo {author}
  {\bibfnamefont {A.}~\bibnamefont {Rubio}}, \bibinfo {author} {\bibfnamefont
  {S.~S.~P.}\ \bibnamefont {Parkin}}, \bibinfo {author} {\bibfnamefont
  {O.}~\bibnamefont {Karis}}, \bibinfo {author} {\bibfnamefont {J.-E.}\
  \bibnamefont {Rubensson}}, \bibinfo {author} {\bibfnamefont {O.}~\bibnamefont
  {Eriksson}}, \ and\ \bibinfo {author} {\bibfnamefont {H.~A.}\ \bibnamefont
  {D\"urr}},\ }\href {\doibase 10.1103/PhysRevResearch.4.L032030} {\bibfield
  {journal} {\bibinfo  {journal} {Phys. Rev. Res.}\ }\textbf {\bibinfo {volume}
  {4}},\ \bibinfo {pages} {L032030} (\bibinfo {year} {2022})}\BibitemShut
  {NoStop}%
\bibitem [{\citenamefont {Slepyan}\ \emph {et~al.}(1999)\citenamefont
  {Slepyan}, \citenamefont {Maksimenko}, \citenamefont {Kalosha}, \citenamefont
  {Herrmann}, \citenamefont {Campbell},\ and\ \citenamefont
  {Hertel}}]{Slepyan1999PRA}%
  \BibitemOpen
  \bibfield  {author} {\bibinfo {author} {\bibfnamefont {G.~Y.}\ \bibnamefont
  {Slepyan}}, \bibinfo {author} {\bibfnamefont {S.~A.}\ \bibnamefont
  {Maksimenko}}, \bibinfo {author} {\bibfnamefont {V.~P.}\ \bibnamefont
  {Kalosha}}, \bibinfo {author} {\bibfnamefont {J.}~\bibnamefont {Herrmann}},
  \bibinfo {author} {\bibfnamefont {E.~E.~B.}\ \bibnamefont {Campbell}}, \ and\
  \bibinfo {author} {\bibfnamefont {I.~V.}\ \bibnamefont {Hertel}},\ }\href
  {\doibase 10.1103/PhysRevA.60.R777} {\bibfield  {journal} {\bibinfo
  {journal} {Phys. Rev. A}\ }\textbf {\bibinfo {volume} {60}},\ \bibinfo
  {pages} {R777} (\bibinfo {year} {1999})}\BibitemShut {NoStop}%
\bibitem [{\citenamefont {Mikhailov}(2007)}]{Mikhailov_2007}%
  \BibitemOpen
  \bibfield  {author} {\bibinfo {author} {\bibfnamefont {S.~A.}\ \bibnamefont
  {Mikhailov}},\ }\href {\doibase 10.1209/0295-5075/79/27002} {\bibfield
  {journal} {\bibinfo  {journal} {Europhysics Letters}\ }\textbf {\bibinfo
  {volume} {79}},\ \bibinfo {pages} {27002} (\bibinfo {year}
  {2007})}\BibitemShut {NoStop}%
\bibitem [{\citenamefont {Wright}\ \emph {et~al.}(2009)\citenamefont {Wright},
  \citenamefont {Xu}, \citenamefont {Cao},\ and\ \citenamefont
  {Zhang}}]{Wright2009}%
  \BibitemOpen
  \bibfield  {author} {\bibinfo {author} {\bibfnamefont {A.~R.}\ \bibnamefont
  {Wright}}, \bibinfo {author} {\bibfnamefont {X.~G.}\ \bibnamefont {Xu}},
  \bibinfo {author} {\bibfnamefont {J.~C.}\ \bibnamefont {Cao}}, \ and\
  \bibinfo {author} {\bibfnamefont {C.}~\bibnamefont {Zhang}},\ }\href
  {\doibase 10.1063/1.3205115} {\bibfield  {journal} {\bibinfo  {journal}
  {Applied Physics Letters}\ }\textbf {\bibinfo {volume} {95}},\ \bibinfo
  {pages} {072101} (\bibinfo {year} {2009})}\BibitemShut {NoStop}%
\bibitem [{\citenamefont {Al-Naib}\ \emph {et~al.}(2014)\citenamefont
  {Al-Naib}, \citenamefont {Sipe},\ and\ \citenamefont
  {Dignam}}]{Dignam2014PRB}%
  \BibitemOpen
  \bibfield  {author} {\bibinfo {author} {\bibfnamefont {I.}~\bibnamefont
  {Al-Naib}}, \bibinfo {author} {\bibfnamefont {J.~E.}\ \bibnamefont {Sipe}}, \
  and\ \bibinfo {author} {\bibfnamefont {M.~M.}\ \bibnamefont {Dignam}},\
  }\href {\doibase 10.1103/PhysRevB.90.245423} {\bibfield  {journal} {\bibinfo
  {journal} {Phys. Rev. B}\ }\textbf {\bibinfo {volume} {90}},\ \bibinfo
  {pages} {245423} (\bibinfo {year} {2014})}\BibitemShut {NoStop}%
\bibitem [{\citenamefont {Al-Naib}\ \emph {et~al.}(2015)\citenamefont
  {Al-Naib}, \citenamefont {Poschmann},\ and\ \citenamefont
  {Dignam}}]{Dignam2015PRB}%
  \BibitemOpen
  \bibfield  {author} {\bibinfo {author} {\bibfnamefont {I.}~\bibnamefont
  {Al-Naib}}, \bibinfo {author} {\bibfnamefont {M.}~\bibnamefont {Poschmann}},
  \ and\ \bibinfo {author} {\bibfnamefont {M.~M.}\ \bibnamefont {Dignam}},\
  }\href {\doibase 10.1103/PhysRevB.91.205407} {\bibfield  {journal} {\bibinfo
  {journal} {Phys. Rev. B}\ }\textbf {\bibinfo {volume} {91}},\ \bibinfo
  {pages} {205407} (\bibinfo {year} {2015})}\BibitemShut {NoStop}%
\bibitem [{\citenamefont {Cheng}\ \emph {et~al.}(2015)\citenamefont {Cheng},
  \citenamefont {Vermeulen},\ and\ \citenamefont {Sipe}}]{Sipe2015PRB}%
  \BibitemOpen
  \bibfield  {author} {\bibinfo {author} {\bibfnamefont {J.~L.}\ \bibnamefont
  {Cheng}}, \bibinfo {author} {\bibfnamefont {N.}~\bibnamefont {Vermeulen}}, \
  and\ \bibinfo {author} {\bibfnamefont {J.~E.}\ \bibnamefont {Sipe}},\ }\href
  {\doibase 10.1103/PhysRevB.91.235320} {\bibfield  {journal} {\bibinfo
  {journal} {Phys. Rev. B}\ }\textbf {\bibinfo {volume} {91}},\ \bibinfo
  {pages} {235320} (\bibinfo {year} {2015})}\BibitemShut {NoStop}%
\bibitem [{\citenamefont {Chizhova}\ \emph {et~al.}(2017)\citenamefont
  {Chizhova}, \citenamefont {Libisch},\ and\ \citenamefont
  {Burgd\"orfer}}]{Chizhova2017PRB}%
  \BibitemOpen
  \bibfield  {author} {\bibinfo {author} {\bibfnamefont {L.~A.}\ \bibnamefont
  {Chizhova}}, \bibinfo {author} {\bibfnamefont {F.}~\bibnamefont {Libisch}}, \
  and\ \bibinfo {author} {\bibfnamefont {J.}~\bibnamefont {Burgd\"orfer}},\
  }\href {\doibase 10.1103/PhysRevB.95.085436} {\bibfield  {journal} {\bibinfo
  {journal} {Phys. Rev. B}\ }\textbf {\bibinfo {volume} {95}},\ \bibinfo
  {pages} {085436} (\bibinfo {year} {2017})}\BibitemShut {NoStop}%
\bibitem [{\citenamefont {Mrudul}\ and\ \citenamefont
  {Dixit}(2021)}]{Dixit2021PRB}%
  \BibitemOpen
  \bibfield  {author} {\bibinfo {author} {\bibfnamefont {M.~S.}\ \bibnamefont
  {Mrudul}}\ and\ \bibinfo {author} {\bibfnamefont {G.}~\bibnamefont {Dixit}},\
  }\href {\doibase 10.1103/PhysRevB.103.094308} {\bibfield  {journal} {\bibinfo
   {journal} {Phys. Rev. B}\ }\textbf {\bibinfo {volume} {103}},\ \bibinfo
  {pages} {094308} (\bibinfo {year} {2021})}\BibitemShut {NoStop}%
\bibitem [{\citenamefont {Sato}\ \emph {et~al.}(2021)\citenamefont {Sato},
  \citenamefont {Hirori}, \citenamefont {Sanari}, \citenamefont {Kanemitsu},\
  and\ \citenamefont {Rubio}}]{Sato2021PRB}%
  \BibitemOpen
  \bibfield  {author} {\bibinfo {author} {\bibfnamefont {S.~A.}\ \bibnamefont
  {Sato}}, \bibinfo {author} {\bibfnamefont {H.}~\bibnamefont {Hirori}},
  \bibinfo {author} {\bibfnamefont {Y.}~\bibnamefont {Sanari}}, \bibinfo
  {author} {\bibfnamefont {Y.}~\bibnamefont {Kanemitsu}}, \ and\ \bibinfo
  {author} {\bibfnamefont {A.}~\bibnamefont {Rubio}},\ }\href {\doibase
  10.1103/PhysRevB.103.L041408} {\bibfield  {journal} {\bibinfo  {journal}
  {Phys. Rev. B}\ }\textbf {\bibinfo {volume} {103}},\ \bibinfo {pages}
  {L041408} (\bibinfo {year} {2021})}\BibitemShut {NoStop}%
\bibitem [{\citenamefont {Baykusheva}\ \emph
  {et~al.}(2021{\natexlab{b}})\citenamefont {Baykusheva}, \citenamefont
  {Chac\'on}, \citenamefont {Kim}, \citenamefont {Kim}, \citenamefont {Reis},\
  and\ \citenamefont {Ghimire}}]{Baykusheva2021PRA}%
  \BibitemOpen
  \bibfield  {author} {\bibinfo {author} {\bibfnamefont {D.}~\bibnamefont
  {Baykusheva}}, \bibinfo {author} {\bibfnamefont {A.}~\bibnamefont
  {Chac\'on}}, \bibinfo {author} {\bibfnamefont {D.}~\bibnamefont {Kim}},
  \bibinfo {author} {\bibfnamefont {D.~E.}\ \bibnamefont {Kim}}, \bibinfo
  {author} {\bibfnamefont {D.~A.}\ \bibnamefont {Reis}}, \ and\ \bibinfo
  {author} {\bibfnamefont {S.}~\bibnamefont {Ghimire}},\ }\href {\doibase
  10.1103/PhysRevA.103.023101} {\bibfield  {journal} {\bibinfo  {journal}
  {Phys. Rev. A}\ }\textbf {\bibinfo {volume} {103}},\ \bibinfo {pages}
  {023101} (\bibinfo {year} {2021}{\natexlab{b}})}\BibitemShut {NoStop}%
\bibitem [{\citenamefont {Kovalev}\ \emph {et~al.}(2020)\citenamefont
  {Kovalev}, \citenamefont {Dantas}, \citenamefont {Germanskiy}, \citenamefont
  {Deinert}, \citenamefont {Green}, \citenamefont {Ilyakov}, \citenamefont
  {Awari}, \citenamefont {Chen}, \citenamefont {Bawatna}, \citenamefont {Ling},
  \citenamefont {Xiu}, \citenamefont {van Loosdrecht}, \citenamefont
  {Sur{\'o}wka}, \citenamefont {Oka},\ and\ \citenamefont
  {Wang}}]{Kovalev2020}%
  \BibitemOpen
  \bibfield  {author} {\bibinfo {author} {\bibfnamefont {S.}~\bibnamefont
  {Kovalev}}, \bibinfo {author} {\bibfnamefont {R.~M.~A.}\ \bibnamefont
  {Dantas}}, \bibinfo {author} {\bibfnamefont {S.}~\bibnamefont {Germanskiy}},
  \bibinfo {author} {\bibfnamefont {J.-C.}\ \bibnamefont {Deinert}}, \bibinfo
  {author} {\bibfnamefont {B.}~\bibnamefont {Green}}, \bibinfo {author}
  {\bibfnamefont {I.}~\bibnamefont {Ilyakov}}, \bibinfo {author} {\bibfnamefont
  {N.}~\bibnamefont {Awari}}, \bibinfo {author} {\bibfnamefont
  {M.}~\bibnamefont {Chen}}, \bibinfo {author} {\bibfnamefont {M.}~\bibnamefont
  {Bawatna}}, \bibinfo {author} {\bibfnamefont {J.}~\bibnamefont {Ling}},
  \bibinfo {author} {\bibfnamefont {F.}~\bibnamefont {Xiu}}, \bibinfo {author}
  {\bibfnamefont {P.~H.~M.}\ \bibnamefont {van Loosdrecht}}, \bibinfo {author}
  {\bibfnamefont {P.}~\bibnamefont {Sur{\'o}wka}}, \bibinfo {author}
  {\bibfnamefont {T.}~\bibnamefont {Oka}}, \ and\ \bibinfo {author}
  {\bibfnamefont {Z.}~\bibnamefont {Wang}},\ }\href {\doibase
  10.1038/s41467-020-16133-8} {\bibfield  {journal} {\bibinfo  {journal}
  {Nature Communications}\ }\textbf {\bibinfo {volume} {11}},\ \bibinfo {pages}
  {2451} (\bibinfo {year} {2020})}\BibitemShut {NoStop}%
\bibitem [{\citenamefont {Paul}\ \emph {et~al.}(2013)\citenamefont {Paul},
  \citenamefont {Chang}, \citenamefont {Thompson}, \citenamefont {Stickel},
  \citenamefont {Wardini}, \citenamefont {Choi}, \citenamefont {Minot},
  \citenamefont {Hou}, \citenamefont {Nees}, \citenamefont {Norris},\ and\
  \citenamefont {Lee}}]{Paul_2013}%
  \BibitemOpen
  \bibfield  {author} {\bibinfo {author} {\bibfnamefont {M.~J.}\ \bibnamefont
  {Paul}}, \bibinfo {author} {\bibfnamefont {Y.~C.}\ \bibnamefont {Chang}},
  \bibinfo {author} {\bibfnamefont {Z.~J.}\ \bibnamefont {Thompson}}, \bibinfo
  {author} {\bibfnamefont {A.}~\bibnamefont {Stickel}}, \bibinfo {author}
  {\bibfnamefont {J.}~\bibnamefont {Wardini}}, \bibinfo {author} {\bibfnamefont
  {H.}~\bibnamefont {Choi}}, \bibinfo {author} {\bibfnamefont {E.~D.}\
  \bibnamefont {Minot}}, \bibinfo {author} {\bibfnamefont {B.}~\bibnamefont
  {Hou}}, \bibinfo {author} {\bibfnamefont {J.~A.}\ \bibnamefont {Nees}},
  \bibinfo {author} {\bibfnamefont {T.~B.}\ \bibnamefont {Norris}}, \ and\
  \bibinfo {author} {\bibfnamefont {Y.-S.}\ \bibnamefont {Lee}},\ }\href
  {\doibase 10.1088/1367-2630/15/8/085019} {\bibfield  {journal} {\bibinfo
  {journal} {New Journal of Physics}\ }\textbf {\bibinfo {volume} {15}},\
  \bibinfo {pages} {085019} (\bibinfo {year} {2013})}\BibitemShut {NoStop}%
\bibitem [{\citenamefont {Tamaya}\ and\ \citenamefont
  {Kato}(2021)}]{Tamaya2021PRB}%
  \BibitemOpen
  \bibfield  {author} {\bibinfo {author} {\bibfnamefont {T.}~\bibnamefont
  {Tamaya}}\ and\ \bibinfo {author} {\bibfnamefont {T.}~\bibnamefont {Kato}},\
  }\href {\doibase 10.1103/PhysRevB.103.205202} {\bibfield  {journal} {\bibinfo
   {journal} {Phys. Rev. B}\ }\textbf {\bibinfo {volume} {103}},\ \bibinfo
  {pages} {205202} (\bibinfo {year} {2021})}\BibitemShut {NoStop}%
\bibitem [{\citenamefont {Nishidome}\ \emph {et~al.}(2020)\citenamefont
  {Nishidome}, \citenamefont {Nagai}, \citenamefont {Uchida}, \citenamefont
  {Ichinose}, \citenamefont {Yomogida}, \citenamefont {Miyata}, \citenamefont
  {Tanaka},\ and\ \citenamefont {Yanagi}}]{Nishidome2020}%
  \BibitemOpen
  \bibfield  {author} {\bibinfo {author} {\bibfnamefont {H.}~\bibnamefont
  {Nishidome}}, \bibinfo {author} {\bibfnamefont {K.}~\bibnamefont {Nagai}},
  \bibinfo {author} {\bibfnamefont {K.}~\bibnamefont {Uchida}}, \bibinfo
  {author} {\bibfnamefont {Y.}~\bibnamefont {Ichinose}}, \bibinfo {author}
  {\bibfnamefont {Y.}~\bibnamefont {Yomogida}}, \bibinfo {author}
  {\bibfnamefont {Y.}~\bibnamefont {Miyata}}, \bibinfo {author} {\bibfnamefont
  {K.}~\bibnamefont {Tanaka}}, \ and\ \bibinfo {author} {\bibfnamefont
  {K.}~\bibnamefont {Yanagi}},\ }\href {\doibase 10.1021/acs.nanolett.0c02717}
  {\bibfield  {journal} {\bibinfo  {journal} {Nano Letters}\ }\textbf {\bibinfo
  {volume} {20}},\ \bibinfo {pages} {6215} (\bibinfo {year}
  {2020})}\BibitemShut {NoStop}%
\bibitem [{\citenamefont {de~Vega}\ \emph {et~al.}(2020)\citenamefont
  {de~Vega}, \citenamefont {Cox}, \citenamefont {Sols},\ and\ \citenamefont
  {Garc\'{\i}a~de Abajo}}]{Javier2020PRR}%
  \BibitemOpen
  \bibfield  {author} {\bibinfo {author} {\bibfnamefont {S.}~\bibnamefont
  {de~Vega}}, \bibinfo {author} {\bibfnamefont {J.~D.}\ \bibnamefont {Cox}},
  \bibinfo {author} {\bibfnamefont {F.}~\bibnamefont {Sols}}, \ and\ \bibinfo
  {author} {\bibfnamefont {F.~J.}\ \bibnamefont {Garc\'{\i}a~de Abajo}},\
  }\href {\doibase 10.1103/PhysRevResearch.2.013313} {\bibfield  {journal}
  {\bibinfo  {journal} {Phys. Rev. Res.}\ }\textbf {\bibinfo {volume} {2}},\
  \bibinfo {pages} {013313} (\bibinfo {year} {2020})}\BibitemShut {NoStop}%
\bibitem [{\citenamefont {Ajiki}\ and\ \citenamefont
  {Ando}(1993)}]{Ando1993JPSJ}%
  \BibitemOpen
  \bibfield  {author} {\bibinfo {author} {\bibfnamefont {H.}~\bibnamefont
  {Ajiki}}\ and\ \bibinfo {author} {\bibfnamefont {T.}~\bibnamefont {Ando}},\
  }\href {\doibase 10.1143/JPSJ.62.1255} {\bibfield  {journal} {\bibinfo
  {journal} {Journal of the Physical Society of Japan}\ }\textbf {\bibinfo
  {volume} {62}},\ \bibinfo {pages} {1255} (\bibinfo {year}
  {1993})}\BibitemShut {NoStop}%
\bibitem [{\citenamefont {Ajiki}\ and\ \citenamefont {Ando}(1994)}]{Ando1994}%
  \BibitemOpen
  \bibfield  {author} {\bibinfo {author} {\bibfnamefont {H.}~\bibnamefont
  {Ajiki}}\ and\ \bibinfo {author} {\bibfnamefont {T.}~\bibnamefont {Ando}},\
  }\href {\doibase https://doi.org/10.1016/0921-4526(94)91112-6} {\bibfield
  {journal} {\bibinfo  {journal} {Physica B: Condensed Matter}\ }\textbf
  {\bibinfo {volume} {201}},\ \bibinfo {pages} {349} (\bibinfo {year}
  {1994})}\BibitemShut {NoStop}%
\bibitem [{\citenamefont {Zaric}\ \emph {et~al.}(2004)\citenamefont {Zaric},
  \citenamefont {Ostojic}, \citenamefont {Kono}, \citenamefont {Shaver},
  \citenamefont {Moore}, \citenamefont {Strano}, \citenamefont {Hauge},
  \citenamefont {Smalley},\ and\ \citenamefont {Wei}}]{Wei2004Science}%
  \BibitemOpen
  \bibfield  {author} {\bibinfo {author} {\bibfnamefont {S.}~\bibnamefont
  {Zaric}}, \bibinfo {author} {\bibfnamefont {G.~N.}\ \bibnamefont {Ostojic}},
  \bibinfo {author} {\bibfnamefont {J.}~\bibnamefont {Kono}}, \bibinfo {author}
  {\bibfnamefont {J.}~\bibnamefont {Shaver}}, \bibinfo {author} {\bibfnamefont
  {V.~C.}\ \bibnamefont {Moore}}, \bibinfo {author} {\bibfnamefont {M.~S.}\
  \bibnamefont {Strano}}, \bibinfo {author} {\bibfnamefont {R.~H.}\
  \bibnamefont {Hauge}}, \bibinfo {author} {\bibfnamefont {R.~E.}\ \bibnamefont
  {Smalley}}, \ and\ \bibinfo {author} {\bibfnamefont {X.}~\bibnamefont
  {Wei}},\ }\href {\doibase 10.1126/science.1096524} {\bibfield  {journal}
  {\bibinfo  {journal} {Science}\ }\textbf {\bibinfo {volume} {304}},\ \bibinfo
  {pages} {1129} (\bibinfo {year} {2004})}\BibitemShut {NoStop}%
\bibitem [{\citenamefont {Matsunaga}\ \emph {et~al.}(2008)\citenamefont
  {Matsunaga}, \citenamefont {Matsuda},\ and\ \citenamefont
  {Kanemitsu}}]{Matsunaga2008PRL}%
  \BibitemOpen
  \bibfield  {author} {\bibinfo {author} {\bibfnamefont {R.}~\bibnamefont
  {Matsunaga}}, \bibinfo {author} {\bibfnamefont {K.}~\bibnamefont {Matsuda}},
  \ and\ \bibinfo {author} {\bibfnamefont {Y.}~\bibnamefont {Kanemitsu}},\
  }\href {\doibase 10.1103/PhysRevLett.101.147404} {\bibfield  {journal}
  {\bibinfo  {journal} {Phys. Rev. Lett.}\ }\textbf {\bibinfo {volume} {101}},\
  \bibinfo {pages} {147404} (\bibinfo {year} {2008})}\BibitemShut {NoStop}%
\bibitem [{\citenamefont {Li}\ \emph {et~al.}(2020)\citenamefont {Li},
  \citenamefont {Golez}, \citenamefont {Mazza}, \citenamefont {Millis},
  \citenamefont {Georges},\ and\ \citenamefont {Eckstein}}]{Li2020PRB}%
  \BibitemOpen
  \bibfield  {author} {\bibinfo {author} {\bibfnamefont {J.}~\bibnamefont
  {Li}}, \bibinfo {author} {\bibfnamefont {D.}~\bibnamefont {Golez}}, \bibinfo
  {author} {\bibfnamefont {G.}~\bibnamefont {Mazza}}, \bibinfo {author}
  {\bibfnamefont {A.~J.}\ \bibnamefont {Millis}}, \bibinfo {author}
  {\bibfnamefont {A.}~\bibnamefont {Georges}}, \ and\ \bibinfo {author}
  {\bibfnamefont {M.}~\bibnamefont {Eckstein}},\ }\href {\doibase
  10.1103/PhysRevB.101.205140} {\bibfield  {journal} {\bibinfo  {journal}
  {Phys. Rev. B}\ }\textbf {\bibinfo {volume} {101}},\ \bibinfo {pages}
  {205140} (\bibinfo {year} {2020})}\BibitemShut {NoStop}%
\bibitem [{\citenamefont {Sch\"uler}\ \emph {et~al.}(2021)\citenamefont
  {Sch\"uler}, \citenamefont {Marks}, \citenamefont {Murakami}, \citenamefont
  {Jia},\ and\ \citenamefont {Devereaux}}]{Michael2021PRB}%
  \BibitemOpen
  \bibfield  {author} {\bibinfo {author} {\bibfnamefont {M.}~\bibnamefont
  {Sch\"uler}}, \bibinfo {author} {\bibfnamefont {J.~A.}\ \bibnamefont
  {Marks}}, \bibinfo {author} {\bibfnamefont {Y.}~\bibnamefont {Murakami}},
  \bibinfo {author} {\bibfnamefont {C.}~\bibnamefont {Jia}}, \ and\ \bibinfo
  {author} {\bibfnamefont {T.~P.}\ \bibnamefont {Devereaux}},\ }\href {\doibase
  10.1103/PhysRevB.103.155409} {\bibfield  {journal} {\bibinfo  {journal}
  {Phys. Rev. B}\ }\textbf {\bibinfo {volume} {103}},\ \bibinfo {pages}
  {155409} (\bibinfo {year} {2021})}\BibitemShut {NoStop}%
\bibitem [{\citenamefont {Murakami}\ and\ \citenamefont
  {Sch\"uler}(2022)}]{Murakami2022PRB}%
  \BibitemOpen
  \bibfield  {author} {\bibinfo {author} {\bibfnamefont {Y.}~\bibnamefont
  {Murakami}}\ and\ \bibinfo {author} {\bibfnamefont {M.}~\bibnamefont
  {Sch\"uler}},\ }\href {\doibase 10.1103/PhysRevB.106.035204} {\bibfield
  {journal} {\bibinfo  {journal} {Phys. Rev. B}\ }\textbf {\bibinfo {volume}
  {106}},\ \bibinfo {pages} {035204} (\bibinfo {year} {2022})}\BibitemShut
  {NoStop}%
\bibitem [{\citenamefont {Hamada}\ \emph {et~al.}(1992)\citenamefont {Hamada},
  \citenamefont {Sawada},\ and\ \citenamefont {Oshiyama}}]{Hamada1992PRL}%
  \BibitemOpen
  \bibfield  {author} {\bibinfo {author} {\bibfnamefont {N.}~\bibnamefont
  {Hamada}}, \bibinfo {author} {\bibfnamefont {S.-i.}\ \bibnamefont {Sawada}},
  \ and\ \bibinfo {author} {\bibfnamefont {A.}~\bibnamefont {Oshiyama}},\
  }\href {\doibase 10.1103/PhysRevLett.68.1579} {\bibfield  {journal} {\bibinfo
   {journal} {Phys. Rev. Lett.}\ }\textbf {\bibinfo {volume} {68}},\ \bibinfo
  {pages} {1579} (\bibinfo {year} {1992})}\BibitemShut {NoStop}%
\bibitem [{Note1()}]{Note1}%
  \BibitemOpen
  \bibinfo {note} {See Supplemental Material at [URL will be inserted by
  publisher] for the details of the model and numerical method; the derivation
  of Eqs.~\protect \textup {\hbox {\mathsurround \z@ \protect \normalfont
  (\ignorespaces \ref {eq:J_t}\unskip \@@italiccorr )}} and ~\protect \textup
  {\hbox {\mathsurround \z@ \protect \normalfont (\ignorespaces \ref
  {eq:sig_3_2}\unskip \@@italiccorr )}}; the doping effects; the dependence of
  the excitation frequency; the dependence on the chiral index; the effects of
  the relaxation times. The Supplemental Material also contains Refs.~\cite
  {Khosravi2009,Dimitrovsk2017,Dvuzhilova2021,Wilhelm2021PRB}.}\BibitemShut
  {Stop}%
\bibitem [{\citenamefont {Yanagi}\ \emph {et~al.}(2008)\citenamefont {Yanagi},
  \citenamefont {Miyata},\ and\ \citenamefont {Kataura}}]{Yanagi_2008}%
  \BibitemOpen
  \bibfield  {author} {\bibinfo {author} {\bibfnamefont {K.}~\bibnamefont
  {Yanagi}}, \bibinfo {author} {\bibfnamefont {Y.}~\bibnamefont {Miyata}}, \
  and\ \bibinfo {author} {\bibfnamefont {H.}~\bibnamefont {Kataura}},\ }\href
  {\doibase 10.1143/APEX.1.034003} {\bibfield  {journal} {\bibinfo  {journal}
  {Applied Physics Express}\ }\textbf {\bibinfo {volume} {1}},\ \bibinfo
  {pages} {034003} (\bibinfo {year} {2008})}\BibitemShut {NoStop}%
\bibitem [{\citenamefont {Huttner}\ \emph {et~al.}(2017)\citenamefont
  {Huttner}, \citenamefont {Kira},\ and\ \citenamefont {Koch}}]{Huttner2017}%
  \BibitemOpen
  \bibfield  {author} {\bibinfo {author} {\bibfnamefont {U.}~\bibnamefont
  {Huttner}}, \bibinfo {author} {\bibfnamefont {M.}~\bibnamefont {Kira}}, \
  and\ \bibinfo {author} {\bibfnamefont {S.~W.}\ \bibnamefont {Koch}},\ }\href
  {\doibase 10.1002/lpor.201700049} {\bibfield  {journal} {\bibinfo  {journal}
  {Las. Photon. Rev.}\ }\textbf {\bibinfo {volume} {11}},\ \bibinfo {pages}
  {1700049} (\bibinfo {year} {2017})}\BibitemShut {NoStop}%
\bibitem [{\citenamefont {Yue}\ and\ \citenamefont
  {Gaarde}(2022)}]{Yue2022tutorial}%
  \BibitemOpen
  \bibfield  {author} {\bibinfo {author} {\bibfnamefont {L.}~\bibnamefont
  {Yue}}\ and\ \bibinfo {author} {\bibfnamefont {M.~B.}\ \bibnamefont
  {Gaarde}},\ }\href {\doibase 10.1364/JOSAB.448602} {\bibfield  {journal}
  {\bibinfo  {journal} {J. Opt. Soc. Am. B}\ }\textbf {\bibinfo {volume}
  {39}},\ \bibinfo {pages} {535} (\bibinfo {year} {2022})}\BibitemShut
  {NoStop}%
\bibitem [{\citenamefont {Habenicht}\ \emph {et~al.}(2007)\citenamefont
  {Habenicht}, \citenamefont {Kamisaka}, \citenamefont {Yamashita},\ and\
  \citenamefont {Prezhdo}}]{Habenicht2007}%
  \BibitemOpen
  \bibfield  {author} {\bibinfo {author} {\bibfnamefont {B.~F.}\ \bibnamefont
  {Habenicht}}, \bibinfo {author} {\bibfnamefont {H.}~\bibnamefont {Kamisaka}},
  \bibinfo {author} {\bibfnamefont {K.}~\bibnamefont {Yamashita}}, \ and\
  \bibinfo {author} {\bibfnamefont {O.~V.}\ \bibnamefont {Prezhdo}},\ }\href
  {\doibase 10.1021/nl0710699} {\bibfield  {journal} {\bibinfo  {journal} {Nano
  Letters}\ }\textbf {\bibinfo {volume} {7}},\ \bibinfo {pages} {3260}
  (\bibinfo {year} {2007})}\BibitemShut {NoStop}%
\bibitem [{\citenamefont {George}\ \emph {et~al.}(2008)\citenamefont {George},
  \citenamefont {Strait}, \citenamefont {Dawlaty}, \citenamefont {Shivaraman},
  \citenamefont {Chandrashekhar}, \citenamefont {Rana},\ and\ \citenamefont
  {Spencer}}]{George2008}%
  \BibitemOpen
  \bibfield  {author} {\bibinfo {author} {\bibfnamefont {P.~A.}\ \bibnamefont
  {George}}, \bibinfo {author} {\bibfnamefont {J.}~\bibnamefont {Strait}},
  \bibinfo {author} {\bibfnamefont {J.}~\bibnamefont {Dawlaty}}, \bibinfo
  {author} {\bibfnamefont {S.}~\bibnamefont {Shivaraman}}, \bibinfo {author}
  {\bibfnamefont {M.}~\bibnamefont {Chandrashekhar}}, \bibinfo {author}
  {\bibfnamefont {F.}~\bibnamefont {Rana}}, \ and\ \bibinfo {author}
  {\bibfnamefont {M.~G.}\ \bibnamefont {Spencer}},\ }\href {\doibase
  10.1021/nl8019399} {\bibfield  {journal} {\bibinfo  {journal} {Nano Letters}\
  }\textbf {\bibinfo {volume} {8}},\ \bibinfo {pages} {4248} (\bibinfo {year}
  {2008})}\BibitemShut {NoStop}%
\bibitem [{\citenamefont {Tani}\ \emph {et~al.}(2012)\citenamefont {Tani},
  \citenamefont {Blanchard},\ and\ \citenamefont {Tanaka}}]{Tani2012PRL}%
  \BibitemOpen
  \bibfield  {author} {\bibinfo {author} {\bibfnamefont {S.}~\bibnamefont
  {Tani}}, \bibinfo {author} {\bibfnamefont {F.~m.~c.}\ \bibnamefont
  {Blanchard}}, \ and\ \bibinfo {author} {\bibfnamefont {K.}~\bibnamefont
  {Tanaka}},\ }\href {\doibase 10.1103/PhysRevLett.109.166603} {\bibfield
  {journal} {\bibinfo  {journal} {Phys. Rev. Lett.}\ }\textbf {\bibinfo
  {volume} {109}},\ \bibinfo {pages} {166603} (\bibinfo {year}
  {2012})}\BibitemShut {NoStop}%
\bibitem [{\citenamefont {Heide}\ \emph {et~al.}(2021)\citenamefont {Heide},
  \citenamefont {Eckstein}, \citenamefont {Boolakee}, \citenamefont {Gerner},
  \citenamefont {Weber}, \citenamefont {Franco},\ and\ \citenamefont
  {Hommelhoff}}]{Heide2021}%
  \BibitemOpen
  \bibfield  {author} {\bibinfo {author} {\bibfnamefont {C.}~\bibnamefont
  {Heide}}, \bibinfo {author} {\bibfnamefont {T.}~\bibnamefont {Eckstein}},
  \bibinfo {author} {\bibfnamefont {T.}~\bibnamefont {Boolakee}}, \bibinfo
  {author} {\bibfnamefont {C.}~\bibnamefont {Gerner}}, \bibinfo {author}
  {\bibfnamefont {H.~B.}\ \bibnamefont {Weber}}, \bibinfo {author}
  {\bibfnamefont {I.}~\bibnamefont {Franco}}, \ and\ \bibinfo {author}
  {\bibfnamefont {P.}~\bibnamefont {Hommelhoff}},\ }\href {\doibase
  10.1021/acs.nanolett.1c02538} {\bibfield  {journal} {\bibinfo  {journal}
  {Nano Letters}\ }\textbf {\bibinfo {volume} {21}},\ \bibinfo {pages} {9403}
  (\bibinfo {year} {2021})}\BibitemShut {NoStop}%
\bibitem [{\citenamefont {Jackson}(1998)}]{Jackson1998Book}%
  \BibitemOpen
  \bibfield  {author} {\bibinfo {author} {\bibfnamefont {J.~D.}\ \bibnamefont
  {Jackson}},\ }\href@noop {} {\emph {\bibinfo {title} {Classical
  Electrodynamics}}}\ (\bibinfo  {publisher} {Wiley},\ \bibinfo {address} {New
  York},\ \bibinfo {year} {1998})\BibitemShut {NoStop}%
\bibitem [{\citenamefont {Neufeld}\ \emph {et~al.}(2019)\citenamefont
  {Neufeld}, \citenamefont {Podolsky},\ and\ \citenamefont
  {Cohen}}]{Cohen2019NatCom}%
  \BibitemOpen
  \bibfield  {author} {\bibinfo {author} {\bibfnamefont {O.}~\bibnamefont
  {Neufeld}}, \bibinfo {author} {\bibfnamefont {D.}~\bibnamefont {Podolsky}}, \
  and\ \bibinfo {author} {\bibfnamefont {O.}~\bibnamefont {Cohen}},\ }\href
  {\doibase 10.1038/s41467-018-07935-y} {\bibfield  {journal} {\bibinfo
  {journal} {Nat. Comm.}\ }\textbf {\bibinfo {volume} {10}},\ \bibinfo {pages}
  {405} (\bibinfo {year} {2019})}\BibitemShut {NoStop}%
\bibitem [{\citenamefont {Ishikawa}(2010)}]{Ishikawa2010PRB}%
  \BibitemOpen
  \bibfield  {author} {\bibinfo {author} {\bibfnamefont {K.~L.}\ \bibnamefont
  {Ishikawa}},\ }\href {\doibase 10.1103/PhysRevB.82.201402} {\bibfield
  {journal} {\bibinfo  {journal} {Phys. Rev. B}\ }\textbf {\bibinfo {volume}
  {82}},\ \bibinfo {pages} {201402} (\bibinfo {year} {2010})}\BibitemShut
  {NoStop}%
\bibitem [{\citenamefont {\'{O}scar Zurr\'{o}n-Cifuentes}\ \emph
  {et~al.}(2020)\citenamefont {\'{O}scar Zurr\'{o}n-Cifuentes}, \citenamefont
  {Boyero-Garc\'{i}a}, \citenamefont {Hern\'{a}ndez-Garc\'{i}a},\ and\
  \citenamefont {Plaja}}]{Zurron-Cifuentes:20}%
  \BibitemOpen
  \bibfield  {author} {\bibinfo {author} {\bibnamefont {\'{O}scar
  Zurr\'{o}n-Cifuentes}}, \bibinfo {author} {\bibfnamefont {R.}~\bibnamefont
  {Boyero-Garc\'{i}a}}, \bibinfo {author} {\bibfnamefont {C.}~\bibnamefont
  {Hern\'{a}ndez-Garc\'{i}a}}, \ and\ \bibinfo {author} {\bibfnamefont
  {L.}~\bibnamefont {Plaja}},\ }\href {\doibase 10.1364/OE.394714} {\bibfield
  {journal} {\bibinfo  {journal} {Opt. Express}\ }\textbf {\bibinfo {volume}
  {28}},\ \bibinfo {pages} {19760} (\bibinfo {year} {2020})}\BibitemShut
  {NoStop}%
\bibitem [{\citenamefont {Malic}\ \emph {et~al.}(2011)\citenamefont {Malic},
  \citenamefont {Winzer}, \citenamefont {Bobkin},\ and\ \citenamefont
  {Knorr}}]{Malic2011}%
  \BibitemOpen
  \bibfield  {author} {\bibinfo {author} {\bibfnamefont {E.}~\bibnamefont
  {Malic}}, \bibinfo {author} {\bibfnamefont {T.}~\bibnamefont {Winzer}},
  \bibinfo {author} {\bibfnamefont {E.}~\bibnamefont {Bobkin}}, \ and\ \bibinfo
  {author} {\bibfnamefont {A.}~\bibnamefont {Knorr}},\ }\href {\doibase
  10.1103/PhysRevB.84.205406} {\bibfield  {journal} {\bibinfo  {journal} {Phys.
  Rev. B}\ }\textbf {\bibinfo {volume} {84}},\ \bibinfo {pages} {205406}
  (\bibinfo {year} {2011})}\BibitemShut {NoStop}%
\bibitem [{\citenamefont {Kemper}\ \emph {et~al.}(2013)\citenamefont {Kemper},
  \citenamefont {Moritz}, \citenamefont {Freericks},\ and\ \citenamefont
  {Devereaux}}]{Kemper2013NJP}%
  \BibitemOpen
  \bibfield  {author} {\bibinfo {author} {\bibfnamefont {A.~F.}\ \bibnamefont
  {Kemper}}, \bibinfo {author} {\bibfnamefont {B.}~\bibnamefont {Moritz}},
  \bibinfo {author} {\bibfnamefont {J.~K.}\ \bibnamefont {Freericks}}, \ and\
  \bibinfo {author} {\bibfnamefont {T.~P.}\ \bibnamefont {Devereaux}},\ }\href
  {http://stacks.iop.org/1367-2630/15/i=2/a=023003} {\bibfield  {journal}
  {\bibinfo  {journal} {New J. Phys.}\ }\textbf {\bibinfo {volume} {15}},\
  \bibinfo {pages} {023003} (\bibinfo {year} {2013})}\BibitemShut {NoStop}%
\bibitem [{\citenamefont {Stefanucci}\ and\ \citenamefont
  {Leeuwen}(2013)}]{stefanucci_nonequilibrium_2013}%
  \BibitemOpen
  \bibfield  {author} {\bibinfo {author} {\bibfnamefont {G.}~\bibnamefont
  {Stefanucci}}\ and\ \bibinfo {author} {\bibfnamefont {R.~v.}\ \bibnamefont
  {Leeuwen}},\ }\href@noop {} {\bibfield  {journal} {\bibinfo  {journal}
  {``Nonequilibrium Many-Body Theory of Quantum Systems: A Modern
  Introduction", Cambridge University Press}\ } (\bibinfo {year}
  {2013})}\BibitemShut {NoStop}%
\bibitem [{\citenamefont {Sch\"{u}ler}\ \emph {et~al.}(2020)\citenamefont
  {Sch\"{u}ler}, \citenamefont {Gole\v{z}}, \citenamefont {Murakami},
  \citenamefont {Bittner}, \citenamefont {Herrmann}, \citenamefont {Strand},
  \citenamefont {Werner},\ and\ \citenamefont {Eckstein}}]{Nessi2020}%
  \BibitemOpen
  \bibfield  {author} {\bibinfo {author} {\bibfnamefont {M.}~\bibnamefont
  {Sch\"{u}ler}}, \bibinfo {author} {\bibfnamefont {D.}~\bibnamefont
  {Gole\v{z}}}, \bibinfo {author} {\bibfnamefont {Y.}~\bibnamefont {Murakami}},
  \bibinfo {author} {\bibfnamefont {N.}~\bibnamefont {Bittner}}, \bibinfo
  {author} {\bibfnamefont {A.}~\bibnamefont {Herrmann}}, \bibinfo {author}
  {\bibfnamefont {H.~U.}\ \bibnamefont {Strand}}, \bibinfo {author}
  {\bibfnamefont {P.}~\bibnamefont {Werner}}, \ and\ \bibinfo {author}
  {\bibfnamefont {M.}~\bibnamefont {Eckstein}},\ }\href {\doibase
  https://doi.org/10.1016/j.cpc.2020.107484} {\bibfield  {journal} {\bibinfo
  {journal} {Computer Physics Communications}\ }\textbf {\bibinfo {volume}
  {257}},\ \bibinfo {pages} {107484} (\bibinfo {year} {2020})}\BibitemShut
  {NoStop}%
\bibitem [{\citenamefont {Ridley}\ \emph {et~al.}(2022)\citenamefont {Ridley},
  \citenamefont {Talarico}, \citenamefont {Karlsson}, \citenamefont {Gullo},\
  and\ \citenamefont {Tuovinen}}]{ridley2022manybody}%
  \BibitemOpen
  \bibfield  {author} {\bibinfo {author} {\bibfnamefont {M.}~\bibnamefont
  {Ridley}}, \bibinfo {author} {\bibfnamefont {N.~W.}\ \bibnamefont
  {Talarico}}, \bibinfo {author} {\bibfnamefont {D.}~\bibnamefont {Karlsson}},
  \bibinfo {author} {\bibfnamefont {N.~L.}\ \bibnamefont {Gullo}}, \ and\
  \bibinfo {author} {\bibfnamefont {R.}~\bibnamefont {Tuovinen}},\ }\href
  {\doibase 10.1088/1751-8121/ac7119} {\bibfield  {journal} {\bibinfo
  {journal} {Journal of Physics A: Mathematical and Theoretical}\ }\textbf
  {\bibinfo {volume} {55}},\ \bibinfo {pages} {273001} (\bibinfo {year}
  {2022})}\BibitemShut {NoStop}%
\bibitem [{\citenamefont {Rostami}\ and\ \citenamefont
  {Cappelluti}(2021)}]{Rostami2021PRB}%
  \BibitemOpen
  \bibfield  {author} {\bibinfo {author} {\bibfnamefont {H.}~\bibnamefont
  {Rostami}}\ and\ \bibinfo {author} {\bibfnamefont {E.}~\bibnamefont
  {Cappelluti}},\ }\href {\doibase 10.1103/PhysRevB.103.125415} {\bibfield
  {journal} {\bibinfo  {journal} {Phys. Rev. B}\ }\textbf {\bibinfo {volume}
  {103}},\ \bibinfo {pages} {125415} (\bibinfo {year} {2021})}\BibitemShut
  {NoStop}%
\bibitem [{\citenamefont {Rubio}\ \emph {et~al.}(1994)\citenamefont {Rubio},
  \citenamefont {Corkill},\ and\ \citenamefont {Cohen}}]{Rubio1994PRB}%
  \BibitemOpen
  \bibfield  {author} {\bibinfo {author} {\bibfnamefont {A.}~\bibnamefont
  {Rubio}}, \bibinfo {author} {\bibfnamefont {J.~L.}\ \bibnamefont {Corkill}},
  \ and\ \bibinfo {author} {\bibfnamefont {M.~L.}\ \bibnamefont {Cohen}},\
  }\href {\doibase 10.1103/PhysRevB.49.5081} {\bibfield  {journal} {\bibinfo
  {journal} {Phys. Rev. B}\ }\textbf {\bibinfo {volume} {49}},\ \bibinfo
  {pages} {5081} (\bibinfo {year} {1994})}\BibitemShut {NoStop}%
\bibitem [{\citenamefont {Chopra}\ \emph {et~al.}(1995)\citenamefont {Chopra},
  \citenamefont {Luyken}, \citenamefont {Cherrey}, \citenamefont {Crespi},
  \citenamefont {Cohen}, \citenamefont {Louie},\ and\ \citenamefont
  {Zettl}}]{Nasreen1995}%
  \BibitemOpen
  \bibfield  {author} {\bibinfo {author} {\bibfnamefont {N.~G.}\ \bibnamefont
  {Chopra}}, \bibinfo {author} {\bibfnamefont {R.~J.}\ \bibnamefont {Luyken}},
  \bibinfo {author} {\bibfnamefont {K.}~\bibnamefont {Cherrey}}, \bibinfo
  {author} {\bibfnamefont {V.~H.}\ \bibnamefont {Crespi}}, \bibinfo {author}
  {\bibfnamefont {M.~L.}\ \bibnamefont {Cohen}}, \bibinfo {author}
  {\bibfnamefont {S.~G.}\ \bibnamefont {Louie}}, \ and\ \bibinfo {author}
  {\bibfnamefont {A.}~\bibnamefont {Zettl}},\ }\href {\doibase
  10.1126/science.269.5226.966} {\bibfield  {journal} {\bibinfo  {journal}
  {Science}\ }\textbf {\bibinfo {volume} {269}},\ \bibinfo {pages} {966}
  (\bibinfo {year} {1995})}\BibitemShut {NoStop}%
\bibitem [{\citenamefont {Tenne}\ \emph {et~al.}(1992)\citenamefont {Tenne},
  \citenamefont {Margulis}, \citenamefont {Genut},\ and\ \citenamefont
  {Hodes}}]{Tenne1992}%
  \BibitemOpen
  \bibfield  {author} {\bibinfo {author} {\bibfnamefont {R.}~\bibnamefont
  {Tenne}}, \bibinfo {author} {\bibfnamefont {L.}~\bibnamefont {Margulis}},
  \bibinfo {author} {\bibfnamefont {M.}~\bibnamefont {Genut}}, \ and\ \bibinfo
  {author} {\bibfnamefont {G.}~\bibnamefont {Hodes}},\ }\href {\doibase
  10.1038/360444a0} {\bibfield  {journal} {\bibinfo  {journal} {Nature}\
  }\textbf {\bibinfo {volume} {360}},\ \bibinfo {pages} {444} (\bibinfo {year}
  {1992})}\BibitemShut {NoStop}%
\bibitem [{\citenamefont {Khosravi}\ \emph {et~al.}(2009)\citenamefont
  {Khosravi}, \citenamefont {Daneshfar},\ and\ \citenamefont
  {Bahari}}]{Khosravi2009}%
  \BibitemOpen
  \bibfield  {author} {\bibinfo {author} {\bibfnamefont {H.}~\bibnamefont
  {Khosravi}}, \bibinfo {author} {\bibfnamefont {N.}~\bibnamefont {Daneshfar}},
  \ and\ \bibinfo {author} {\bibfnamefont {A.}~\bibnamefont {Bahari}},\ }\href
  {\doibase 10.1364/OL.34.001723} {\bibfield  {journal} {\bibinfo  {journal}
  {Opt. Lett.}\ }\textbf {\bibinfo {volume} {34}},\ \bibinfo {pages} {1723}
  (\bibinfo {year} {2009})}\BibitemShut {NoStop}%
\bibitem [{\citenamefont {Dimitrovski}\ \emph {et~al.}(2017)\citenamefont
  {Dimitrovski}, \citenamefont {Madsen},\ and\ \citenamefont
  {Pedersen}}]{Dimitrovsk2017}%
  \BibitemOpen
  \bibfield  {author} {\bibinfo {author} {\bibfnamefont {D.}~\bibnamefont
  {Dimitrovski}}, \bibinfo {author} {\bibfnamefont {L.~B.}\ \bibnamefont
  {Madsen}}, \ and\ \bibinfo {author} {\bibfnamefont {T.~G.}\ \bibnamefont
  {Pedersen}},\ }\href {\doibase 10.1103/PhysRevB.95.035405} {\bibfield
  {journal} {\bibinfo  {journal} {Phys. Rev. B}\ }\textbf {\bibinfo {volume}
  {95}},\ \bibinfo {pages} {035405} (\bibinfo {year} {2017})}\BibitemShut
  {NoStop}%
\bibitem [{\citenamefont {Dvuzhilova}\ \emph {et~al.}(2021)\citenamefont
  {Dvuzhilova}, \citenamefont {Dvuzhilov}, \citenamefont {Konobeeva},\ and\
  \citenamefont {Belonenko}}]{Dvuzhilova2021}%
  \BibitemOpen
  \bibfield  {author} {\bibinfo {author} {\bibfnamefont {Y.~V.}\ \bibnamefont
  {Dvuzhilova}}, \bibinfo {author} {\bibfnamefont {I.~S.}\ \bibnamefont
  {Dvuzhilov}}, \bibinfo {author} {\bibfnamefont {N.~N.}\ \bibnamefont
  {Konobeeva}}, \ and\ \bibinfo {author} {\bibfnamefont {M.~B.}\ \bibnamefont
  {Belonenko}},\ }\href {\doibase 10.1142/S021797922150017X} {\bibfield
  {journal} {\bibinfo  {journal} {International Journal of Modern Physics B}\
  }\textbf {\bibinfo {volume} {35}},\ \bibinfo {pages} {2150017} (\bibinfo
  {year} {2021})}\BibitemShut {NoStop}%
\bibitem [{\citenamefont {Wilhelm}\ \emph {et~al.}(2021)\citenamefont
  {Wilhelm}, \citenamefont {Gr\"ossing}, \citenamefont {Seith}, \citenamefont
  {Crewse}, \citenamefont {Nitsch}, \citenamefont {Weigl}, \citenamefont
  {Schmid},\ and\ \citenamefont {Evers}}]{Wilhelm2021PRB}%
  \BibitemOpen
  \bibfield  {author} {\bibinfo {author} {\bibfnamefont {J.}~\bibnamefont
  {Wilhelm}}, \bibinfo {author} {\bibfnamefont {P.}~\bibnamefont {Gr\"ossing}},
  \bibinfo {author} {\bibfnamefont {A.}~\bibnamefont {Seith}}, \bibinfo
  {author} {\bibfnamefont {J.}~\bibnamefont {Crewse}}, \bibinfo {author}
  {\bibfnamefont {M.}~\bibnamefont {Nitsch}}, \bibinfo {author} {\bibfnamefont
  {L.}~\bibnamefont {Weigl}}, \bibinfo {author} {\bibfnamefont
  {C.}~\bibnamefont {Schmid}}, \ and\ \bibinfo {author} {\bibfnamefont
  {F.}~\bibnamefont {Evers}},\ }\href {\doibase 10.1103/PhysRevB.103.125419}
  {\bibfield  {journal} {\bibinfo  {journal} {Phys. Rev. B}\ }\textbf {\bibinfo
  {volume} {103}},\ \bibinfo {pages} {125419} (\bibinfo {year}
  {2021})}\BibitemShut {NoStop}%
\end{thebibliography}%

\clearpage

\section{Detail of numerical simulation}
\subsection{Tight Binding Model}
\begin{figure}[t]
  \centering
    \hspace{-0.cm}
    \vspace{0.0cm}
\includegraphics[width=70mm]{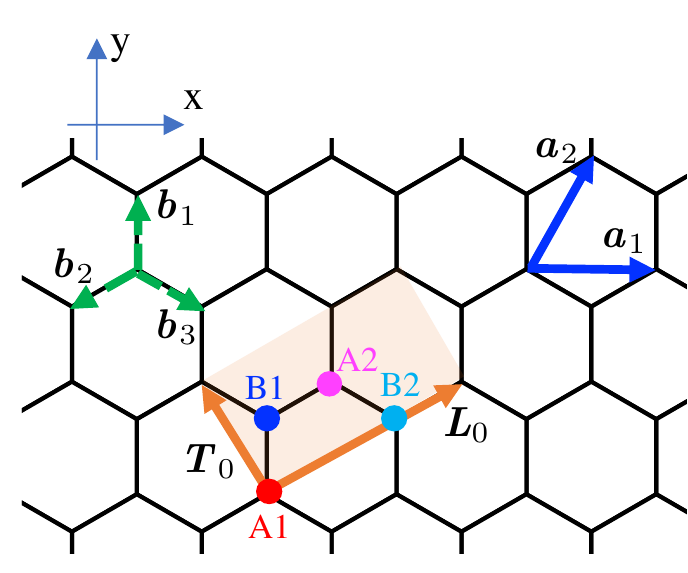} 
  \caption{Schematic figure of a graphite sheet. The orange shaded area corresponds to the primitive cell for armchair CNTs. We also show the name of four sites in the cell. }
  \label{fig:schematic2}
\end{figure}

As explained in the main text, the tight-binding (TB) model for the armchair CNT with the static magnetic field and the AC electric field along the tube can be constructed on a honeycomb (graphite) sheet;
\eqq{
\hH(t) = -t_{\rm hop} \sum_{\langle ij\rangle} e^{i\frac{q}{\hbar}\bA(t)\cdot {\bm r}_{ij}} \hc^\dagger_{i} \hc_{j} -\mu \sum_i \hn_i ~\label{eq:Hamiltonian},
}
where the periodic boundary condition (PBC) with respect to the chiral vector $\bL(=n{\bm a}_1 + n{\bm a}_2)$ is imposed. We set $\bA(t) = \bA_B + \bA_E(t)$.
In this TB model for CNTs, the effects of the curvature of the tube are ignored for simplicity~\cite{Hamada1992PRL}. Due to the curvature of the tube, the Dirac points may shift. 
Still, for the armchair CNT, which we focus on in this paper, the shift is along $k_{\parallel}$ and the system remains metallic~\cite{Hamada1992PRL}. 
Thus, the physics of the manipulation of HHG discussed in this paper is hardly affected.
The primitive cell is defined by
\eqq{
\bL_0 \equiv {\bm a}_1 + {\bm a}_2, \;\; \bT_0 \equiv {\bm a}_2 - {\bm a}_1,
}
see Fig.~\ref{fig:schematic2}.
Note that $\bL = n\bL_0 \propto {\bm e}_{\perp}$ and $\bT_0 \propto {\bm e}_\parallel$. We have four sites (atoms) in the primitive cell.
Due to the PBC with respect to $\bL$, we have $n$ primitive cells along ${\bm e}_{\perp}$.
We also impose the PBC along ${\bm e}_{\parallel}$ and take $N_T(\rightarrow \infty)$ primitive cells along ${\bm e}_{\parallel}$.
Using the reciprocal vectors ${\bm l}_0 \equiv 2\pi \frac{\bL_0}{|\bL_0|^2}$ and ${\bm t}_0 \equiv 2\pi \frac{\bT_0}{|\bT_0|^2}$, the wave vectors can be defined as
\eqq{
\bk = \frac{i_L}{n} {\bm l}_0 + \frac{i_T}{N_T} {\bm t}_0, \label{eq:k_expression}
}
where $i_L = 0,1\cdots n-1$ and $i_T = 0,1\cdots N_T-1$.
We name the four sites in the primitive cell as $A_1,B_1,A_2,$ and $B_2$ as in Fig.~\ref{fig:schematic2}.

Now, we consider the expression of the Hamiltonian in the momentum (Fourier) space.
We introduce the electron operators in the momentum space as 
$ \hc^\dagger_{\bk\alpha } = \frac{1}{\sqrt{N_c}} \sum_{i\in \alpha } e^{i\bk\cdot {\bm r}_i}  \hc^\dagger_{i}$ 
for $\alpha = A_1,A_2,B_1,B_2$. Here $\br_i$ is the position vector of the $i$th site, $N_c=n\cdot N_T$ is the total number of the cells.
The inverse tranform is  $\hc^\dagger_{\bm r_i} = \frac{1}{\sqrt{N_c}} \sum_\bk e^{-i\bk \cdot \br_i} \hc^\dagger_{\bk\alpha}$ for ${\bm r}_i\in \alpha$.
Then the Hamiltonian \eqref{eq:Hamiltonian} is expressed as
\small
\eqq{
\hH(t) = \tilde{\sum}_\bk
\begin{bmatrix}
\hc^\dagger_{\bk A_1} & \hc^\dagger_{\bk A_2} & \hc^\dagger_{\bk B_1} & \hc^\dagger_{\bk B_2}
\end{bmatrix}
\tilde{\bf h}(\bk (t))
\begin{bmatrix}
\hc_{\bk A_1} \\ \hc_{\bk A_2} \\ \hc_{\bk B_1} \\ \hc_{\bk B_2}
\end{bmatrix},
}
\normalsize
where
\eqq{
\tilde{\bf h}(\bk )= 
\begin{bmatrix}
-\mu & 0 & \alpha(\bk)^* &  \gamma(\bk)^* \\
0 & -\mu & \gamma(\bk)^* &  \alpha(\bk)^* \\
 \alpha(\bk) &  \gamma(\bk) & -\mu & 0 \\
 \gamma(\bk) &  \alpha(\bk) & 0 & -\mu
\end{bmatrix},
}
$\bk(t)=\bk -\frac{q}{\hbar} \bA(t)$ and 
\eqq{
\alpha(\bk) = -t_{\rm hop} [e^{-i{\bm b}_1\cdot \bk} + e^{-i{\bm b}_3\cdot \bk}], \;\;
\gamma(\bk) = -t_{\rm hop} e^{-i{\bm b}_2\cdot \bk}.
}
Here, $\tilde{\sum}_\bk$ means to calculate the sum over $\bk$ expressed as Eq.~\eqref{eq:k_expression} with $i_L = 0,1\cdots n-1$ and $i_T = 0,1\cdots N_T-1$.

Next, we show that the Hamiltonian for each $\bk$ can be block diagonalized into $2\times2$ matrices, each of which corresponds to the Hamiltonian of graphene at a certain momentum.
We introduce the new basis for electrons at each $\bk$ as \cite{Murakami2022PRB}
\eqq{
\begin{bmatrix}
\hc_{\bk A_+} \\ \hc_{\bk B_+} \\ \hc_{\bk A_-} \\ \hc_{\bk B_-}
\end{bmatrix}
\equiv \frac{1}{\sqrt{2}}
\begin{bmatrix}
1 & 1 & 0 & 0 \\
0 & 0 & 1 &  1 \\
e^{i\frac{\pi}{6}} & -e^{i\frac{\pi}{6}} & 0 & 0 \\
0 &  0 & e^{-i\frac{\pi}{6}} & -e^{-i\frac{\pi}{6}}
\end{bmatrix}
\begin{bmatrix}
\hc_{\bk A_1} \\ \hc_{\bk A_2} \\ \hc_{\bk B_1} \\ \hc_{\bk B_2}
\end{bmatrix} \label{eq:transformation}.
}
Then, we have
\eqq{
&\hH(t) =\nonumber \\
& \tilde{\sum}_\bk
\begin{bmatrix}
\hc^\dagger_{\bk A_+} & \hc^\dagger_{\bk B_+} & \hc^\dagger_{\bk A_-} & \hc^\dagger_{\bk B_-}
\end{bmatrix}
{\bf h}'(\bk(t))
\begin{bmatrix}
\hc_{\bk A_+} \\ \hc_{\bk B_+} \\ \hc_{\bk A_-} \\ \hc_{\bk B_-}
\end{bmatrix}
}
with 
\eqq{
{\bf h}'(\bk) = 
\begin{bmatrix}
-\mu &F(\bk) & 0 & 0  \\
F(\bk)^* & -\mu &0 &  0 \\
 0&  0 & -\mu & F(\bk+{\bm l}_0)  \\
0 &0  &  F(\bk+{\bm l}_0)^*& -\mu
\end{bmatrix}.
}
We note that each $2\times2$ matrix in ${\bf h}'(\bk)$ has the same form as that of graphene at momentum $\bk$ and $\bk+{\bm l}_0$~\cite{Murakami2022PRB}.
If we extend the region of $\bk$ for CNTs as Eq.~\eqref{eq:k_expression}  with $i_L = 0,1\cdots 2n-1$ and $i_T = 0,1\cdots N_T-1$ and remove $\pm$ in $\hc_{\bk A_\pm}$ and  $\hc_{\bk B_\pm}$,
we have the expression used in the main text;
\eqq{
\hH(t)=\sum_\bk 
\begin{bmatrix}
\hc^\dagger_{\bk A} & \hc^\dagger_{\bk B}
\end{bmatrix} 
{\bf h}(\bk (t)) 
\begin{bmatrix}
\hc_{\bk A} \\ \hc_{\bk B}
\end{bmatrix} \label{eq:H_graphene}
}
with 
\eqq{
 {\bf h}(\bk ) = 
\begin{bmatrix}
-\mu  & F( \bk ) \\
 F^*( \bk ) & -\mu
\end{bmatrix}.  \label{eq:Graphene_TB_k}
}
In the following, we express the Hamiltonian~\eqref{eq:H_graphene} for each $\bk$ as $\hH_\bk(t)$.
The expression tells that for CNTs one needs to choose 2n lines along ${\bm t}_0$ ($\bk_\parallel$ in the main text) from the momentum space of graphene
and the corresponding Hamiltonian for each point is the same as graphene.
The magnetic field $\bA_B$ shifts the position of lines in the direction of ${\bm l}_0$ ($\bk_\perp$ in the main text).

The explicit expression of the current is 
\eqq{
\hat{J}_\parallel(t) &=  \frac{q t_{\rm hop}}{\hbar N}\sum_{\alpha=1,2,3}  i({\bm e}_{||}\cdot{\bm b}_{\alpha})\Bigl\{ \nonumber\\
&\sum_{\bk }
\begin{bmatrix}
\hc^\dagger_{\bk A} & \hc^\dagger_{\bk B}
\end{bmatrix}
\begin{bmatrix}
0 &- e^{i\bk (t)\cdot {\bm b}_\alpha} \\
 e^{-i\bk (t)\cdot {\bm b}_\alpha} & 0
\end{bmatrix}
\begin{bmatrix}
\hc_{\bk A} \\ \hc_{\bk B}
\end{bmatrix}
\Bigl\},
}
where $N=2nN_T=2N_c$.
In the following, we express the contribution from each $\bk$ as $\hat{J}_{\parallel,\bk}(t)$, i.e. $\hat{J}_\parallel(t)= \frac{1}{N}\sum_\bk \hat{J}_{\parallel,\bk}(t)$.

\subsection{Dirac model around $K$ and $K'$}
The TB model for graphene hosts two Dirac points ${\bm K}$ and ${\bm K}'$, where
$F({\bm K})=0$ and $F({\bm K}')=0$. To be specific, ${\bm K}={\bm l}_0 + \frac{1}{3}{\bm t}_0$ and ${\bm K}'= \frac{2}{3}{\bm t}_0$.
The Hamiltonian ~\eqref{eq:H_graphene} around these points can be approximated by the Dirac model.
By expanding $F(\bk )=0$ around ${\bm K}$ or ${\bm K}'$ and  regarding $e^{-i{\bm K}\cdot {\bm b}_1}\hc^\dagger_{{\bm k}B}$ ($e^{-i{\bm K}'\cdot {\bm b}_1}\hc^\dagger_{{\bm k}B}$ ) as new $\hc^\dagger_{{\bm k}B}$, 
$\hH_\bk(t)$ is approximated as 
\eqq{
\hH^{({\bm K},{\bm K}')}_{\bk} (t) = 
\begin{bmatrix}
\hc^\dagger_{\bk A} & \hc^\dagger_{\bk B}
\end{bmatrix} 
{\bf h}_\pm(\bk (t)) 
\begin{bmatrix}
\hc_{\bk A} \\ \hc_{\bk B}
\end{bmatrix} 
}
with
\eqq{
{\bf h}_\pm(\bk )  = 
\begin{bmatrix}
-\mu  &-\frac{\sqrt{3} a t_{\rm hop}}{2} [\pm k_x + i k_y] \\
-\frac{\sqrt{3} a t_{\rm hop}}{2} [\pm k_x - i k_y] & -\mu
\end{bmatrix}.
}
Here, "+" is for ${\bm K}$ and "-" is for ${\bm K}'$, $k_x$ and $k_y$ are the x and y components of $\bk$.
In this expression, we shift the origin of the momentum to ${\bm K}$ or ${\bm K}'$.
The corresponding expression of $\hat{J}_{\parallel,\bk}$ is 
\eqq{
&\hat{J}^{({\bm K}, {\bm K}')}_{\parallel,\bk} = \mp\frac{\sqrt{3}aqt_{\rm hop}}{2\hbar} ({\bm e}_\parallel\cdot{\bm e}_x) \;
\begin{bmatrix}
\hc^\dagger_{\bk A} & \hc^\dagger_{\bk B}
\end{bmatrix}
\begin{bmatrix}
0 &1 \\
1 & 0
\end{bmatrix}
\begin{bmatrix}
\hc_{\bk A} \\ \hc_{\bk B}
\end{bmatrix}  \nonumber \\
&\;\;-\frac{\sqrt{3}aqt_{\rm hop}}{2\hbar} ({\bm e}_\parallel\cdot{\bm e}_y) \;
\begin{bmatrix}
\hc^\dagger_{\bk A} & \hc^\dagger_{\bk B}
\end{bmatrix}
\begin{bmatrix}
0 &i \\
-i & 0
\end{bmatrix}
\begin{bmatrix}
\hc_{\bk A} \\ \hc_{\bk B}
\end{bmatrix}.
}
Here, ${\bm e}_x$ and ${\bm e}_y$ are the unit vectors for the x and y directions.

\subsection{Semiconductor Bloch Equation}
In the dipolar gauge~\cite{Li2020PRB,Michael2021PRB,Murakami2022PRB}, the semiconductor Bloch equation (SBE, i.e. the von Neumann equation for the single particle density matrix) is expressed as
\eqq{
 \hbar \partial_t \brho_\bk(t) &= i[\brho_\bk(t),\bh(\bk(t))]  +\hbar  [\partial_t \brho_\bk(t)]_{\rm corr}, \label{eq:vN_eq_T1T2_DW}
}
where $ [\partial_t \brho_\bk(t)]_{\rm corr}$ indicates the collision integral. 
The collision integral expresses the scattering, relaxation and dephasing processes originating from electron-electron interactions, electron-phonon couplings and/or impurity scatterings.
In principle, these terms can be evaluated microscopically~\cite{stefanucci_nonequilibrium_2013,Nessi2020,ridley2022manybody}.
However, the direct microscopic evaluation is computationally expensive.
Therefore, in this paper, we employ the relaxation time approximation as  
\eqq{
& [\partial_t \brho_\bk(t)]_{\rm corr} =  -\frac{\brho_\bk(t) - \brho_{{\rm eq},\bk(t)}}{T_1} \nonumber\\
&\;\;\;\;\; + \Bigl(\frac{1}{T_1}-\frac{1}{T_2}\Bigl) {\bf U}(\bk(t))\brho^{\rm H}_{{\rm off},\bk}(t){\bf U}^\dagger(\bk(t)). \label{eq:T1T2_DW}
}
Here $T_1$ corresponds to the relaxation of the band population, while $T_2$ represents the interband dephasing time.
We assume that initially ($t=0$) the system is in equilibrium at temperature $T$ and there is no electric field $\bA_E(0)=0$.
$\brho_{{\rm eq},\bk}$ indicates the equilibrium single particle density matrix for $\bk$ at this initial temperature.
The meaning of these terms becomes clear if one expresses the SBE using the basis that diagonalizes the Hamiltonian at each time  (the Houston basis).
For the expression, we introduce a unitary matrix ${\bf U}(\bk)$ that diagonalizes ${\bf h}(\bk)$ as ${\bf U}^\dagger(\bk){\bf h}(\bk) {\bf U}(\bk) = {\boldsymbol \epsilon}(\bk)$ 
with $ {\boldsymbol \epsilon}(\bk)= {\rm diag}[\epsilon_0(\bk),\epsilon_1(\bk)]$. The single-particle density matrix in the new basis is $\brho^{\rm H}_{\bk}(t)\equiv \bU^\dagger(\bk(t))\brho_\bk(t) \bU(\bk(t))$.
The diagonal component of $\brho^{\rm H}_\bk$ indicates the occupation of the bands, and the off-diagonal component means the interband hybridization (coherence).
In equilibrium, $\brho^{\rm H}_{{\rm eq}, \bk}$ only hosts the diagonal terms.
In this basis, the SBE equivalent to Eq.~\eqref{eq:T1T2_DW} is 
\eqq{
\hbar \partial_t \brho^{\rm H}_{\bk}(t) \nonumber &= i[\brho^{\rm H}_{\bk}(t), \beps(\bk(t)) -q\sum_{a=x,y} E_a(t) \bd_a(\bk(t)) ] \nonumber \\
&+ \hbar[\partial_t \brho_\bk^{\rm H}(t)]_{\rm corr}, \label{eq:vN_T1T2_H}
}
with 
\eqq{
[\partial_t \brho_\bk^{\rm H}(t)]_{\rm corr} &=  
- \frac{\brho^{\rm H}_{{\rm diag},\bk}(t) - \brho^{\rm H}_{{\rm eq},\bk(t)}}{T_1} - \frac{\brho^{\rm H}_{{\rm off},\bk}(t)}{T_2}.
}
Here, $\brho_{{\rm diag}}$ indicates a matrix consisting of diagonal components of $\brho$, while $\brho_{{\rm off}}$ indicates a matrix consisting of off-diagonal components of $\brho$.
${\bf d}_a(\bk) = i \bU^\dagger (\bk)[\partial_{a} \bU(\bk)]$ is the (non-abelian) Berry connection for the direction $a$, which plays the role of dipole matrix elements. 
 $\partial_{a}=\partial/\partial k_a$ is the momentum derivate for the direction $a$.
Note that Eqs.~\eqref{eq:vN_eq_T1T2_DW} and \eqref{eq:vN_T1T2_H} are equivalent to the conventional expression of SBE in the length gauge~\cite{Wilhelm2021PRB,Yue2022tutorial,Murakami2022PRB}.

As explain in the main text, the origin of the radiation is the time-dependent change in the electric polarization $P\;(\equiv q\sum_i{\bm e}_\parallel \cdot {\bm r}_i \langle \hn_i\rangle)$, or the corresponding current~\cite{Huttner2017,Yue2022tutorial}.
As in previous works~\cite{Ishikawa2010PRB,Dimitrovsk2017}, we introduce the intraband and interband currents based on the Houston basis as 
\eqq{
J_{\parallel,\rm intra}(t) &= \frac{q}{\hbar N}  \sum_\bk {\rm Tr}\Bigl[\brho^{\rm H}_\bk(t)\partial_{k_\parallel}{\boldsymbol \epsilon}(\bk(t))\Bigl],\\
J_{\parallel,\rm inter}(t) &= -i\frac{q}{\hbar N}  \sum_{\bk,a} e_{\parallel,a}  {\rm Tr}\Bigl[\brho^{\rm H}_\bk(t) \;[ {\bf d}_a(\bk(t)),\beps(\bk(t))] \Bigl] .
}
The intraband current is related to the diagonal components of $\brho^{\rm H}_\bk(t)$. Hence this current originates from the change in the band occupation.
On the other hand, the interband current originates from the off-diagonal components of $\brho^{\rm H}_\bk(t)$ (the interband components). 
Given that the current is the time derivative of the polarization, the interband current is closely related the change in the interband polarization $P_{\rm inter}$.
It is the polarization originating from the interband components as 
\eqq{
& P_{\rm inter} = \frac{q}{\hbar N}  \sum_{\bk,a} e_{\parallel,a}  {\rm Tr}\Bigl[\brho^{\rm H}_\bk(t) {\bf d}^{\rm off}_a(\bk(t)) \Bigl], \\
&  {\bf d}^{\rm off}_a(\bk) = \begin{bmatrix}
 0 & d_{01,a}(\bk) \\
 d_{10,a}(\bk) & 0
 \end{bmatrix} \nonumber.
}

In the actual implementation, we evaluate currents ($J_\parallel,J_{\parallel,\rm intra},J_{\parallel,\rm inter}$) using the difference of the single-particle density matrix from equilibrium $\delta {\brho}_\bk(t) = \brho_\bk(t) -\brho_{{\rm eq},\bk(t)}$.
Note that these currents are zero in equilibrium.
This strategy has two benefits. Firstly, one can save the calculation cost since away from the Dirac points  $\delta {\brho}_\bk(t)$ is essentially zero and we do not need to evaluate them.
Secondly, this can be directly applied for the Dirac model.
Namely, the currents for CNTs and graphene can be evaluated from the Dirac models using $\delta \brho_{\bk}(t)$ at ${\bm K}$ and ${\bm K}'$ and summing up these contributions.
If we evaluate the currents from the Dirac models using $\brho_{\bk}(t)$ the results depend on the momentum cutoff.

\subsection{Comments on previous studies}
Although the manipulation of HHG in a single-wall CNT using the Aharanov-Bohm effect is our main point, there have been some studies on the effects of the static magnetic field on HHG in CNTs~\cite{Khosravi2009,Dvuzhilova2021}. In Ref.~\onlinecite{Khosravi2009}, they consider the mangetic field applied to the perpendicular direction of the tube and numerically show the sensitivity of the HHG spectrum 
for the excitation with the light of the wave length 800nm. Thus, the Aharanov-Bohm effect and the Dirac physics are not relevant here.
In Ref.~\onlinecite{Dvuzhilova2021}, they study the propagation of the Airy beam with 1 fs duration in the photonic crystal consisting of many CNTs under magnetic field.
We note that the setup and the timescale involved are very different from our case.

\section{Derivation of Eq.~(4) in the main text}
We focus on the case of $B=0$ and the contribution from the bands passing the Dirac points.
The AC electric field is applied along ${\bm e}_\parallel (= \begin{bmatrix}
\cos\theta \\
\sin\theta
\end{bmatrix})$. The momentum on the line passing $K$ or $K'$ can be expressed as $\bk = k {\bm e}_\parallel$.
The corresponding Hamiltonian of the Dirac model is
\small
\eqq{
&\hH_\bk(t)  =\\  \nonumber &
\begin{bmatrix}
\hc_{\bk A}^\dagger & \hc_{\bk B}^\dagger
\end{bmatrix} 
\begin{bmatrix}
-\mu  &\frac{\sqrt{3} a t_{\rm hop}}{2} e^{i\theta_\pm} k(t) \\
\frac{\sqrt{3} a t_{\rm hop}}{2} e^{-i\theta_\pm} k(t) & -\mu
\end{bmatrix}
\begin{bmatrix}
\hc_{\bk A} \\ \hc_{\bk B}
\end{bmatrix},
}
\normalsize
where $k(t) \equiv k-\frac{q}{\hbar}A_E(t) $, $\theta_+ = \theta + \pi$ for ${\bm K}$ and  $\theta_+ = -\theta$ for ${\bm K}'$.
We introduce a new basis (the Houston basis) as 
\eqq{
\begin{bmatrix}
\hc_{k R} \\ \hc_{k L}
\end{bmatrix}
\equiv\frac{1}{\sqrt{2}}
\begin{bmatrix}
e^{-i\frac{\theta_\pm}{2}} & e^{i\frac{\theta_\pm}{2}} \\
-e^{-i\frac{\theta_\pm}{2}} & e^{i\frac{\theta_\pm}{2}}
\end{bmatrix}
\begin{bmatrix}
\hc_{k A} \\ \hc_{k B}
\end{bmatrix}.
}
Then we can express the Hamiltonian as
\eqq{
&\hH_k(t)  =\\  \nonumber 
& \begin{bmatrix}
\hc_{k R}^\dagger & \hc_{k L}^\dagger
\end{bmatrix} 
\begin{bmatrix}
\frac{\sqrt{3}a t_{\rm hop}}{2} k(t)-\mu  & 0  \\
0 & -\frac{\sqrt{3} a t_{\rm hop}}{2}  k(t)-\mu
\end{bmatrix}
\begin{bmatrix}
\hc_{k R} \\ \hc_{k L}
\end{bmatrix},
}
and the current as
\eqq{
\hat{J}_{k,\parallel }(t)= \frac{\sqrt{3}aqt_{\rm hop}}{2\hbar}
 \begin{bmatrix}
\hc_{k R}^\dagger & \hc_{k L}^\dagger
\end{bmatrix} 
\begin{bmatrix}
1  & 0  \\
0 & -1
\end{bmatrix}
\begin{bmatrix}
\hc_{k R} \\ \hc_{k L}
\end{bmatrix}.
}
We introduce the single-particle density matrix in the present basis as $\rho^R_k(t) = \langle \hc_{k R}^\dagger(t) \hc_{k R}(t) \rangle$ and $\rho^L_k(t) = \langle \hc_{k L}^\dagger(t) \hc_{k L}(t) \rangle$.
The corresponding SBE becomes 
\eqq{
\partial_t \rho_k^X(t) =  [\partial_t \rho_k^X(t)]_{\rm corr}
}
with 
\eqq{
[\partial_t \rho_k^X(t) ]_{\rm corr} = -\frac{\rho_k^X(t) -\rho^X_{\rm eq}(k(t),T(t))}{T_1}. \label{eq:relax_approx}
}
Here $X=R,L$ and $\rho^X_{\rm eq}(k(t),T(t))$ represents the equilibrium single-particle density matrix at the momentum $k(t)$ and the temperature $T(t)$.
Note that this expression is slightly extended from Eq.~\eqref{eq:T1T2_DW}.
Namely, we consider the time-dependent temperature for the equilibrium state that the system is relaxed to at each time.
Furthermore, the SBE tells that no interband component in the density matrix is induced by the electric field, which already indicates the absence of the interband current. 
Using the integrated form of the SBE,  the deviation of the single-particle density matrix from equilibrium $\delta  \rho_\bk^X(t) =  \rho_\bk^X(t)-\rho^X_{\rm eq}(k(t),T(t)) $ can be expressed as 
\eqq{
 \delta\rho_k^X(t) &=  \rho^X_{\rm eq}(k,T(0))- \rho^X_{\rm eq}(k(t),T(t)) + \int^t_0  dt [\partial_t \rho_k^X(t)]_{\rm corr}.
} 
Note that, at $t=0$, the system is in equilibrium at the temperature $T(0)$ and there is no electric field.

Now, if we focus on the contributions to the current from the line passing $K$ and $K'$ points, we have 
\eqq{
J_{\parallel}(t) &= \frac{1}{N} \sum_\bk J_{\bk\parallel}(t) \nonumber \\
&= \frac{1}{n} \frac{\sqrt{3}a^2qt_{\rm hop}}{4\pi\hbar} \int dk [ \delta\rho_k^R(t)- \delta\rho_k^L(t)],
}
where
\eqq{
&\int dk [ \delta\rho_\bk^R(t)- \delta\rho_\bk^L(t)] = \nonumber \\
&- \int dk [\rho^R_{\rm eq}(k(t),T(t))- \rho^R_{\rm eq}(k,T(0))] \nonumber \\
&+ \int dk [\rho^L_{\rm eq}(k(t),T(t))- \rho^L_{\rm eq}(k,T(0))] \nonumber \\
&+ \int dk \int^t_0  dt \{[\partial_t \rho_k^R(t)]_{\rm corr}-[\partial_t \rho_k^L(t)]_{\rm corr}\} \\
& = -2\frac{q}{\hbar}A_E(t) + \int dk \int^t_0  dt \{[\partial_t \rho_k^R(t)]_{\rm corr}-[\partial_t \rho_k^L(t)]_{\rm corr}\}. \nonumber
}
Note that the temperature effect vanishes in the integral of the first two terms in the second line.
Finally, we obtain 
\eqq{
J_{\parallel}(t) = - \frac{1}{n} \frac{\sqrt{3}a^2q^2t_{\rm hop}}{2\pi\hbar}A_E(t) + \int^t_0 dt [\partial_t J_{\parallel}(t)]_{\rm corr} 
}
with $[\partial_t J_{\parallel}(t)]_{\rm corr} \equiv \frac{1}{n} \frac{\sqrt{3}a^2qt_{\rm hop}}{4\pi\hbar}\int dk \{[\partial_t \rho_k^R(t)]_{\rm corr}-[\partial_t \rho_k^L(t)]_{\rm corr}\}$.
Taking the derivative and substituting Eq.~\eqref{eq:relax_approx} to $[\partial_t \rho_k^X(t) ]_{\rm corr}$, we have 
\eqq{
\partial_t J_{\parallel}(t)  = \frac{1}{n} \frac{\sqrt{3}a^2q^2t_{\rm hop}}{2\pi\hbar^2}E(t) -\frac{J_{\parallel}(t)}{T_1}.
}
This is Eq.~(4) in the main text.
Our derivation tells that, even if we improve the relaxation time approximation considering the time-dependent temperature, HHG is absent in the gapless 1d Dirac system.
\section{Derivation of $\sigma^{(3)}$}
We derive the third-order optical conductivity by perturbatively solving the SBE for the Dirac model as in Ref.~\onlinecite{Sipe2015PRB}.
To this end, we change the gauge from the dipolar gauge, which is mainly used in this paper, to the length gauge expressed with the band basis~\cite{Murakami2022PRB}.
From the Dirac Hamiltonian Eq. (3) in the main text, we reach the SBE in the length gauge
\eqq{
\partial_t \brho^{\rm LB}_k(t) &= -\frac{q}{\hbar} E(t) \partial_k \brho^{\rm LB}_k(t) + \frac{i}{\hbar} [\brho^{\rm LB}_k(t), {\boldsymbol \epsilon}(k) -q E(t){\bf d}(k) ] \nonumber \\
&- \frac{\brho^{\rm LB}_{{\rm diag},k}(t) - \brho^{\rm LB}_{{\rm eq},k}}{T_1} - \frac{\brho^{\rm LB}_{{\rm off},k}(t)}{T_2}, \label{eq:SBE_length}
}
where 
\eqq{
{\boldsymbol \epsilon}(k)  &= 
\begin{bmatrix}
\tilde{t}_{\rm hop}\sqrt{k^2 + K_B^2}-\mu  & 0  \\
0 & -\tilde{t}_{\rm hop}\sqrt{k^2 + K_B^2}-\mu
\end{bmatrix}, \nonumber \\
{\bf d}(k)
&= -\frac{K_B}{2(k^2 + K_B^2)^2}
\begin{bmatrix}
0 & -i \\
i & 0
\end{bmatrix}.
}
Here, $k\in[-\infty,\infty]$ represents the momentum along ${\bf k}_\parallel$, $\brho^{\rm LB}_k$ is the $2\times2$ single-particle density matrix in the length gauge with the band basis, $\tilde{t}_{\rm hop}\equiv \frac{\sqrt{3} a t_{\rm hop}}{2}$ and $K_B \equiv \frac{q}{\hbar} A_B$. In the following, we set $\kappa \equiv \sqrt{k^2 + K_B^2}$, $r_k\equiv \frac{K_B}{2\kappa^2}$, denote the band index of the matrix as $s=0,1$, and omit "LB" from $\brho^{\rm LB}_k$.

To solve the SBE perturbatively, we expand $\brho_k$ as 
\eqq{
\rho_{s_1s_2,k}(t) = \sum_{n=0}^\infty \rho_{s_1s_2,k}^{(l)}(t) \label{eq:rho_expand}
}
with $\rho^{(l)}_{s_1s_2,k}(t) \propto E^l$. Note that we have $\rho_{s_1s_2,k}^{(0)}(t) = \delta_{s_1,s_2} n_{s_1,k}$, where $n_{s,k}=[e^{\epsilon_{s,k}/{k_B T}}+1]^{-1}$ and $\epsilon_{s,k}=(-)^s \tilde{t}_{\rm hop}\kappa-\mu$.

By substituting Eq.~\eqref{eq:rho_expand} into Eq.~\eqref{eq:SBE_length}, we have 
\eqq{ \label{eq:SBE_order}
\begin{split}
\hbar \partial_t \rho^{(l)}_{ss,k}(t) &= -(-)^s qE(t) r_k (\rho^{(l-1)}_{\bar{s}s,k} + \rho^{(l-1)}_{s\bar{s},k}) \\
&-qE(t) \partial_k \rho^{(n-1)}_{ss,k} -\Gamma_1 \rho^{(l)}_{ss,k}, \\
\hbar \partial_t \rho^{(l)}_{s\bar{s},k}(t) &= -i(-)^s \epsilon_{g,k} \rho_{s\bar{s},k}^{(l)}(t) -(-)^s qE(t) r_k (\rho^{(n-1)}_{\bar{s}\bar{s},k} - \rho^{(l-1)}_{ss,k}) \\
& -qE(t) \partial_k \rho^{(l-1)}_{s\bar{s},k} -\Gamma_2 \rho^{(l)}_{s\bar{s},k},
\end{split}
}
where $\Gamma_1 \equiv \hbar/T_1$, $\Gamma_2 \equiv \hbar/T_2$ and $\epsilon_{g,k} \equiv \epsilon_{0,k}-\epsilon_{1,k}$.
We express the SPDM as 
\eqq{
\begin{split}
\rho^{(1)}_{s_1s_2,k}(t) & = \int \frac{d\omega_3}{2\pi} (-q) E_{\omega_3} e^{-i\omega_3 t} \mathcal{P}^{(1)}_{s_1s_2,k}(\omega_3), \\
\rho^{(2)}_{s_1s_2,k}(t) & = \int \frac{d\omega_2 d\omega_3}{(2\pi)^2} (-q)^2 E_{\omega_2} E_{\omega_3} e^{-i\omega_0 t} \mathcal{P}^{(2)}_{s_1s_2,k}(\omega_2,\omega_3), \\
\rho^{(3)}_{s_1s_2,k}(t) & = \int \frac{d\omega_1 d\omega_2 d\omega_3}{(2\pi)^3} (-q)^3 E_{\omega_1} E_{\omega_2} E_{\omega_3} \times \\
& \;\;\;\;\;\;\;\;\;\;\;\;\;\;\;\;\;\;\;\;\;\;\;\;\;\;\;\;\;\;\;\;\;\;\;\; e^{-i\omega t} \mathcal{P}^{(3)}_{s_1s_2,k}(\omega_1,\omega_2,\omega_3),
\end{split}
}
where $\omega_0 \equiv \omega_2 + \omega_3$ and $\omega  \equiv \omega_1 + \omega_2 + \omega_3$.
Using Eq.~\eqref{eq:SBE_order}, one can evaluate $\mathcal{P}^{(l)}$ sequentially from $l=1$.
For $l=1$, we have 
\eqq{
\begin{split}
\mathcal{P}^{(1)}_{ss,k}(\omega_3) &= \frac{i}{\nu_3} \partial_k n_{s,k},\\
\mathcal{P}^{(1)}_{s\bar{s},k}(\omega_3) &= \frac{-ir_k \Delta n_k}{\vartheta_3 - (-)^s \epsilon_{g,k}},
\end{split}
}
where $\nu_3 \equiv  \hbar \omega_3 +  i\Gamma_1$, $\vartheta_3\equiv \hbar \omega_3 + i\Gamma_2$, $\Delta n_k \equiv n_{0,k}-n_{1,k}$.
For $l=2$, we have 
\eqq{
\begin{split}
\mathcal{P}^{(2)}_{ss,k}(\omega_2,\omega_3) &= \frac{i}{\nu_0} \Bigl[\frac{i}{\nu_3} \frac{\partial^2 n_{s,k}}{\partial k^2}  \\
&\;\;\;\;\;\;\;\;\;\;\;\; - i (-)^s r_k^2 \Delta n_k \bigl(\frac{1}{\vartheta_3 + \epsilon_{g,k} } + \frac{1}{\vartheta_3 - \epsilon_{g,k} }\bigl) \Bigl],\\
 \mathcal{P}^{(2)}_{s\bar{s},k}(\omega_2, \omega_3) &= \frac{1}{\vartheta_0 - (-)^s \epsilon_{g,k}} \Bigl[ \frac{r_k}{\nu_3} \frac{\partial \Delta n_k}{\partial k} + \partial_k \Bigl( \frac{r_k \Delta n_k}{\vartheta_3 -(-)^s \epsilon_{g,k}} \Bigl) \Bigl],
 \end{split}
}
where  $\nu_0 \equiv \hbar \omega_0 +  i\Gamma_1$ and  $\vartheta_0 \equiv \hbar \omega_0 + i\Gamma_2$.

For $l=3$, we have 
\eqq{
\mathcal{P}^{(3)}_{ss,k}(\omega_1,\omega_2, \omega_3) &= \frac{1}{\nu \nu_0 \nu_3} P_{1;s,k} +  \frac{1}{\nu \nu_0} P_{2;s,k} (\vartheta_3) \\
& +  \frac{1}{\nu \nu_3} P_{3;s,k} (\vartheta_0) + \frac{1}{\nu} P_{4;s,k} (\vartheta_0,\vartheta_3) \nonumber
}
with
\eqq{
\begin{split}
P_{1;s,k} &= -i \frac{\partial^3 n_{s,k}}{\partial k^3}, \\
P_{2;s,k} (\vartheta_3)  &= i (-)^s \partial_k \Bigl[ r_k^2 \Delta n_k \Bigl(\frac{1}{\vartheta_3 + \epsilon_{g,k}} + \frac{1}{\vartheta_3 - \epsilon_{g,k}}\Bigl)\Bigl],\\
P_{3;s,k} (\vartheta_0) &= i (-)^s  r_k^2 \frac{\partial \Delta n_k}{\partial k} \Bigl(\frac{1}{\vartheta_0 + \epsilon_{g,k}} + \frac{1}{\vartheta_0 - \epsilon_{g,k}}\Bigl),\\
P_{4;s,k} (\vartheta_0,\vartheta_3) &=  i (-)^s  r_k \Bigl[\frac{1}{\vartheta_0 + \epsilon_{g,k} } \partial_k \Bigl(\frac{r_k \Delta n_k}{\vartheta_3 + \epsilon_{g,k}} \Bigl)\\
&\;\;\;\;\;\;\;\;\;\;\;+ \frac{1}{\vartheta_0 - \epsilon_{g,k} } \partial_k \Bigl(\frac{r_k \Delta n_k}{\vartheta_3 - \epsilon_{g,k}} \Bigl) \Bigl],
\end{split}
}
and 
\eqq{
\mathcal{P}^{(3)}_{s\bar{s},k}(\omega_1,\omega_2, \omega_3) &=  \frac{1}{\nu_0 \nu_3} P_{5;s,k}(\vartheta) +  \frac{1}{\nu_0} P_{6;s,k} (\vartheta,\vartheta_3) \\
& +  \frac{1}{\nu_3} P_{7;s,k} (\vartheta, \vartheta_0) + P_{8;s,k} (\vartheta, \vartheta_0,\vartheta_3) \nonumber
}
with
\eqq{
\begin{split}
P_{5;s,k}(\vartheta) &= \frac{i r_k}{\vartheta - (-)^s \epsilon_{g,k}} \frac{\partial^2 \Delta n_{k}}{\partial k^2}, \\
P_{6;s,k} (\vartheta,\vartheta_3)   &= -2i\frac{ r_k^3 \Delta n_k}{\vartheta - (-)^s \epsilon_{g,k}}  \Bigl(\frac{1}{\vartheta_3 + \epsilon_{g,k}} + \frac{1}{\vartheta_3 - \epsilon_{g,k}}\Bigl),\\
P_{7;s,k} (\vartheta, \vartheta_0)  &= \frac{i}{\vartheta - (-)^s \epsilon_{g,k}} \partial_k \Bigl[ \frac{r_k}{\vartheta_0 - (-)^s \epsilon_{g,k}} \frac{\partial \Delta n_k}{\partial k}\Bigl],\\
P_{8;s,k} (\vartheta, \vartheta_0,\vartheta_3) &= \frac{i}{\vartheta - (-)^s \epsilon_{g,k}} \times \\
& \partial_k \Bigl[ \frac{1}{\vartheta_0 - (-)^s \epsilon_{g,k}}\Bigl\{ \partial_k \Bigl( \frac{r_k \Delta n_k}{\vartheta_3 - (-)^s \epsilon_{g,k}} \Bigl) \Bigl\} \Bigl].
\end{split}
}
Here $\nu \equiv \hbar \omega +  i\Gamma_1$ and  $\vartheta \equiv \hbar \omega + i\Gamma_2$.

Now we evaluate the third order conductivity. By taking into account the contribution from the two Dirac points ($K$ and $K'$), the third order current is expressed as
\eqq{
J^{(3)}_{\parallel} (t) =  \frac{q a}{n\hbar}\sum_{s_1,s_2} \int \frac{dk}{2\pi} v_{s_2 s_1,k} \rho^{(3)}_{s_1s_2,k}(t),
}
where $v_{s_2 s_1,k}$ denotes an element of the velocity operator
\eqq{
{\bm v}_k \equiv \frac{\tilde{t}_{\rm hop}}{\kappa}
\begin{bmatrix}
k & -K_B \\
-K_B & -k
\end{bmatrix}.
}
We define the unsymmetrized third order conductivity as 
\eqq{
\sigma^{(3)}(\omega_1,\omega_2,\omega_3) \equiv -\frac{q^4 a}{n\hbar} \sum_{s_1,s_2} \int \frac{dk}{2\pi} v_{s_2 s_1,k} \mathcal{P}^{(3)}_{s_1s_2,k}(\omega_1,\omega_2,\omega_3), \label{eq:sig_3}
}
where we have
\eqq{
J^{(3)}_{\parallel} (t) = \int \frac{d\omega_1 d\omega_2 d \omega_3}{(2\pi)^3} E_{\omega_1}E_{\omega_2}E_{\omega_3} e^{-i\omega t} \sigma^{(3)}(\omega_1,\omega_2,\omega_3).
}
Note that in Eq.~\eqref{eq:sig_3} the contribution from $s_1 \neq s_2$ corresponds to the interband current ($\sigma^{(3)}_{\rm er}$), while that from $s_1 = s_2$  corresponds to  the intraband current ($\sigma^{(3)}_{\rm ra}$).

In the following, we focus on $T=0$ and half filling ($\mu=0$). In this case, we have $\Delta n_k  = -1$ and $P_{1;s,k}=P_{3;s,k}=P_{5;s,k}=P_{7;s,k}=0$.
Then, we have $\sigma^{(3)}_{\rm ra}=S_2 + S_4$ and $\sigma^{(3)}_{\rm er}=S_6 + S_8$, where 
\eqq{
\begin{split}
S_2(\omega_1,\omega_2,\omega_3) &\equiv -\frac{q^4 a}{n\hbar} \frac{1}{\nu \nu_0} 2\int \frac{dk}{2\pi} v_{00,k} P_{2;0,k}(\vartheta_3),\\
S_4(\omega_1,\omega_2,\omega_3) &\equiv -\frac{q^4 a}{n\hbar} \frac{1}{\nu} 2\int \frac{dk}{2\pi} v_{00,k} P_{4;0,k}(\vartheta_3), \\
S_6(\omega_1,\omega_2,\omega_3) &\equiv -\frac{q^4 a}{n\hbar} \frac{1}{\nu_0} \int \frac{dk}{2\pi} v_{01,k} [P_{6;0,k}(\vartheta, \vartheta_3) +P_{6;1,k} (\vartheta, \vartheta_3)],\\
S_8(\omega_1,\omega_2,\omega_3) &\equiv -\frac{q^4 a}{n\hbar} \int \frac{dk}{2\pi} v_{01,k} [P_{8;0,k}(\vartheta, \vartheta_0,\vartheta_3) + P_{8;1,k}(\vartheta, \vartheta_0,\vartheta_3)].
\end{split}
}
The concrete expressions for these terms are
\begin{widetext}
\begin{subequations}
\eqq{
&S_2(\omega_1,\omega_2,\omega_3) = -\frac{i q^4 a\tilde{t}_{\rm hop}}{4\pi n\hbar} \frac{1}{\nu \nu_0} \int dk \frac{K_B^4}{\kappa^7} \Bigl( \frac{1}{\vartheta_3 + 2\tilde{t}_{\rm hop}\kappa}+ \frac{1}{\vartheta_3 - 2\tilde{t}_{\rm hop}\kappa}\Bigl), \\
&S_4(\omega_1,\omega_2,\omega_3) = -\frac{i q^4 a\tilde{t}_{\rm hop}}{2\pi n\hbar} \frac{1}{\nu } \int dk \frac{K_B^2 k^2}{\kappa^7}  
\Bigl[ \frac{(\vartheta_3 + 3\tilde{t}_{\rm hop}\kappa)}{(\vartheta_0 + 2\tilde{t}_{\rm hop}\kappa)(\vartheta_3 + 2\tilde{t}_{\rm hop}\kappa)^2} 
+ \frac{(\vartheta_3 - 3\tilde{t}_{\rm hop}\kappa)}{(\vartheta_0 - 2\tilde{t}_{\rm hop}\kappa)(\vartheta_3 - 2\tilde{t}_{\rm hop}\kappa)^2} \Bigl] , \\
&S_6(\omega_1,\omega_2,\omega_3) = \frac{i q^4 a\tilde{t}_{\rm hop}}{8\pi n\hbar} \frac{1}{\nu_0} \int dk \frac{K_B^4}{\kappa^7} \Bigl( \frac{1}{\vartheta + 2\tilde{t}_{\rm hop}\kappa}+ \frac{1}{\vartheta - 2\tilde{t}_{\rm hop}\kappa}\Bigl)  \Bigl( \frac{1}{\vartheta_3 + 2\tilde{t}_{\rm hop}\kappa}+ \frac{1}{\vartheta_3 - 2\tilde{t}_{\rm hop}\kappa}\Bigl), \\
& S_8(\omega_1,\omega_2,\omega_3) = \frac{i q^4 a\tilde{t}_{\rm hop}}{2\pi n\hbar} \int dk \frac{K_B^2 k^2}{\kappa^7}\times \\
&\;\;\;\;\;\;\;\;\;\;\;\;\;\;\;\;\;\;\;\;\;\;\;\;\Bigl[ \frac{(\vartheta + 4\tilde{t}_{\rm hop}\kappa)(\vartheta_3 + 3\tilde{t}_{\rm hop}\kappa)}{(\vartheta + 2\tilde{t}_{\rm hop}\kappa)^2(\vartheta_0 + 2\tilde{t}_{\rm hop}\kappa)(\vartheta_3 + 2\tilde{t}_{\rm hop}\kappa)^2} 
+ \frac{(\vartheta - 4\tilde{t}_{\rm hop}\kappa)(\vartheta_3 - 3\tilde{t}_{\rm hop}\kappa)}{(\vartheta - 2\tilde{t}_{\rm hop}\kappa)^2(\vartheta_0 - 2\tilde{t}_{\rm hop}\kappa)(\vartheta_3 - 2\tilde{t}_{\rm hop}\kappa)^2} \Bigl]. \nonumber 
}
\end{subequations}
\end{widetext}
The third harmonic generation corresponds to the case of $\omega_1 = \omega_2 = \omega_3 = \Omega$.
In the main text, we denote $\sigma^{(3)}(\Omega,\Omega,\Omega)$ simply as $\sigma^{(3)}$.

Now we consider the limit of $B\rightarrow 0$ ($K_B\rightarrow 0$). One can show that the leading terms are 
\begin{subequations}
\eqq{
&S_2(\omega_1,\omega_2,\omega_3) = -\frac{i q^4 a\tilde{t}_{\rm hop}}{2\pi n\hbar} \frac{1}{\nu \nu_0 \vartheta_3} \int dk \frac{K_B^4}{\kappa^7}, \\
&S_4(\omega_1,\omega_2,\omega_3) = -\frac{i q^4 a\tilde{t}_{\rm hop}}{\pi n\hbar} \frac{1}{\nu \vartheta_0 \vartheta_3} \int dk \frac{K_B^2 k^2}{\kappa^7}, \\
&S_6(\omega_1,\omega_2,\omega_3) = \frac{i q^4 a\tilde{t}_{\rm hop}}{2 \pi n\hbar} \frac{1}{\nu_0 \vartheta \vartheta_3} \int dk \frac{K_B^4}{\kappa^7}, \\
& S_8(\omega_1,\omega_2,\omega_3) = \frac{i q^4 a\tilde{t}_{\rm hop}}{\pi n\hbar} \frac{1}{\vartheta_0 \vartheta \vartheta_3} \int dk \frac{K_B^2 k^2}{\kappa^7}.
}
\end{subequations}
These terms diverge as $B\rightarrow 0$, and lead to the expression of $\sigma^{(3)}$ shown in the main text.
The rest contributions do not diverge (regular).
The expression of $\sigma^{(3)}$ tells that the intraband component strongly depends on $T_1$, while the interband component strongly depends on $T_2$.
This can be understood by the following points. 
The intraband component is related to the occupation of electrons in the conduction and valence bands, and $T_1$ is the time scale 
of the occupation returning to its equilibrium value.
On the other hand, the interband component is related to the modification of the interband polarization, i.e. the off-diagonal (interband) component of the density matrix in the Houston basis. 
These off-diagonal components are zero in equilibrium, and $T_2$ is the time scale that the components vanish. 

 \begin{figure}[t]
  \centering
    \hspace{-0.cm}
    \vspace{0.0cm}
\includegraphics[width=70mm]{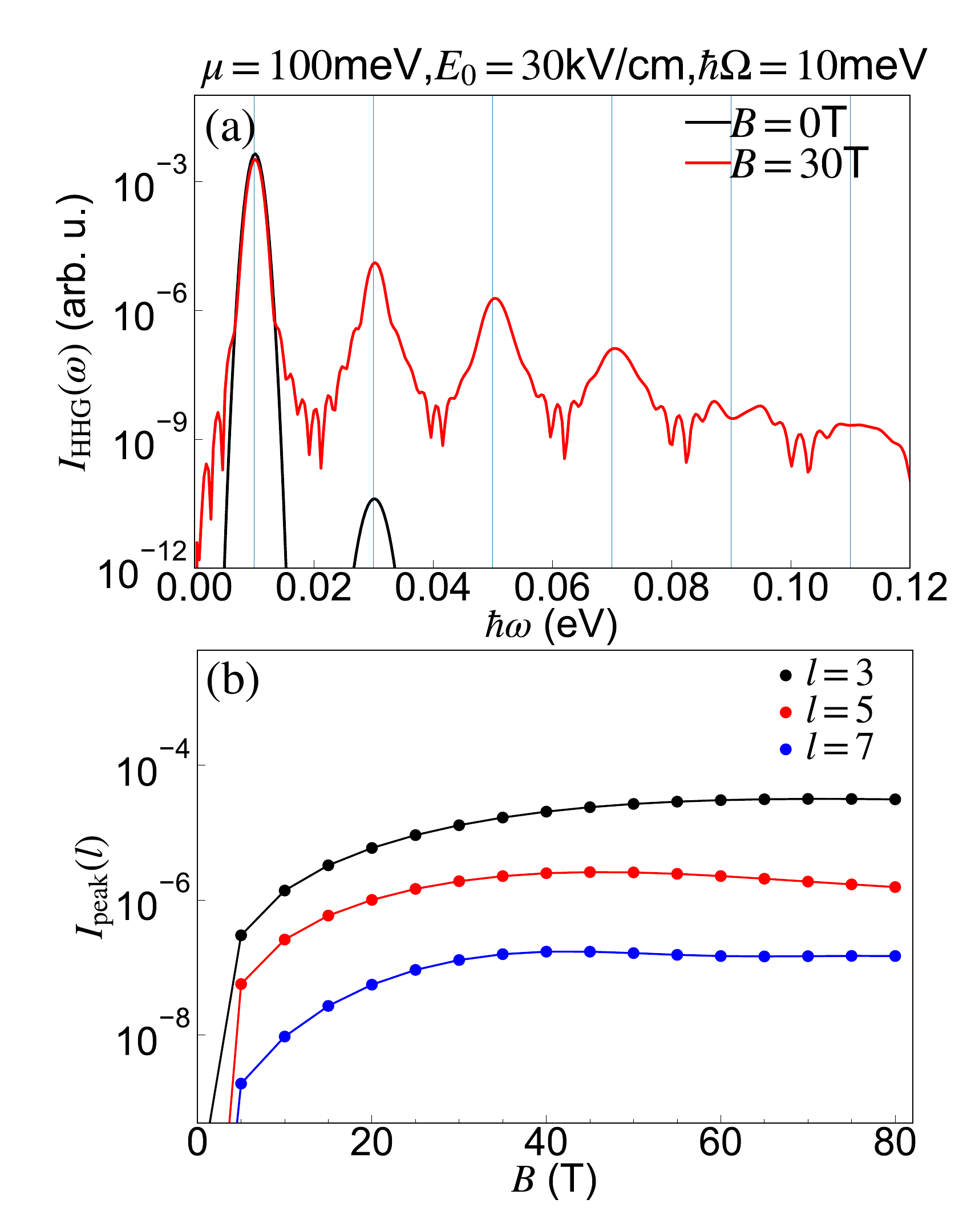} 
  \caption{(a) HHG spectra $I_{\rm HHG}$ of the doped armchair CNT with and without the static magnetic field simulated with the tight-binding model. (b) The intensity at the HHG peaks $I_{\rm peak}(l)$ as a function of $B$. We set the chiral index $(n,m)=(15,15)$ and consider doped systems with $\mu=100$meV. 
  We set $T=11.6$K, $T_1=98.8$fs and $T_2=19.8$fs. The parameters of the AC electric field are $\hbar\Omega = 10$meV, $E_0=30$kV/cm, $t_0=3.95$ps and $\sigma={658}$fs.}
  \label{fig:HHG_B_dep_mu01}
\end{figure}

\section{Supplementary results }

 \begin{figure}[t]
  \centering
    \hspace{-0.cm}
    \vspace{0.0cm}
\includegraphics[width=70mm]{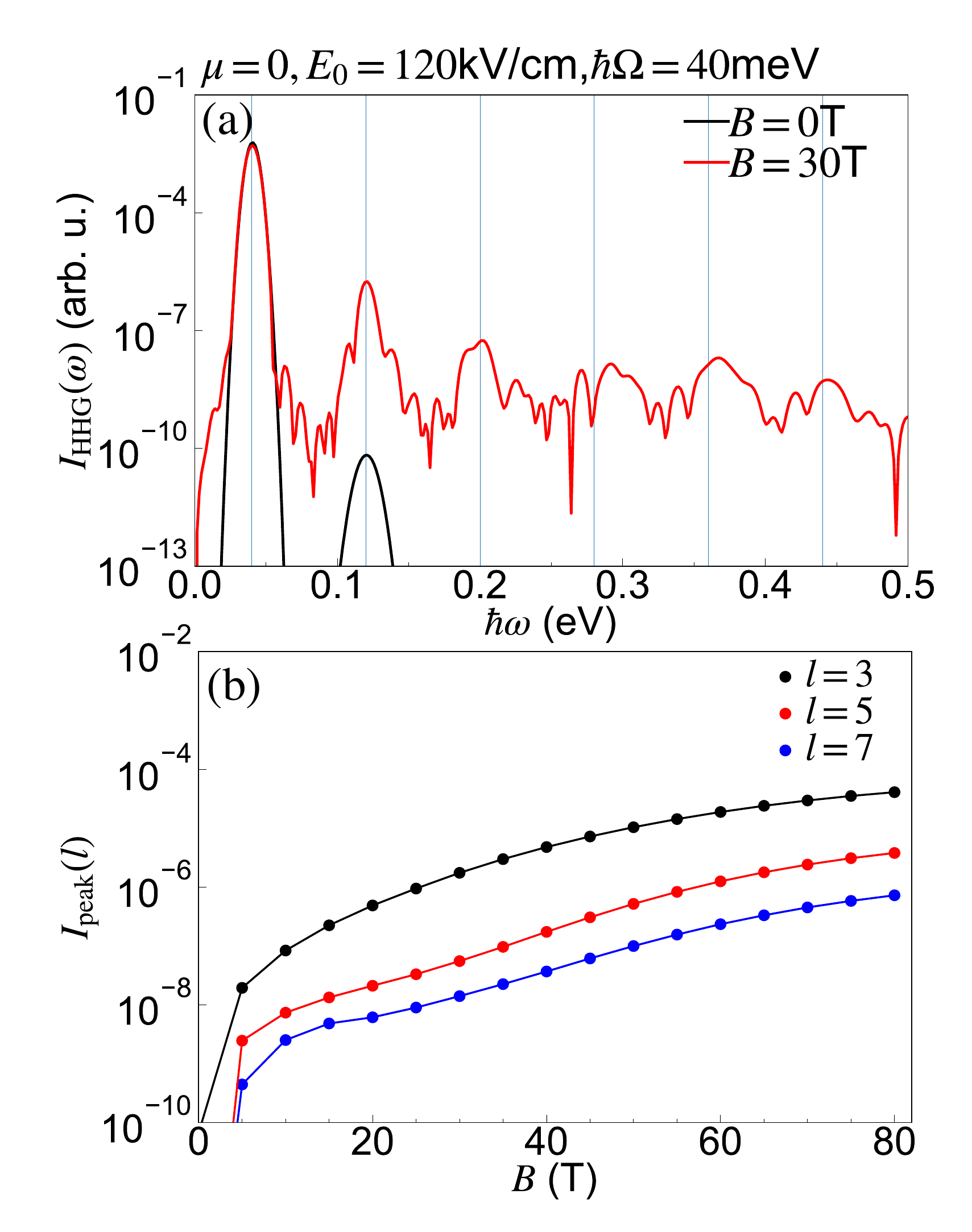} 
  \caption{(a) HHG spectra $I_{\rm HHG}$ of the armchair CNT with and without the static magnetic field simulated with the tight-binding model. 
  (b) The intensity at the HHG peaks $I_{\rm peak}(l)$ as a function of $B$.  We set the chiral index $(n,m)=(15,15)$ and consider half filling ($\mu=0$).  
  We set $T=11.6$K, $T_1=98.8$fs and $T_2=19.8$fs. The parameters of the AC electric field are $\hbar\Omega = 40$meV, $E_0=120$kV/cm, $t_0=986$fs and $\sigma={164}$fs.}
  \label{fig:HHG_B_dep_Ome004}
\end{figure}

In this section, we show the supplementary results to support the discussion in the main text.

\subsection{Doping effects}
In Fig.~\ref{fig:HHG_B_dep_mu01}(a), we show the HHG spectra of the armchair CNT with $(n,m)=(15,15)$ away from half filling ($\mu=0.1$eV) with and without the magnetic field $B$.
The excitation condition is the same as that in the main text.
Although the position of the chemical potential can be much larger than the gap and the excitation frequency, 
the drastic enhancement of HHG caused by the magnetic field is observed.
In Fig.~\ref{fig:HHG_B_dep_mu01}(b), we show the intensity of the HHG peaks as a function of $B$. Unlike in the half-filled case, there is no clear optimal value of $B$ which maximizes the peak intensity.
This may be attributed to that, in the doped system, charge carriers contributing to the intraband current exist from the beginning.
This situation is different from the half-filled system, where charge carriers need to be created by the AC electric field.

\subsection{Excitation with higher frequency}
In Fig.~\ref{fig:HHG_B_dep_Ome004}(a), we show the HHG spectra of the armchair CNT with $(n,m)=(15,15)$ at half filling for the excitation frequency $\hbar\Omega=40$meV.
Figure~\ref{fig:HHG_B_dep_mu01}(b) shows the corresponding intensity of the HHG peaks as a function of $B$.
The drastic enhancement of HHG can be still observed even when the gap is smaller than $\hbar\Omega$. For example, the gap for $B=10$T is about $10$meV.
Note that the peaks in the HHG spectrum is less sharp for $\hbar\Omega=40$meV than for $\hbar\Omega=10$meV, since the relaxation and dephasing times become relatively large with respect to the period of the excitation $\frac{2\pi}{\Omega}$.
In the range of $B$ in Fig.~\ref{fig:HHG_B_dep_Ome004}(b) , we cannot observe the optimal value of $B$ unlike in the case of $\hbar\Omega=10$meV. However, it should exist in the larger-$B$ regime.
Note that, for $\hbar\Omega=10$meV, the optimal $B$ is around $30\sim40$T, where the gap is about $30\sim40$meV.

 \begin{figure}[t]
  \centering
    \hspace{-0.cm}
    \vspace{0.0cm}
\includegraphics[width=80mm]{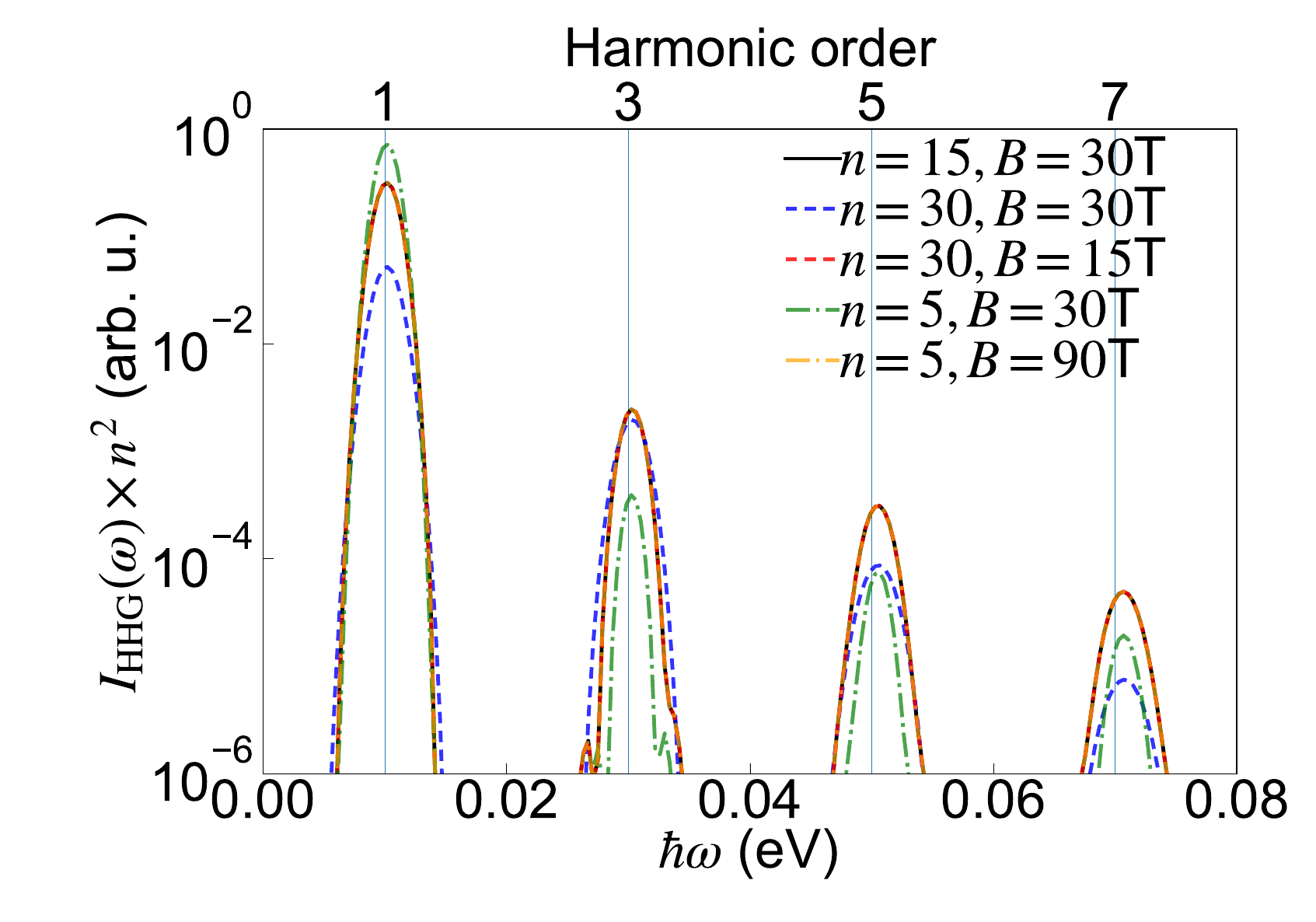} 
  \caption{HHG spectra $I_{\rm HHG}\times n^2$ of the armchair CNT simulated with the tight-binding model for indicated values of the chiral index $n$ and magnetic field $B$. 
  Note that the cases of $(n,B)=(15, 30)$,  $(n,B)=(5, 90)$ and $(n,B)=(30, 15)$ overlap each other.  We set $T=11.6$K, $T_1=98.8$fs and $T_2=19.8$fs. The parameters of the electric field are $\hbar\Omega = 10$meV, $E_0=30$kV/cm, $t_0=3.95$ps and $\sigma={658}$fs.}
  \label{fig:HHG_n_dep}
\end{figure}

 \begin{figure}[t]
  \centering
    \hspace{-0.cm}
    \vspace{0.0cm}
\includegraphics[width=85mm]{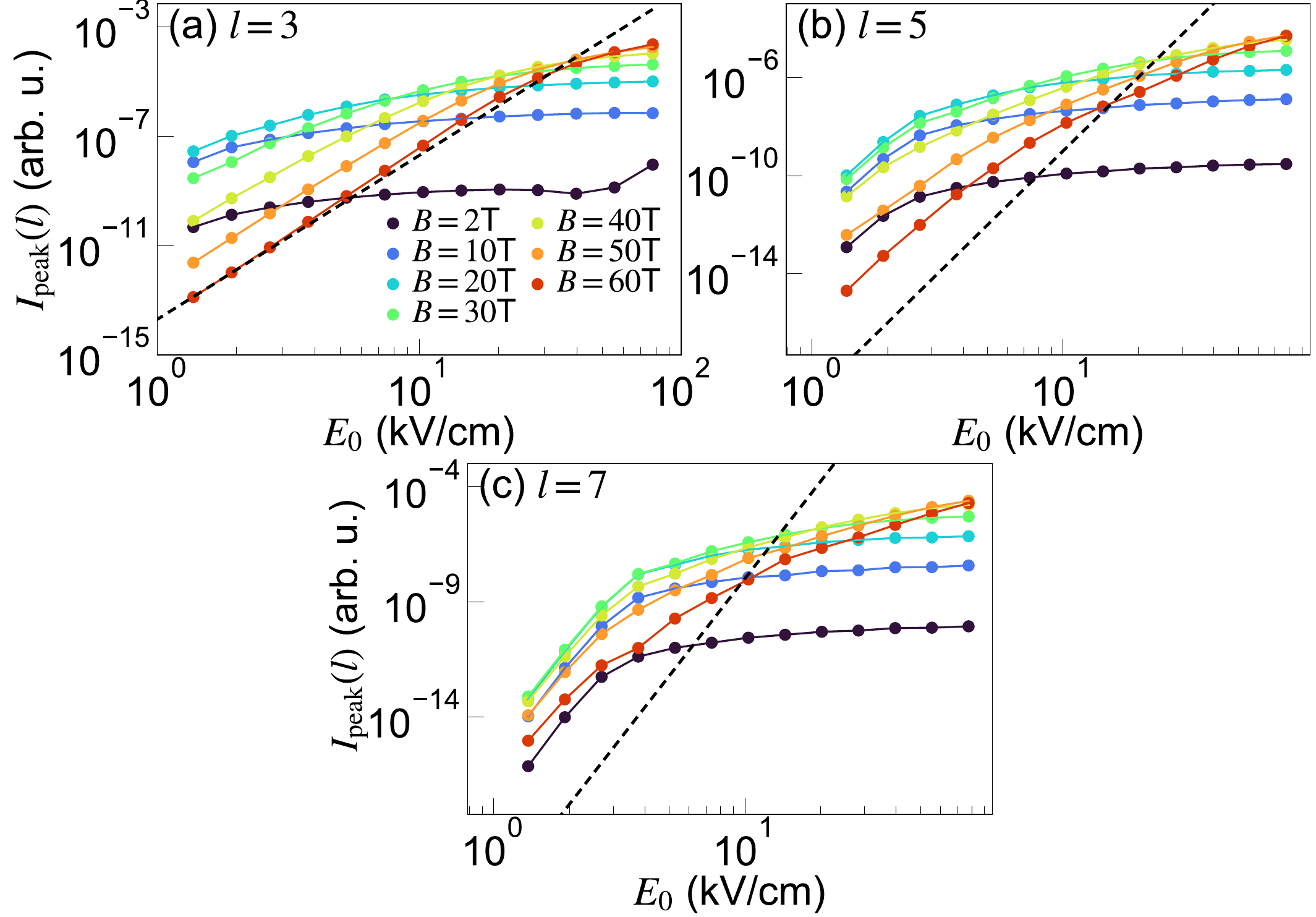} 
  \caption{The intensity at the peaks of the HHG spectra $I_{\rm peak}(l)$ as a function of the strength of the AC electric field $E_0$ for (a) $l=3$, (b) $l=5$ and (c) $l=7$.
  The dashed lines are proportional to $(E_0)^{2l}$ for each panels.  We set $(n,m)=(15,15)$, $\mu=0$, $T=11.6$K, $T_1=98.8$fs and $T_2=98.8$fs. The parameters of the electric field are $\hbar\Omega = 10$meV, $t_0=3.95$ps and $\sigma={658}$fs. }
  \label{fig:HHG_Edep_T1T2}
\end{figure}

 \begin{figure}[t]
  \centering
    \hspace{-0.cm}
    \vspace{0.0cm}
\includegraphics[width=70mm]{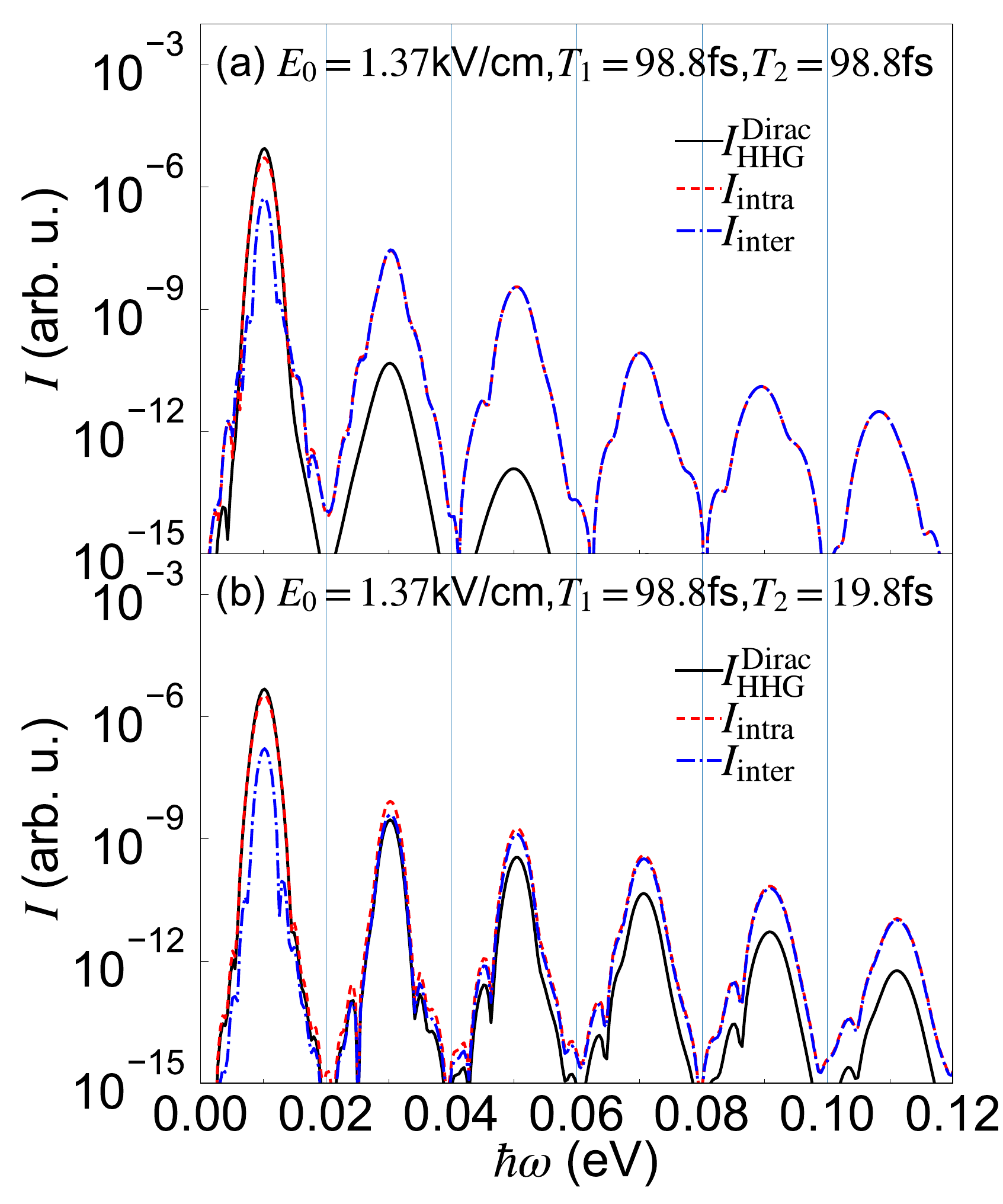} 
  \caption{Comparison of the HHG spectra obtained in different ways for (a) $(T_1,T_2)=(98.8{\rm fs},98.8{\rm fs})$ and (b) $(T_1,T_2)=(98.8{\rm fs},19.8{\rm fs})$.   
  All of $I_{\rm HHG}^{\rm Dirac}$, $I_{\rm intra}$ and $I_{\rm inter}$  are evaluated from the Dirac model considering only bands closest to $K$ or $K'$. We set $(n,m)=(15,15)$, $\mu=0$, $T=11.6$K, and $B=2$T. The parameters of the electric field are $\hbar\Omega = 10$meV, $E_0=1.37$kV/cm, $t_0=3.95$ps and $\sigma={658}$fs. }
  \label{fig:HHG_small_E_decompose}
\end{figure}

\subsection{Dependence on chiral index $n$}
As we discussed in the main text, in order to achieve the same amount of momentum shift due to the AB effect and thus to open the same size of the gap, an armchair CNT with larger $n$ requires weaker magnetic field $B$.
To exemplify this, we compared the HHG spectra for different values of the chiral index $n$ and magnetic field in Fig.~\ref{fig:HHG_n_dep}. To compare the structure of the HHG spectra for different $n$,  $I_{\rm HHG}(\omega)$ is multiplied by $n^2$. Remember that  $I_{\rm HHG}(\omega)$ is defined as the HHG intensity normalized by the number of atoms of a CNT.
In Fig.~\ref{fig:HHG_n_dep}, we take the case of $n=15$ and $B=30$T as a reference. For $n=5$, the case with the magnetic field $B=90$T shows essentially the same HHG spectrum as the reference.
As we discussed in the main text, this is because i) in the present excitation condition, only the bands closest to $K$ and $K'$ contributes to HHG and ii) the gap of the corresponding massive Dirac system is controlled by $A_B\propto B\cdot n$.
With $B=30$T, the gap is smaller compared to the reference case. For $n=30$, the case with $B=15$T shows essentially the same HHG spectrum as the reference due to the same reason. 
With $B=30$T, the gap becomes larger than the reference case. 

\subsection{Effects of $T_1$ and $T_2$}

In Fig.~\ref{fig:HHG_Edep_T1T2}, we show the intensity at the HHG peaks $I_{\rm peak}(l)$ as a function of the electric-field strength $E_0$ for $(T_1,T_2)=(98.8{\rm fs},98.8{\rm fs})$.
This figure can be directly compared with Fig.~4 in the main text, which shows the results for $(T_1,T_2)=(98.8{\rm fs},19.8{\rm fs})$.
The generic behavior is similar to Fig.~4 in the main text.
Namely, the higher order corrections lead to the saturation of the HHG intensity.
(The additional increase of $I_{\rm peak}(3)$ around $E_0\simeq 100$kV/cm for $B=2$T originates from the deviation from the linear band dispersion.)
The deviation from the perturbative regime happens at smaller $E_0$ for smaller $B$, i.e. for smaller gap.
The optimal value of $B$ that maximizes $I_{\rm peak}$ shifts to larger values with increasing $E_0$.
On the other hand, unlike the results for $(T_1,T_2)=(98.8{\rm fs},19.8{\rm fs})$, the optimal value of $B$ does not decrease continuously with decreasing $E_0$.
For small $E_0$, the HHG intensity is largest around $B=20$T, where the gap is almost twice of $\Omega$.

In order to understand the origin of the different behavior in the weak-electric-field regime between the cases of $(T_1,T_2)=(98.8{\rm fs},98.8{\rm fs})$ and  $(T_1,T_2)=(98.8{\rm fs},19.8{\rm fs})$, we compare the contributions to HHG from the interaband and intraband currents, i.e. $I_{\rm inter}$ and $I_{\rm intra}$, for a weak magnetic field, see Fig.~\ref{fig:HHG_small_E_decompose}.
When  $T_1=T_2$, $J_{\rm inter}$ and $J_{\rm intra}$ cancel each other very efficiently, and thus the total intensity $I_{\rm HHG}$ becomes much smaller than $I_{\rm inter}$ and $I_{\rm intra}$, see Fig.~\ref{fig:HHG_small_E_decompose}(a).
On the other hand, when $T_1 > T_2$, the cancelation between $J_{\rm inter}$ and $J_{\rm intra}$ is less efficient and $I_{\rm HHG}$ becomes closer to $I_{\rm inter}$ and $I_{\rm intra}$, see Fig.~\ref{fig:HHG_small_E_decompose}(b).
Thus, the total HHG intensity becomes much larger in the latter case, although both of $I_{\rm inter}$ and $I_{\rm intra}$ in the former case can be larger than those in the latter case.
This result is consistent with the expression of $\sigma^{(3)}$.
Namely, for $T_1 \neq T_2$ it diverges with $B\rightarrow 0$ due to the absence of the cancelation between $J_{\rm inter}$ and $J_{\rm intra}$.

\end{document}